\documentclass[preprint,12pt]{elsarticle}

\usepackage{amssymb}
\usepackage{amsthm}
\usepackage{mathtools}
\usepackage{graphicx}
\usepackage{float}
\usepackage{graphicx}
\usepackage{bm}
\usepackage{lineno}
\usepackage{color}
\usepackage{amsmath}
\usepackage{graphicx}
\usepackage{subcaption}
\usepackage{url}

\newcommand{\be}{\begin{equation}}
\newcommand{\ee}{\end{equation}}

\journal{XXX}

\begin{document}

\begin{frontmatter}
\title{Surrogate-based global sensitivity analysis\\
for turbulence and fire-spotting effects in\\ 
regional-scale wildland fire modeling}

\author[bcam,upv]{A. Trucchia\corref{cor1}}
\cortext[cor1]{Email: atrucchia@bcamath.org}
\author[bcam]{V. Egorova}
\author[bcam,iker]{G. Pagnini}
\author[cerfacs]{M. C. Rochoux\corref{cor2}}
\cortext[cor2]{Email: melanie.rochoux@cerfacs.fr}

\address[bcam]{BCAM -- Basque Center for Applied Mathematics,\\ 
Alameda de Mazarredo 14, E-48009 Bilbao, Basque Country, Spain}
\address[upv]{University of the Basque Country UPV/EHU,\\ 
Barrio Sarriena s/n, 48940 Leioa, Basque Country, Spain}
\address[iker]{Ikerbasque -- Basque Foundation for Science,\\ 
Calle de Mar\'ia D\'iaz de Haro 3, E-48013 Bilbao,  Basque Country, Spain}
\address[cerfacs]{CECI, University of Toulouse, CNRS, CERFACS,\\ 
42 Avenue Gaspard Coriolis, 31057 Toulouse cedex 1, France} 
           
\begin{abstract}
In presence of strong winds, wildfires feature nonlinear behavior, possibly inducing fire-spotting. We present a global sensitivity analysis of a new submodel for turbulence and fire-spotting included in a wildfire spread model based on a stochastic representation of the fireline. To limit the number of model evaluations, fast surrogate models based on generalized Polynomial Chaos (gPC) and Gaussian Process are used to identify the key parameters affecting topology and size of burnt area. This study investigates the application of these surrogates to compute Sobol' sensitivity indices in an idealized test case. The wind is known to drive the fire propagation. The results show that it is a more general leading factor that governs the generation of secondary fires. This study also compares the performance of the surrogates for varying size and type of training sets as well as for varying parameterization and choice of algorithms. The best performance was achieved using a gPC strategy based on a sparse least-angle regression (LAR) and a low-discrepancy Halton's sequence. Still, the LAR-based gPC surrogate tends to filter out the information coming from parameters with large length-scale, which is not the case of the cleaning-based gPC surrogate. For both algorithms, sparsity ensures a surrogate can be built using an affordable number of forward model evaluations, while the model response is highly multi-scale and nonlinear. Using a sparse surrogate is thus a promising strategy to analyze new models and its dependency on input parameters in wildfire applications.
\end{abstract}


\end{frontmatter}
\section*{Nomenclature}
\begin{table}[H]
		\label{tbl:main_list_abbr}
		\caption{List of abbreviations}
		\begin{center}
			\begin{tabular}{l|l}
				\hline
				\bf{Abbreviation} & \bf{Meaning} \\
				\hline
				ABL   & Atmospheric Boundary Layer \\
				FT      & Free Atmosphere \\
    			    GP      & Gaussian Process \\
				gPC   & generalized Polynomial Chaos \\
   			    LAR    & Least Angle Regression \\
   			    LSM   & Level Set Method \\
    			    MSR   & Minimum Spanning Rectangle\\
			    PDF    & Probability Density Function \\
				ROS   & Rate of Spread \\ 
				SLS    & Standard Least Squares \\
				STD   & STandard Deviation \\
			\end{tabular}
		\end{center}
\end{table}	

\newpage
\begin{table}
		\label{tbl:main_list}
		\caption{List of important static and dynamic model parameters.}
		\begin{center}
			\begin{tabular}{l|l}
				\hline
				\bf{Model quantities} & \bf{Units} \\
				\hline
   			     $\mathcal{B}(t)$, burnt area at time $t$ & -- \\ 
				$f$, PDF of the random process & $\rm m^{-2}$ \\  	
				$G(\mathrm{x};t)$, isotropic bivariate Gaussian PDF of turbulence & $\rm m^{-2}$\\
				$q(l)$, lognormal PDF of firebrand landing distance & $\rm m^{-1}$ \\					     
   				$\mathrm{x}=(x_1,x_2)$, horizontal space variable &  $\rm m$ \\
   				$\mathrm{n}_{\text{fr}}$, normal direction to the fireline & -- \\
   				$\mathrm{n}_{\text{U}}$, unit vector aligned with the mean wind direction & -- \\
  			    $t$, time  & $\rm s$ \\
				$\phi$, level-set function &  --   \\
				$\Omega$, 2--D computational domain & -- \\
				$|\Omega|$, area of the computational domain & $\rm m^{2}$ \\
				\hline
				\bf{Physical Model Parameters} & \bf{Value/Units} \\
				\hline
				$C_d$, drag coefficient & -- \\
				$D$, turbulent diffusion coefficient & $\rm m^2\,s^{-1}$ \\
  			    $g$, acceleration due to gravity & 9.8 $\rm m\,s^{-2}$\\
  			    $h$, dimension of convective cell  & 100~m \\
    				$H$, fire plume height & m \\
   				$I$, fireline intensity & $\rm kW\,m^{-1}$ \\
		      	$P_{\text{f0}}$, reference fire power &  $\rm 10^6\,W$ \\
   				$\mathrm{U}$, horizontal wind vector field at mid-flame height &  $\rm m\,s^{-1}$ \\
 			     $\left\lVert\mathrm{U}\right\rVert$, horizontal wind magnitude &  $\rm m\,s^{-1}$\\
   				$\mathcal{V}$, rate of spread & $\rm m\,s^{-1}$\\
				$z_p$, $p$th percentile  & 0.45 \\
 			     $\Delta h_c$, heat of combustion of wildland fuels & 18,620 $\rm kJ\,kg^{-1}$\\
				$(\mu, \sigma)$ , parameters of the log-normal PDF $q(l)$ & --  \\
				$\rho_a$, air density & 1.2 $\rm kg\,m^{-3}$\\
				$\rho_f^*$, wildland fuel density (\emph{Pinus Ponderosa})& 542 $\rm kg\,m^{-3}$\\
				$\omega_0$, oven-dry mass of wildland fuel & 2.243 $\rm kg\,m^{-2}$\\
				$\tau$, ignition delay of firebrands & $\rm s$ \\
				$\chi$, air thermal diffusivity & $ 2 \cdot 10^{-5}\,{\rm m^2\,s^{-1}}$ \\
			    $\Delta T$, temperature difference of convective cell & 800-923~K \\
			    $\ell$, firebrand landing distance & $\rm m$ \\
			    $\nu$, kinematic viscosity & $1.5 \cdot 10^{-5} \, {\rm m^2\,s^{-1}}$\\
			    $\gamma$, thermal expansion coefficient & $\rm K^{-1}$ \\
				$\alpha_H, \beta_H, \gamma_H, \delta_H$, coefficients for fire plume height $H$ & -- \\
			\end{tabular}
		\end{center}
\end{table}		

\begin{table}
		\label{tbl:main_list2}
		\caption{List of important algorithmic parameters.}
		\begin{center}
			\begin{tabular}{l}
			    $A_t$, burnt area ratio at time $t$ \\
   				$d$, dimension of the stochastic space ($d = 3$) \\
				$\mathcal{D}_N$, training set of size $N$ \\
				$\mathcal{M}$, forward model \\
				$\mathcal{M}_{\text{pc}}$, gPC-expansion \\
				$N$, size of the training set \\
				$P$, total polynomial order \\
				$q$, hyperbolic truncation parameter \\
				$r$, number of terms in the surrogate basis \\
				$S_t$, minimum spanning rectangle ratio at time $t$ \\			
				$\mathrm{y}$, quantity of interest \\	 
			    $\mathrm{\widehat{y}}$, estimate of the quantity of interest $\mathrm{y}$ \\	 
			    	$\mathrm{y}^{(k)}$, $k$th realization of the quantity of interest $\mathrm{y}$ \\	 
			    	$\mathcal{A}$, set of selected multi-indices in gPC-expansion \\
   				$\boldsymbol{\alpha}$, multi-index for gPC-expansion\\
   				$\delta$, Kronecker delta-function \\
				$\boldsymbol{\theta} = (\theta_1, \cdots, \theta_d)$, vector of uncertain input parameters, $\left[ \left\lVert\mathrm{U}\right\rVert, I, \tau \right] $ or $\left[ \mu,  \sigma, D \right]$ \\
				$\boldsymbol{\theta}^{(k)}$, $k$th realization of the uncertain input vector $\boldsymbol{\theta}$ \\
				$\boldsymbol{\zeta} = (\zeta_1, \cdots, \zeta_d)$, vector $\boldsymbol{\theta}$ in standard probabilistic space \\
				$\rho_{\theta_i}$, marginal PDF of $i$th input parameter in $\boldsymbol{\theta}$ \\  
			    $\boldsymbol{\rho}_{\boldsymbol{\zeta}}$, joint PDF of $\boldsymbol{\theta}$ in standard probabilistic space \\  
				$\Psi_{\boldsymbol{\alpha}}$, $\boldsymbol{\alpha}$th basis function for surrogate model \\
   			    $\Phi_{\alpha_i}$, $i$th one-dimensional basis function \\
				$\gamma_{\boldsymbol{\alpha}}$, $\boldsymbol{\alpha}$th coefficient in the surrogate basis \\
				$\boldsymbol{\gamma}$, vector of surrogate coefficients \\
				$\left(\omega^{(k)},\boldsymbol{\zeta}^{(k)}\right)$, $k$th quadrature weight and root \\
				$\ell_{\text{gp}}$, correlation length-scale for GP-model \\
				$\sigma_{\text{gp}}$, observable standard deviation for GP-model  \\
				$\tau_{\text{gp}}$, nugget effect for GP-model \\			
				$\pi(\boldsymbol{\theta},\boldsymbol{\theta}')$, correlation kernel for GP-model \\		
				$\epsilon_{\text{emp}}$, empirical training error \\
				$Q_2$, cross-validation predictive coefficient \\
			\end{tabular}
		\end{center}
\end{table}

\clearpage
\section{Introduction}\label{sec:intro}

Despite our recent progress in computer-based wildland fire spread modeling and remote sensing technology, our general understanding of wildland fire behavior remains limited. This is mainly due to the complexity of wildfire dynamics that results from multi-scale interactions between biomass pyrolysis, combustion and turbulent flow dynamics, heat transfer as well as atmospheric dynamics~\cite{viegas1998,linn2002,mell2007,strada2012,finney2013,mcallister2014}. Turbulence plays an important role: wildland fires release large amounts of heat that lead to the development of a turbulent flow in the vicinity of the flame zone and thereby enhance the heat transfer to unburnt fuel, boosting biomass fuel ignition, combustion and fire spread. There is therefore a strong coupling between wildland fires and micrometeorology~\cite{clark1996,potter2002,potter2012a,potter2012b,mandel2011,filippi2013}. When extreme conditions are met in complex terrain such as canyons in combination with strong winds and severe drought, highly destructive fires referred to as ``megafires" can develop~\cite{viegas2004,viegas2010,nijhuis2012,cruz2012}. For such fires, a massive buoyant smoke plume forms above the flame zone modifying micro-meteorological conditions~\cite{paugam2016} and thereby fire spread conditions. Windborne embers can be transported over large distances, causing fire spotting and further ignitions downstream from the current fire, leading to multiple ``spot fires" that are difficult to stop by firefighters and that dramatically increase fire danger. Turbulence and fire-spotting result from very nonlinear effects that are still poorly understood and that have been identified as a valuable research target with direct applications in fire emergency response, especially at wildland-urban interface~\cite{koo2010}.

The representation of these processes is beyond the scope of current operational wildfire spread models. At regional scales (i.e.~at scales ranging from a few tens of meters up to several hectares), a wildland fire is indeed represented as a two-dimensional propagating interface (referred to as the ``fire front" or ``fireline") separating the burnt area to the unburnt vegetation; the local propagation speed is called the ``rate of spread" (ROS). This front representation is the dominant approach in current wildfire spread simulators such as FARSITE~\cite{finney1998}, FOREFIRE~\cite{filippi2009,filippi2013}, PROMETHEUS~\cite{tymstra2010}, PHOENIX RapidFire~\cite{chong2013}, SFIRE~\cite{mandel2011} or ELMFIRE~\cite{lautenberger2013}. These simulators rely on an empirical parameterization of the ROS that is derived from steady-state assumption and that is an analytic function of biomass fuel properties, topographical properties and micro-meteorological conditions~\cite{sullivan2009}. The ROS submodel is included in an Eulerian or Lagrangian front-tracking solver to simulate the fireline propagation. This approach is limited in scope~\cite{gollner2015,cruz2013b,cruz2018} due to the large uncertainties associated with the input parameters of the ROS model~\cite{jimenez2008,liu2015}, which can be partially reduced by integrating real-time fire front measurements through data assimilation~\cite{mandel2011,denham2012,rochoux2013,rochoux2014,artes2014,rochoux2015,zhang2017,rochoux2017}. This approach is also limited due to the lack of knowledge on the physics of the fire problem~\cite{finney2013}, in particular on the processes associated with turbulence and fire-spotting.

These modeling limitations at regional scales have motivated investigation of turbulence and fire-spotting effects both from experimental and modeling viewpoints~\cite{manzello2008,sardoy2008,kortas2009,perryman2009,koo2010,perryman2013,tohidi2015,tohidi2016phd,tohidi2017}. To better characterize these nonlinear processes, there is a need to develop new submodels including the effects of random processes such as turbulence and fire-spotting in operationally-oriented wildfire spread models. This is one of the objectives of the work proposed in~\cite{pagnini2012,pagnini2014_smai,pagnini2014_nhess,kaur2016}, which introduces a randomized representation of the fireline. A novel family of reaction-diffusion equations have been developed to link front models to reaction-diffusion ones and thereby integrate the effects of random processes in fire models. The front propagation is randomized by adding to the driving function, a random displacement distributed according to a probability density function (PDF) corresponding to heat turbulent transport and fire-spotting landing distance. The driving equation of the resulting averaged process is analogous to an evolution equation of the reaction-diffusion type, where the ROS controls the source term. In absence of random process, the model is identical to the one given by the standard wildfire spread model, which is only driven by the ROS analytic function.

Including new modeling components in wildfire spread simulators adds some complexity and in particular introduces new model parameters. There is therefore a strong need to perform sensitivity analysis to analyze in a rigorous way the model structure, i.e.~the dependency between the input parameters and the simulated quantities of interest (the topology and the extension of the burnt area at a given time in the present study). The objective in such an extensive global sensitivity analysis is two-fold. First, sensitivity analysis identifies the most influential parameters on the model predictions over a wide range of values for the model parameters, ranks them by order of importance and spots unimportant parameters~\cite{saltelli2007,storlie2009,lamboni2011}. This is helpful to provide hints and guidelines about the physical processes that are essential to account for to track wildland fire behavior. Second, sensitivity analysis is a mandatory step to select which are the estimation targets to consider when the wildfire spread model is integrated in a data assimilation framework to produce short-term predictions of wildfire behavior; the model parameters shall indeed be uncertain and the quantities of interest shall be sensitive to changes in these model parameters to ensure data assimilation is efficient~\cite{ruiz2013,rochoux2014,zhang2017}.
  
When relying on stochastic non-intrusive methods (meaning that no modification of the physical model, also referred to as the ``forward model", is required), global sensitivity analysis requires the use of an ensemble of model evaluations. This procedure can be divided into three steps: (1)~characterization of the variability in the model parameters based on available information and statistical sampling to obtain an ensemble of parameter values; (2)~multiple evaluations of the forward model while accounting for the identified uncertainties to obtain an ensemble of quantities of interest (the forward model is used as a ``black-box"); and (3)~computing Sobol' sensitivity indices~\cite{sobol1993} that provides a relative measure of how the variability of the model response is affected by the variability in each uncertain parameter (this variability is measured in terms of variance). Computing these Sobol' indices therefore requires to have access to an accurate mapping between the uncertain inputs and the quantities of interest. This is computationally intensive when using standard Monte Carlo sampling method since this method features a slow convergence rate and thus requires a large ensemble to obtain reliable statistics. The cost of global sensitivity analysis is significantly reduced when the forward model is replaced by a surrogate model that mimics its response for the considered range of the model parameters.  The formulation of such a surrogate requires a limited number of model evaluations, referred to as the ``training set". Then the surrogate can be evaluated multiples times at almost no cost to evaluate uncertainties in the quantities of interest and/or perform sensitivity analysis~\cite{sudret2008,marrel2009,iooss2016,legratiet2017,owen2017}.

There are various ways of formulating a surrogate. In the present work, we focus our attention on generalized polynomial chaos (gPC) expansions~\cite{ciriello2013,despres2013,dubreuil2014,sudret2008,xiu2010} and Gaussian process (GP) models~\cite{delozzo2017,legratiet2014,legratiet2017,marrel2009,marrel2015,oakley2004,rasmussen2006}. The gPC-approach formulates a polynomial expansion, in which the basis is defined according to the PDF of the uncertain parameters and in which the associated weights directly relate to the statistics of the quantities of interest. This implies that by construction the quantities of interest are projected upon the same basis as the input parameters. The GP-approach adopts a different viewpoint by considering the simulated quantities of interest as a realization of a Gaussian stochastic process conditioned by the training set. This stochastic process is fully characterized with mean and covariance kernel functions, which rely on the estimation of hyperparameters. Both gPC and GP surrogates are compared in the literature for uncertainty quantification and sensitivity analysis studies~\cite{legratiet2017,owen2017,roy2018a,schoebi2015}. In particular, \cite{legratiet2017} showed that for a given size of the training set, gPC and GP surrogates feature similar predictive quality for application in structural mechanics. Still, they emphasized that the ranking between gPC and GP approaches remains problem-dependent. It is thus of great interest to compare these approaches for application in wildland fires.

In wildland fire applications, the performance of the gPC-approach has already been demonstrated within the framework of data assimilation to reduce the computational cost of sequential parameter estimation~\cite{rochoux2014,rochoux2017}. However, the gPC-algorithm relied on the use of a full basis and a standard spectral projection method. Building the surrogate this way may be too costly for high-dimensional problems, i.e.~when the number of uncertain parameters increases. There exists more advanced gPC-strategies in the literature to reduce the number of elements in the gPC basis and thus reduce the required size of the training set. These strategies explore alternative ways of selecting the polynomials, for instance by limiting high-order interaction terms (e.g.~hyperbolic truncation scheme; see~\cite{blatman2009phd}) or through the construction of sparse bases using least-square regression projection methods~\cite{blatman2011,migliorati2013}. Due to the multiple sources of uncertainty in wildland fire models, there is a strong need to evaluate the performance of gPC and GP approaches, i.e.~for varying size and type of the training set as well as for varying parameterization and choice of the surrogate algorithms. In the present study, the objective is to determine what is the best surrogate strategy to compute Sobol' sensitivity indices and thereby examine the relevance of the parameters that are part of the turbulence and fire-spotting submodel included in the wildfire spread model~\cite{pagnini2014_nhess}. Our objective is to identify the key parameters affecting the topology and the size of the burnt area that is simulated by an Eulerian-type fire spread model ({\tt LSFire+}) and that corresponds to an idealized test case. For this purpose, we compare the performance of gPC-expansion and GP-model in their standard and sparse versions for a fixed size of the training set with different designs of experiment (Monte Carlo random sampling, quasi-random Halton's sequence, quadrature rule); a convergence study is carried out to determine the required size of the training set to ensure accuracy.

The structure of the paper is as follows. Section~\ref{sec:model} introduces the wildfire spread model, the main sources of uncertainty, the quantities of interest and the idealized test case study. The gPC and GP approaches are detailed in Section~\ref{sec:uq} along with statistical analysis tools and error metrics. Section~\ref{sec:results} presents the results of the comparative study between gPC and GP algorithms for different types of truncation, projection and training set. Conclusions and perspectives are given in Section~\ref{sec:ccl}.

\section{Wildland Fire Model and Sources of Uncertainties}\label{sec:model}

\subsection{Forward Model}\label{sec:model_forward}

We focus the present study on Eulerian-type wildfire spread model ({\tt LSFire+}) based on level-set methods~\cite{sethian1999,osher2003,mallet2009}. This is similar to the approach adopted in the ELMFIRE fire simulator~\cite{lautenberger2013,lautenberger2017} or the WRF-SFIRE coupled fire-atmosphere system~\cite{mandel2011}.

\subsubsection{Deterministic Front Propagation}\label{sec:model_forward_deterministic}

To represent the time-evolving burning active areas over the computational domain $\Omega \subset \mathbb{R}^2$, we introduce an implicit function $\phi \equiv \phi(\mathrm{x},t)$ as the fireline marker with $\phi: \Omega \times [0; +\infty[ \rightarrow \mathbb{R}$. The fireline is identified as the contour line $\phi(\mathrm{x},t) = \phi^*$ referred to as the ``level set". We thus denote the time-evolving two-dimensional burnt area as $\mathcal{B}(t) = \lbrace \mathrm{x} = (x_1,x_2) \in \Omega \, | \, \phi(\mathrm{x},t) > \phi^*\rbrace$. Note that at a given time $t$, $\mathcal{B}(t)$ can represent more than one independently-evolving bounded area.

The temporal evolution of the level set $\phi(\mathrm{x},t) = \phi^*$ is governed by the Eikonal equation
\begin{equation}
\label{eq:levelset}
\frac{\partial \phi}{\partial t}(\mathrm{x},t) = \mathcal{V}(\mathrm{x},t)\,\left\lVert\nabla \phi(\mathrm{x},t)\right\rVert, \quad \phi(\mathrm{x},t_0) = \phi_0(\mathrm{x}), \quad \mathrm{x} \in \Omega,\; t \geq t_0,
\end{equation}
where $\mathcal{V}$ corresponds to the ROS parameterization that is a function of the wind field $\mathrm{U}(\mathrm{x},t)$, orography and biomass fuel conditions, and where $\phi_0(\mathrm{x})$ is the initial condition at time $t_0$. The propagation of the fireline is assumed to be directed towards the normal direction to the front $\mathrm{n}_{\text{fr}} \equiv \mathrm{n}_{\text{fr}}(\mathrm{x},t) = - \nabla\phi(\mathrm{x},t)/\left\lVert \nabla \phi(\mathrm{x},t)\right\rVert$.

\subsubsection{Random Front Formulation}\label{sec:model_forward_random}

The stochastic approach that is adopted in the present study is based on the idea of splitting the motion of the fireline into a drifting part and a fluctuating part~\cite{pagnini2014_nhess,mentrelli2015,kaur2016}. The drifting part corresponds to the resolution of the deterministic problem in Eq.~(\ref{eq:levelset}). The fluctuating part results from a comprehensive statistical description of the dynamic system, which includes random effects in agreement with the physics of the system. As a consequence, the fluctuating part can have a non-zero mean, implying that the drifting part does not correspond to the average motion.

The motion of each burning point can be random due to the effect of turbulence and/or fire-spotting. The effective indicator function, $\phi_{\text{e}}(\mathrm{x},t):\mathcal{B} \times [0,+\infty[ \rightarrow [0,1]$ emerges from the superposition of the front weighted by the distribution of fluctuations around the deterministic front, i.e.
\begin{equation}
\phi_{\text{e}}(\mathrm{x},t)  = \int_{\mathcal{B}}\,\phi(\overline{\mathrm{x}},t)\,f(\mathrm{x}; t | \overline{\mathrm{x}})\,d\overline{\mathrm{x}},
\label{eq:phi_e}
\end{equation}
where $f(\mathrm{x}; t | \overline{\mathrm{x}})$ denotes the PDF of the displacement of the active burning points around the mean position $\overline{\mathrm{x}}$. An arbitrary threshold value $\phi_{\text{e,fr}}$ is used as the criterion to separate burnt area and unburnt area. The effective burnt area is therefore defined as $\mathcal{B}_{\text{e}}(\mathrm{x},t) = \left\lbrace \mathrm{x} \in \mathcal{B} \mid \phi_{\text{e}}(\mathrm{x},t) > \phi_{\text{e,fr}}\right\rbrace$.

Note that the PDF $f(\mathrm{x}; t | \overline{\mathrm{x}})$ is associated with two independent random variables representing turbulence and fire-spotting, with fire-spotting a downwind phenomenon acting along the wind direction. $f(\mathrm{x}; t | \overline{\mathrm{x}})$ is expressed as
\begin{equation}
f(\mathrm{x}; t | \overline{\mathrm{x}}) = 
\begin{cases}
\displaystyle
\int_{0}^{\infty} G(\mathrm{x} - \overline{\mathrm{x}} - \ell\,\mathrm{n}_{U}; t)\,q(\ell; t)\,d\ell\,, &  \mathrm{n}\cdot \mathrm{n}_{\text{U}} \geq 0 \,,\\ \\
G(\mathrm{x} - \overline{\mathrm{x}}; t), & \text{otherwise} \,,
\end{cases}
\label{eq:pdf}
\end{equation}
where $\mathrm{n}_{\text{U}}$ is the unit vector aligned with the mean wind direction, where $G(\mathrm{x} - \overline{\mathrm{x}}; t)$ is the PDF associated with turbulent diffusion, and where $q(\ell; t)$ is the PDF associated with firebrand landing distance $\ell$. We follow the same choices as in \cite{pagnini2014_nhess,mentrelli2015,kaur2016}. Hence, we assume that turbulent diffusion is isotropic and represented as a bivariate Gaussian PDF
\begin{equation}
G(\mathrm{x} - \overline{\mathrm{x}}; t) = \frac{1}{4\pi\,D\,t}\exp{\left\lbrace \frac{(x_1 - \overline{x}_1)^{2} + (x_2 - \overline{x}_2)^{2}}{4\,D\,t}\right\rbrace}, 
\label{eq:gaussian}
\end{equation}
where $D$ is the turbulent diffusion coefficient. We also assume that the downwind distribution of the firebrands follows a log-normal distribution
\begin{equation}
q(\ell;t) = \frac{1}{\sqrt{2\pi}\,\sigma\,\ell}\exp\left\lbrace -\frac{(\ln \ell/\ell_0 - \mu)^{2}}{2\,\sigma^{2}}\right\rbrace, 
\label{eq:lognormal}
\end{equation}
where $\mu \equiv \mu(t) = \left\langle \ln \ell/\ell_0\right\rangle $ and $\sigma \equiv \sigma(t) = \left\langle(\ln \ell/\ell_0 -\mu)^{2}\right\rangle$ are the mean and the standard deviation (STD) of $\ln \ell/\ell_0$, respectively, and where $\ell_0$ is a unit reference length.

Since fuel ignition due to hot air and firebrands is not instantaneous, a suitable criterion related to ignition delay is introduced. 
This criterion is based on  heating-before-burning mechanism as follows:
\begin{equation}
\psi(\mathrm{x},t) = \int_{0}^{t}\,\phi_{\text{e}}(\mathrm{x},\eta)\,\frac{d\eta}{\tau}, 
\label{eq:psi}
\end{equation}
where $\psi(\mathrm{x},0) = 0$ corresponds to the initial unburnt biomass fuel, and where $\tau$ is a reference time for ignition delay. A point $\mathrm{x}$ is considered ignited at time $t$ when $\psi(\mathrm{x},t) = 1$. In this case, $\mathrm{x} \in \mathcal{B}(t)$. 

\subsubsection{Rate of Spread Submodel and Test Case Study}\label{sec:model_testcase}

Since the focus is here on sensitivity analysis methodology, we consider a simplified version of the ROS parameterization required in Eq.~(\ref{eq:levelset}). The maximum value of the ROS, $\mathcal{V}(\mathrm{x},t)$, is specified by means of Byram's formula~\cite{byram1959,alexander1982}:
\begin{equation}
\mathcal{V}_0 = \frac{I}{\Delta h_c\,\omega_0},
\label{eq:byram}
\end{equation}
where $I$~[kW\,m$^{-1}$] is the fireline intensity, $\Delta h_c$~[kJ\,kg$^{-1}$] is the fuel heat of combustion and $\omega_0$~[kg\,m$^{-2}$] is the oven-dry mass of fuel consumed per unit area in the active flaming zone. By analogy to the approach adopted in~\cite{pagnini2014_nhess}, the effect of the near-surface wind $\mathrm{U}$ on the ROS is accounted for through a corrective factor $f_{\text{w}}$ as follows:
\begin{equation}
\mathcal{V} = \mathcal{V}_{0}\, \frac{(1+f_{\text{w}})}{\alpha_{\text{w}}}, 
\label{eq:ros_wind}
\end{equation} 
where $f_{\text{w}}$ is computed following the choices made in the {\tt fire-Lib} and {\tt Fire Behaviour SDK} libraries (\url{http://fire.org}; see also~\cite{mandel2011}, in the case of the NFFL -- Northern Forest Fire Laboratory -- Model 9), and where $\alpha_{\text{w}}$ is a suitable angle parameter for ensuring that the maximum ROS in the upwind direction is equal to the ROS prescribed by Byram's formula (\ref{eq:byram}).This choice makes the ROS dependent on the wind direction rather than on its magnitude to constrain the well-known dominant role of the wind in the fire propagation and to allow for the emergence, if they exist, of second-order effects due to other factors.

In the present study, we consider an idealized test case of wildland fire. The computational domain is $7,200~\text{m} \times 6,000~\text{m}$. Terrain is flat. Vegetation is homogeneous. The wind is uniform and constant. Fire ignition is represented as a circular front characterized by a radius $r_c = 130$~m and a center located at $\mathrm{x}_c = (1,500~\text{m}; 3,000~\text{m})$.

\subsection{Model Input Description}\label{sec:model_var_in}

The set of uncertain parameters is noted $\boldsymbol{\theta} \in \mathbb{R}^{d}$, where $d$ is the number of parameters to consider for sensitivity analysis. We consider two different sets of uncertain model parameters in the present work with $d = 3$. To carry out sensitivity analysis, we need to prescribe a PDF representing the statistics of each parameter and thereby its variability; this corresponds to step.~(1) discussed in the Introduction.

\subsubsection{Sensitivity analysis for macroscopic/microscopic quantities}

The first set of parameters mixes macroscopic and microscopic quantities: the wind speed magnitude $\left\lVert U \right\rVert$, the fireline intensity $I$ and the ignition delay $\tau$. Sensitivity analysis with $\boldsymbol{\theta} = \left(\left\lVert\mathrm{U}\right\rVert, I, \tau\right)^T$ corresponds to a preliminary step: we consider uniform marginal distributions that spanned around the mean values adopted in previous work~\cite{pagnini2014_nhess,mentrelli2015,kaur2016}, see Table~\ref{tab:inputs1_pdf}.
\begin{table}[h]
\centering
\caption{Ranges of variation and uniform marginal PDFs for $\boldsymbol{\theta} = \left(\left\lVert\mathrm{U}\right\rVert, I, \tau\right)^T$. Note that the uniform distribution is formulated as $\mathcal{U}\left(a; b\right)$ with $a$ the minimum value and $b$ the maximum value of the parameter.}
\begin{tabular}{cc}
 \hline
Parameter & Uniform distribution \\
\hline
 Wind $\left\lVert\mathrm{U}\right\rVert$~[m\,s$^{-1}$]  & $\mathcal{U}\left(6; 14\right)$ \\ 
Fireline intensity $I$~[kW\,m$^{-1}$] & $\mathcal{U}\left(15,000; 25,000\right)$ \\ 
Reference time for ignition delay $\tau$~[s] & $\mathcal{U}\left(0.6; 1.4\right)$ \\ 
 \hline
\end{tabular} 
\label{tab:inputs1_pdf}
\end{table}

\subsubsection{Sensitivity analysis for microscopic parameters}

The focus of the present work is to explore the dependence of the wildfire spread model on a set of microscopic variables. We therefore determine a suitable Bayesian description for the uncertain parameters $\boldsymbol{\theta} = \left(\mu, \sigma, D\right)^T$, which relate exclusively to the fluctuating part of the forward model. Recall that $\mu$ and $\sigma$ are two parameters of the log-normal PDF $q(\ell;t)$ (Eq.~\ref{eq:lognormal}) that describes the ember landing position. Recall also that $D$ is the diffusive coefficient of turbulent hot air involved in the Gaussian PDF $G(\mathrm{x} - \overline{\mathrm{x}}; t)$ (Eq.~\ref{eq:gaussian}) that describes turbulent diffusion. Some functional dependence is explored for each parameter and their marginal PDFs are determined using a Monte Carlo random sampling. The resulting Beta-distributions are summarized in Table~\ref{tab:inputs2_pdf}.

\paragraph{Physical parameterization.}

We assume that all turbulent processes are represented in the forward model through the standalone turbulent diffusion coefficient $D$. We only consider turbulent fluctuations, implying that the estimation of $D$ is independent of the wind $\mathrm{U}$. Since we consider a flat terrain and an extension of the wildland fire that is not limited to the computational domain $\Omega$ under consideration, we assume horizontal isotropy. Even though an exact estimation of $D$ is beyond the scope of the present study, a quantitative estimation of $D$ is required to carry out sensitivity analysis related to turbulence and fire-spotting. $D$ corresponds to the turbulent heat convection generated by the fire. The ratio between the total heat transfer and the heat molecular conduction is widely known as the Nusselt number, ${\rm Nu}=(D+\chi)/\chi$, where $\chi$ is the air thermal diffusivity. The relation between the Nusselt number ${\rm Nu}$ and the Rayleigh number ${\rm Ra}$ (i.e.~ratio between convection and heat conduction) is given by the experimental correlation ${\rm Nu} \simeq 0.1\,{\rm Ra}^{\beta_{\text{RA}}}$ with ${\rm Ra} = \gamma\,\Delta T \,g \,h^3/(\nu \chi)$ ($\gamma$ is the thermal expansion coefficient, $\Delta T$ is the temperature difference in the convective cell, $h$ is the dimension of the convective cell, $g$ is the gravity constant and $\nu$ is the kinematic viscosity). Thus, the turbulent diffusion coefficient $D$ is computed in this work as
\begin{equation}
D \simeq 0.1 \, \chi \, \left[\frac{\gamma\,\Delta T\,g\,h^3}{\nu \chi}\right]^{1/3} - \chi,
\label{eq:diffcoeff}
\end{equation} 
with $\chi = 2 \cdot 10^{-5}\,{\rm m^2\,s^{-1}}$, $\gamma = 3.4 \times 10^{-3}\,{\rm K}^{-1}$, $g=9.8 \, {\rm m\,s^{-2}}$ and $\nu=1.5 \times 10^{-5} \, {\rm m^2\,s^{-1}}$. To define the range of variation of $D$, we introduce some assumptions. The heat transfer is considered in the horizontal plane, perpendicular to the vertical ``heating wall" embodied by the fire; the length scale of the convective cell is assumed to be $h = 100\,\text{m}$~\cite{kaur2016}, and $\Delta T$ varies from 800 to 923~K. Note that the relation between the Rayleigh number and the Nusselt number is highly sensitive to the scaling exponent $\beta_{\text{RA}}$ due to the power-law. Libchaber's experiments found $\beta_{\text{RA}} \thickapprox 2/7$ instead of $1/3$. In~\cite{niemela_etal-nat-2000}, the relation $\text{Nu} = 0.146\,\text{Ra}^{0.299}$ is proposed for $\text{Ra} > 5 \cdot 10^{7}$; for higher values of $\text{Ra}$, it is recommended to use $\beta_{\text{RA}} = 0.3$.

The fire-spotting parameterization introduced in~\cite{kaur2016} is adopted in this work. So firebrand transport is characterized through the log-normal parameters $\mu$ and $\sigma$. $\mu$ describes firebrand lofting inside the convective column. The relative density and the atmospheric drag impact the buoyant forces acting on the firebrands; hence, it is appropriate to include these quantities in the definition of $\mu$ to describe the maximum allowable height for each firebrand for varying fireline intensity. The density ratio $\rho_{\text{a}}/\rho_{\text{f}}$ also limits the maximum allowable height for each firebrand. $\mu$ is thus defined as
\begin{equation}
\mu = H\,\left(\frac{3\,\rho_{\text{a}}\,C_{\text{d}}}{2\,\rho^*_{\text{f}}}\right)^{1/2},
\label{eq:mu}
\end{equation}
where $H$~[m] represents the plume height, which is related to the maximum loftable height $H_p$ via the relation  $H_p = \lambda\,H$, and where $\rho^*_f = \rho_f / \lambda^2$~[kg\,m$^{-3}$] is the biomass fuel density that accounts for the correlation factor $\lambda$ between smoke plume height and maximum allowable height for firebrands. We adopt the analytic formulation of $H$ with respect to the fireline intensity $I$ used in~\cite{sofiev2012}, i.e.
\begin{equation}
H = \alpha_H\,H_{\text{ABL}} + \beta_H\,\left(\frac{I}{dP_{\text{f0}}}\right)^{\gamma_H}\,\exp\left(\delta_H\,\frac{N^2_{\text{FT}}}{N^2_0}\right),
\label{eq:height}
\end{equation}
where $\alpha_H$, $\beta_H$, $\gamma_H$ and $\delta_H$ are empirical constant parameters, $P_{\text{f0}}$~[W] is the reference fire power ($P_{\text{f0}} =10^6\,W$), $H_{\text{abl}}$~[m] is the height of the atmospheric boundary layer (ABL), and the subscript FT stands for free troposphere.

The parameter $\sigma$ characterizes the wind-aided transport of firebrands after they are ejected from the convective column. In a wind-driven regime of fire-spotting, the flight path of the firebrands is affected by their size, and firebrands beyond a critical size cannot be steered by the prevailing wind. This critical size is defined as the maximum liftable radius $r_{\text{max}} = ||\mathrm{U}||^2/g$. It is interesting to note that the dimensionless ratio $||\mathrm{U}||^2/(rg)$ ($r$ is the brand radius) is also known as the Froude number: it quantifies the balance between inertial and gravitational forces applying on firebrands. So $\sigma$ is computed as
\begin{equation}
\sigma = \frac{1}{2 z_{\text{p}}}\ln\left(\frac{\left\lVert\mathrm{U}\right\rVert^2}{rg}\right).
\label{eq:sigma}
\end{equation}
Note that $z_p$ corresponds to the $p$th percentile and can be estimated from the $z$-tables ({\tt http://www.itl.nist.gov/div898/handbook/eda/section3/eda3671.htm}). We assume that the $p$th percentile represents the maximum landing distance for firebrands under different situations and no ignition is possible beyond this cut-off. The cut-off criteria is chosen empirically so that $z_p = 0.45$ as in~\cite{kaur2016}, which corresponds to the 67th percentile point.

\paragraph{Statistical Description.}

The following strategy is adopted to obtain a statistical description of these three parameters $\lbrace D, \sigma, \mu \rbrace$, which depend on a large set of subparameters.

The subparameters are perturbed around their nominal values found in the literature following uniform PDFs. $D$ is computed following Eq.~(\ref{eq:diffcoeff}). To obtain a range of variation for $D$, we modify the temperature difference in the convective cell $\Delta T$ and the dimension of the convective cell $h$. As for parameters $\sigma$ and $\mu$, they are computed following Eqs.~(\ref{eq:mu})--(\ref{eq:sigma}). We modify the following parameters: $\alpha_H$, $\beta_H$, $\gamma_H$ , $\delta_H$, $H_{\text{abl}}$ in Eq.~(\ref{eq:height}); $\rho_{\text{a}}$, $\rho_{\text{f}}$ in Eq.~(\ref{eq:mu}); $z_{\text{p}}$ and $r$ in Eq.~(\ref{eq:sigma}). All the identified subparameters are associated with a uniform PDF. For the parameters $\alpha_H$, $\beta_H$, $\gamma_H$ and $\delta_H$, the extrema of the uniform PDF correspond to the highest and lowest values encountered in all the possible configurations described in~\cite{sofiev2012}, accounting for both ABL and FT regimes. $\Delta T$ varies in the range $[800; 923~\text{K}]$. For all other parameters, we use a uniform PDF, where the extrema are defined such as adding a perturbation of $20~\%$ to the values adopted in~\cite{kaur2016}. 

\begin{figure}[h!]
\centering
\begin{subfigure}{0.5\textwidth}
\includegraphics[width=0.9\linewidth,]{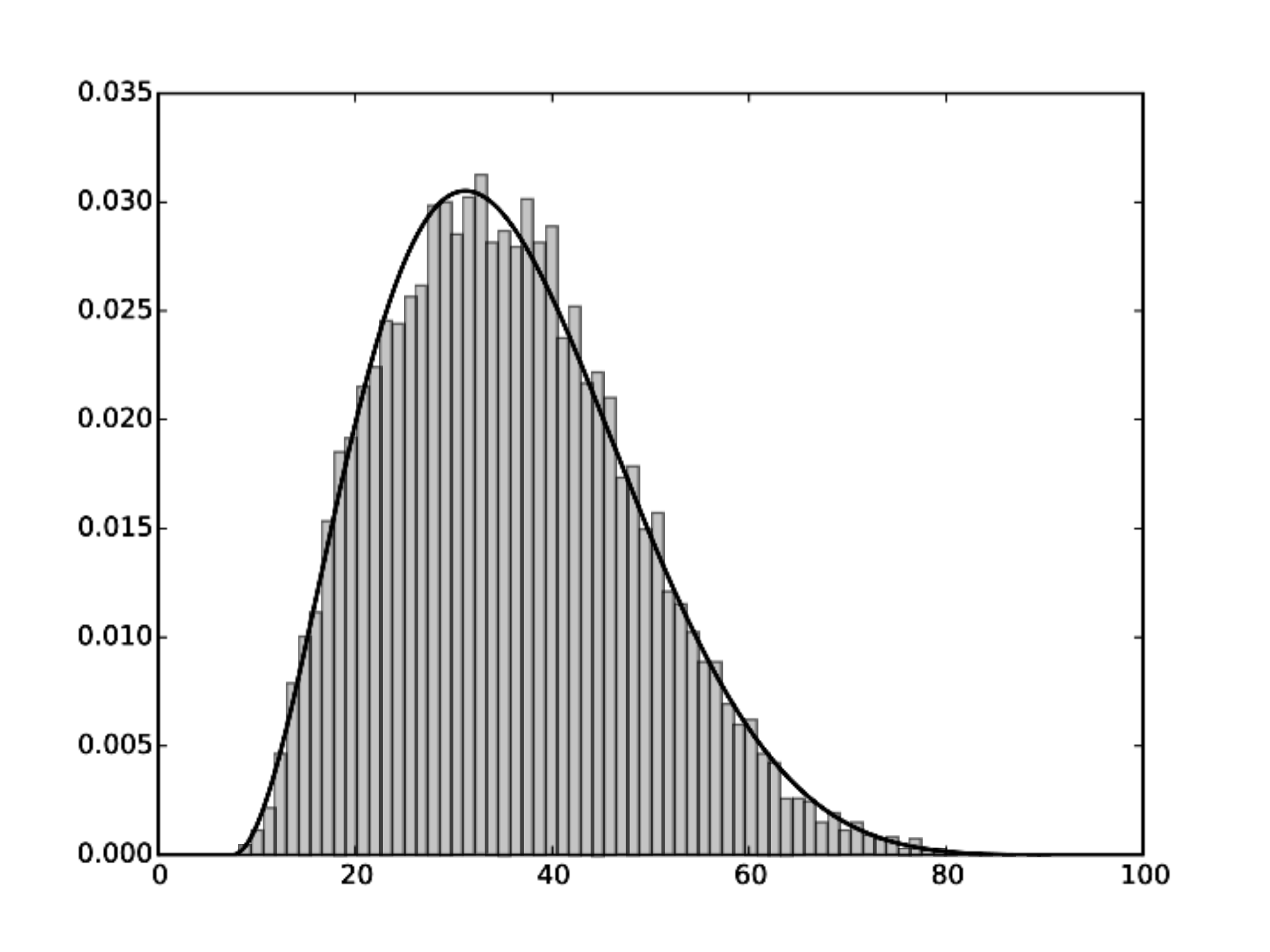} 
\caption{Fire-spotting parameter $\mu$.}
\label{fig:fitbetaMU}
\end{subfigure}
\begin{subfigure}{0.5\textwidth}
\includegraphics[width=0.9\linewidth, height=5cm]{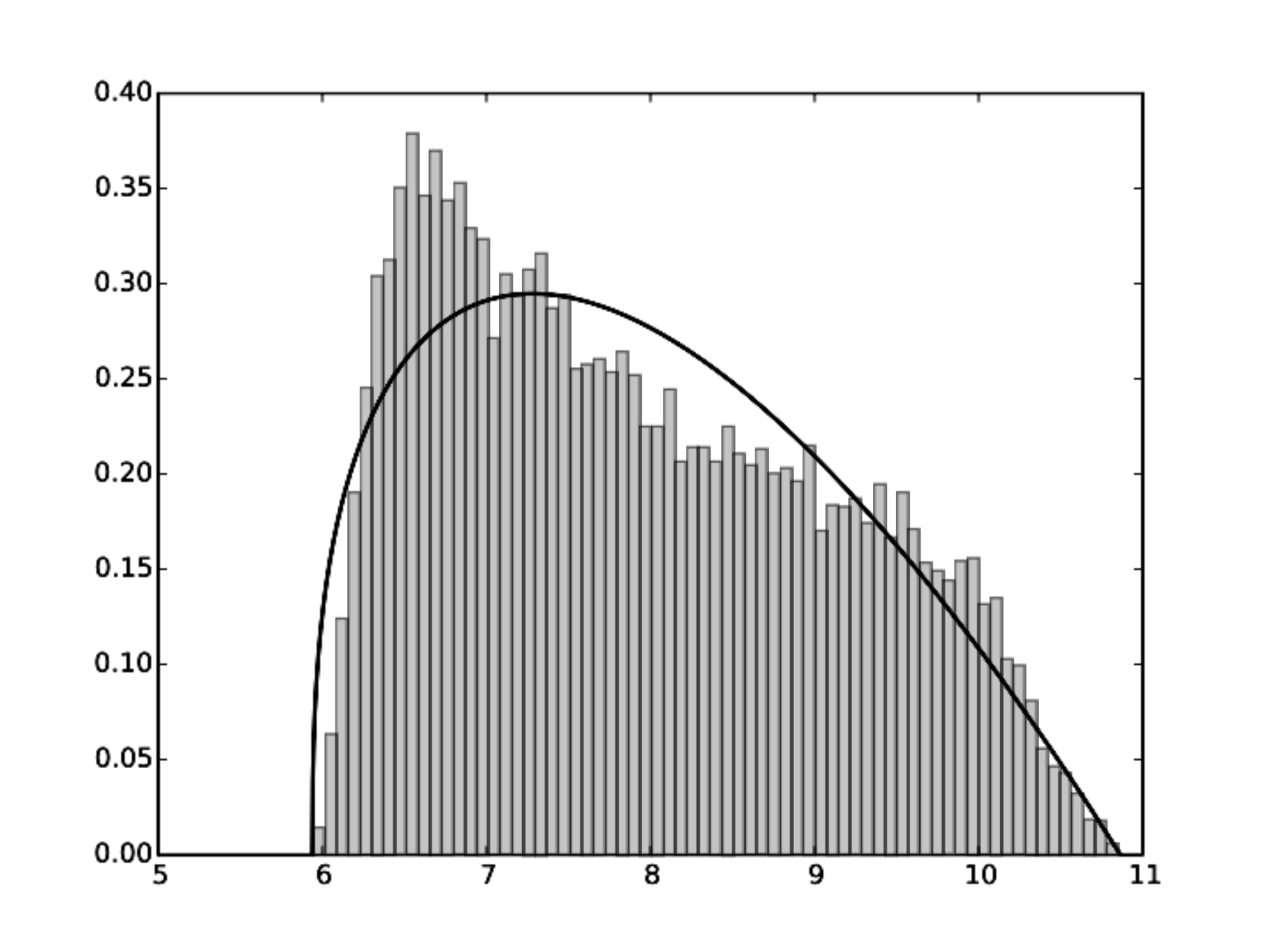}
\caption{Fire-spotting parameter $\sigma$.}
\label{fig:fitbetaSIGMA}
\end{subfigure}
\begin{subfigure}{0.5\textwidth}
\includegraphics[width=0.9\linewidth, height=5cm]{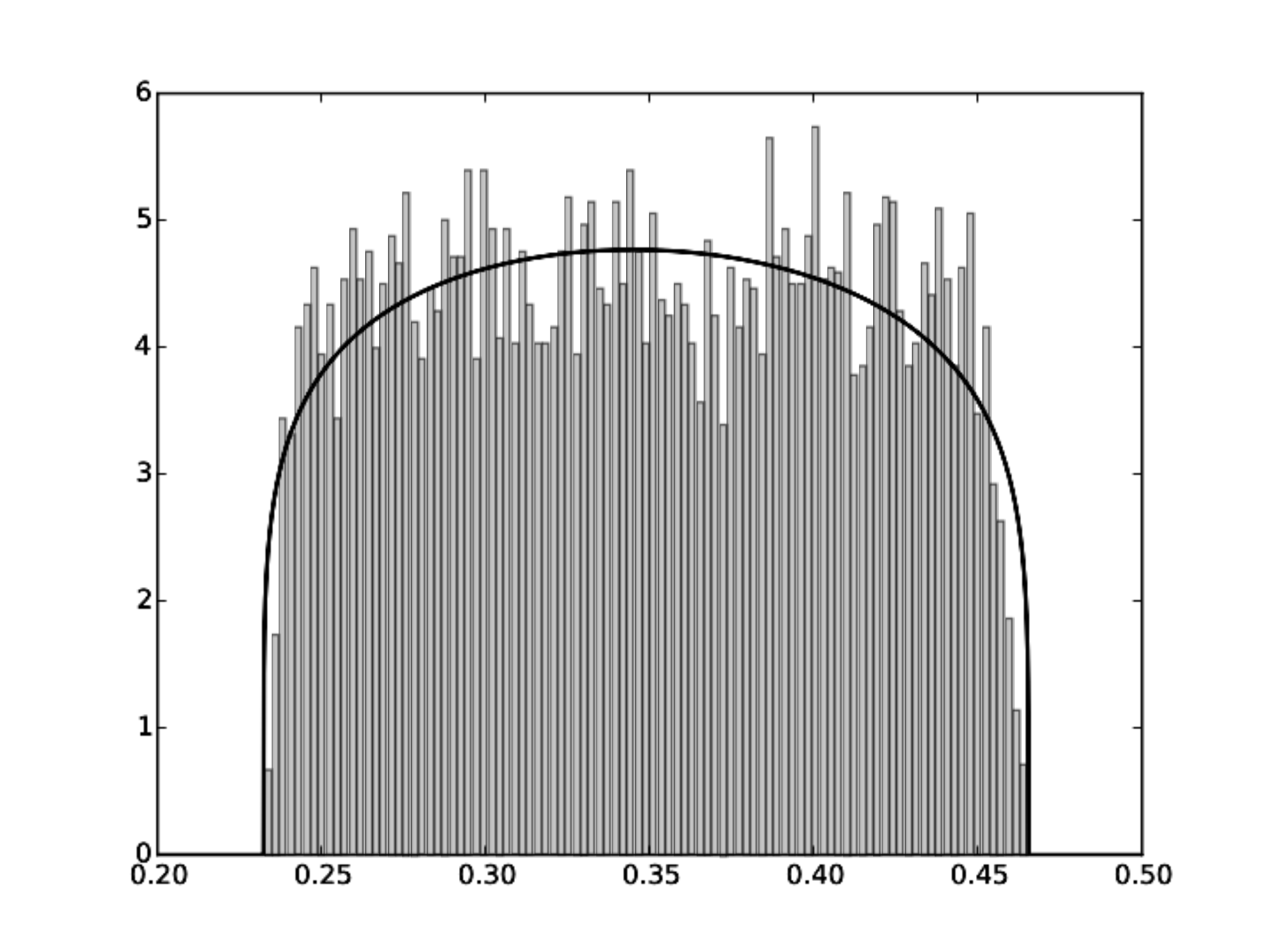}
\caption{Turbulent diffusion parameter $D$~$[\rm m^2\,s^{-1}]$.}
\label{fig:fitbetaD}
\end{subfigure}
\caption{Histograms and corresponding fits with Beta-distribution (solid lines) for the three parameters $\mu$, $\sigma$ (fire-spotting effects) and $D$ (turbulence effect) following a Monte Carlo random sampling with 10,000 realizations in the ensemble.}
\label{fig:betafits}
\end{figure}

Once uniform PDFs are defined for each subparameter, we sample them through a Monte Carlo random sampling. The size of the sample (or ``ensemble") is 10,000 to obtain converged statistics. Based on Eqs.~(\ref{eq:diffcoeff})--(\ref{eq:sigma}), we thus obtain 10,000 realizations of the three parameters of interest $\lbrace D, \sigma, \mu \rbrace$. We can then analyze their empirical statistical distribution by fitting the resulting histograms with different types of PDF. Figure~\ref{fig:betafits} presents the good fits obtained when using a Beta-distribution for each sample. We adopt such distribution due to the requirement for positiveness, limitlessness, and compatibility with the available surrogates. Table~\ref{tab:inputs2_pdf} presents the characteristics of each Beta-distribution and the associated range of variation for each parameter in $\boldsymbol{\theta} = \left(\mu, \sigma, D\right)^T$. We recall the analytic formulation for the Beta-distribution denoted by Beta, with $a$ and $b$ ($a,b > 0$) the ``shape parameters":
\begin{equation}
\label{eq:beta_dist}
\text{Beta}(\mathrm{x}; a, b) = \frac{\Gamma(a+b)\,\mathrm{x}^{a-1}\,(1-\mathrm{x})^{b-1}}{\Gamma(a)\Gamma(b)},
\end{equation}
for $\mathrm{x} \in (0, 1)$, with $\Gamma(\mathrm{x})$ the Gamma function. To shift and/or scale the distribution, the ``location" and ``scale" parameters are introduced. More specifically, $\text{Beta}(\mathrm{x}, a, b, \text{location} , \text{scale})$ is equivalent to $\text{Beta}(\mathrm{y}, a, b)/\text{scale}$ with $\mathrm{y} = (\mathrm{x} - \text{location})/\text{scale}$.

\begin{table}
\caption{Range of variations and Beta-distribution for $\boldsymbol{\theta} = \left(\mu, \sigma, D\right)^T$. Note that the parameters of the Beta-distribution (Eq.~\ref{eq:beta_dist}) are given in the following order: shape parameters $a$ and $b$, location and scale.}
\hspace{-2.0truecm}
\begin{tabular}{ccc}
\hline
Parameter & Minimum/maximum values & Beta-distribution parameters \\
\hline 
Log-normal parameter $\sigma$ & 5.49--12.69 & 1.37 1.99 5.94 4.93 \\ 
Log-normal parameter  $\mu$ & 7.25--98.16 & 3.18 7.49 7.43 94.73 \\ 
Turbulent diffusion coef. $D$~[m$^2$\,s$^{-1}$] & 0.23--0.47 & 1.19 1.20 0.23 0.23 \\ 
\hline 
\end{tabular} 
\label{tab:inputs2_pdf}
\end{table}
  
\subsection{Simulated Quantities of Interest}\label{sec:model_var_out}

\begin{figure}[h!]
\centering
\includegraphics[width=0.7\linewidth]{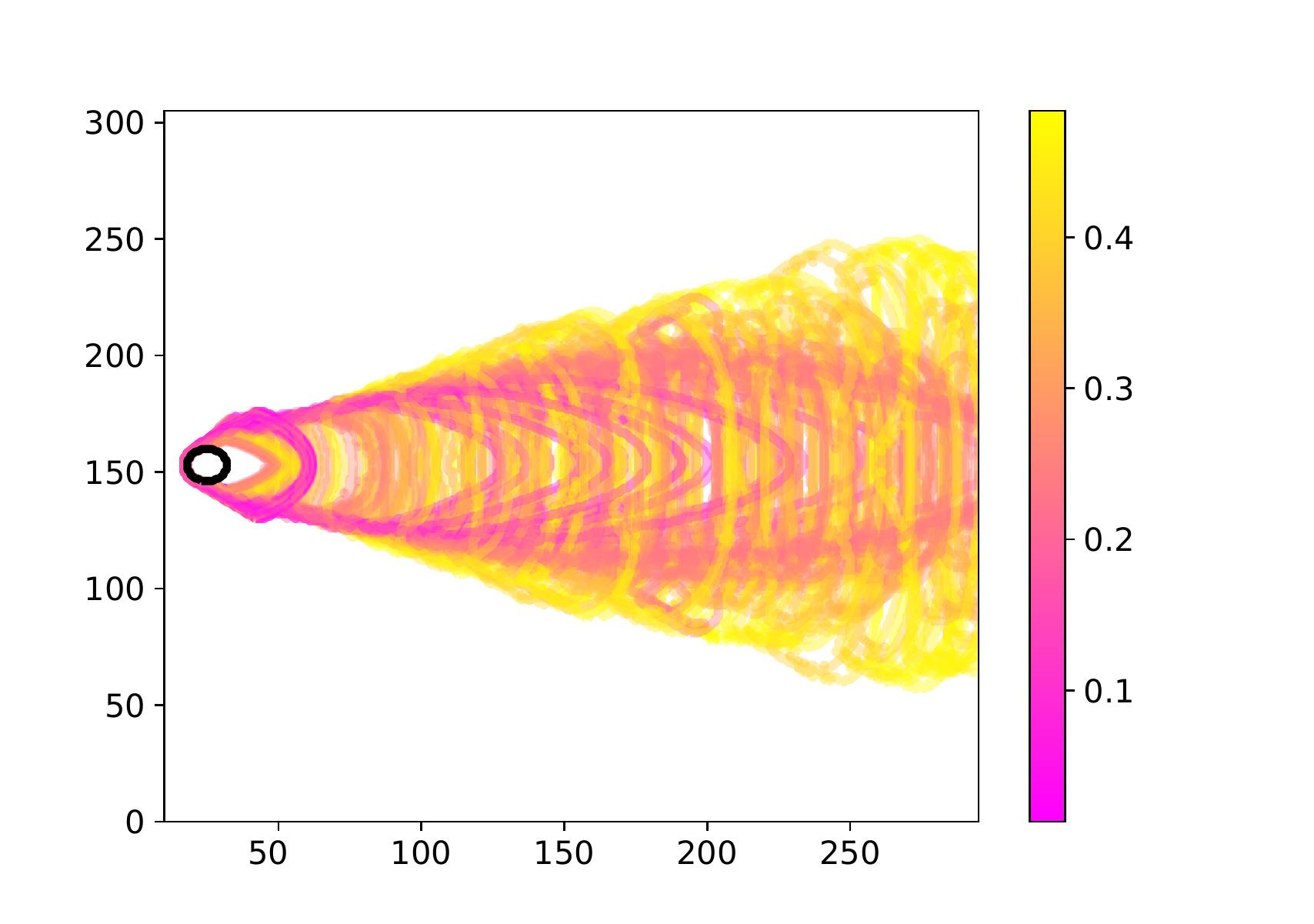} 
\caption{Ensemble of 100 fireline positions over the 2-D computational domain $\Omega$ after 50~min of {\tt LSFire+} model integration obtained when varying $D$, $\mu$ and $\sigma$ as presented in Table~\ref{tab:inputs2_pdf}. The black circle is the initial fireline that is the same for all simulations. The colormap corresponds to the normalized MSR $S_t$ at time $t = 50$~min (Eq.~\ref{eq:qoi_msr}).}
\label{fig:model_ex}
\end{figure}

We now define two scalar indices to represent the evolution of a fire over a time period [0; $T$]. We consider first the percentage of the computational domain $\Omega$ that is burnt at a given time $t$:
\begin{equation}
A_t = \frac{\int_{\Omega} \mathcal{I}_{\mathcal{B}(t)} (\mathrm{x}, t)\,d\mathrm{x}}{|\Omega|}, 
\label{eq:qoi_area}
\end{equation}
where $|\Omega|$~[m$^2$] corresponds to the area of the computational domain and $\mathcal{I}_{\mathcal{B}(t)}$ is the indicator function of the burnt area, which returns $1$ inside of the  burnt area and $0$ elsewhere. $A_t$ corresponds to a normalized burnt area. However, this quantity does not give information on the topology of the fire, which can be complex in the case of fire-spotting. To overcome this limitation, we also consider an indicator $S_t$ that describes the minimum spanning rectangle (MSR) of the burnt area over the area of the computational domain $|\Omega|$ at a given time:
\begin{equation}
S_t = \frac{|\text{MSR}(t)|}{|\Omega|}.
\label{eq:qoi_msr}
\end{equation}
The MSR is a geometrical quantity that corresponds to the smallest rectangle within which all burnt grid points lie at a given time $t$. So $|\text{MSR}(t)|$~[m$^2$] measures the area of this rectangle. As an example, Fig.~\ref{fig:model_ex} presents an ensemble of 100 firelines at time 50~min, where each fireline corresponds to a different set of parameters $D$, $\mu$ and $\sigma$ (i.e.~a different realization of $\boldsymbol{\theta} = \left(\mu, \sigma, D\right)^T$) obtained by sampling the Beta-distributions given in Table~\ref{tab:inputs2_pdf}. For each fireline, Fig.~\ref{fig:model_ex} shows the corresponding normalized MSR as defined in Eq.~(\ref{eq:qoi_msr}) at time 50~min. Low MSR values (rose colors) indicate simple topology of the fireline, while for high MSR values (yellow colors) the fireline presents more irregularities and a more complex propagation induced by turbulence and fire-spotting.

In this work, we analyze the time dependency of the quantities $A_t$ and $S_t$ by comparing them at two different times, $t_1 = 26~\text{min}$ and $t_2 = 34~\text{min}$. The resulting scalar quantities (or ``observables") are noted $A_1$, $A_2$, $S_1$ and $S_2$.

\subsection{Numerical Implementation}

The code {\tt LSFire+} is developed in C and Fortran, where the turbulence and fire-spotting parametrization routines, labeled as {\tt RandomFront 2.3b}, act as a post-processing routine at each time step in a level-set-method (LSM) code for the front propagation implemented through the library {\tt LSMLIB} \cite{lsmlib} and the ROS is computed by using the library {\tt FireLib} \cite{firelib}.
The numerical library {\tt LSMLIB} is written in Fortran2008/OpenMP. It advects the fireline through standard algorithms for the LSM, including also fast marching method algorithms. The aforementioned routines are freely available at the official git repository of BCAM, Bilbao, {\url{https://gitlab.bcamath.org/atrucchia/randomfront-wrfsfire-lsfire}}.

\section{Surrogate Modeling}\label{sec:uq}

\subsection{Principles and Notations}\label{sec:uq_notations}

The objective of the present paper is to build surrogate models (or ``response surfaces") that represent how the normalized burnt area $A_t$ or the normalized MSR $S_t$ (the generic scalar output is noted $\mathrm{y} \in \mathbb{R}$) changes with respect to a selection of the most relevant input parameters (the set of uncertain parameters is noted $\boldsymbol{\theta} \in \mathbb{R}^d$). The input stochastic space is defined either by $\boldsymbol{\theta} = \left(U, I ,D\right)^T$ or $\boldsymbol{\theta} = \left(\mu, \sigma, D\right)^T$ (see Sec.~\ref{sec:model_var_in}); the size of the input stochastic space is $d = 3$.

The key idea of a surrogate is to replace the fire spread model $\mathrm{y} = \mathcal{M}(\boldsymbol{\theta})$ by a weighted finite sum of basis functions that can be generally expressed as
\begin{equation}
\widehat{\mathrm{y}}\left(\boldsymbol{\theta}\right) = \displaystyle\sum_{\boldsymbol{\alpha} \in \mathcal{A}}\,\gamma_{\boldsymbol{\alpha}}\,\Psi_{\boldsymbol{\alpha}}\left(\boldsymbol{\theta}\right),
\label{eq:surrogate_form}
\end{equation}
where the coefficients $\gamma_{\boldsymbol{\alpha}}$ and the basis functions $\Psi_{\boldsymbol{\alpha}}$ are to be determined, $\mathcal{A}$ being the set of indices that defines the basis size. In practice, the coefficients and basis functions are calibrated by the training set (or ``database") $\mathcal{D}_N$ that corresponds to a limited number $N$ of forward model integrations (or ``training set") such that 
\begin{equation}
\mathcal{D}_N = (\Theta, \mathcal{Y}) = \left\lbrace\left(\boldsymbol{\theta}^{(k)},\mathrm{y}^{(k)}\right)_{1\leq k \leq N}\right\rbrace,
\label{eq:surrogate_trainingset}
\end{equation}
where $\mathrm{y}^{(k)} = \mathcal{M}(\boldsymbol{\theta}^{(k)})$ corresponds to the integration of the forward model $\mathcal{M}$ ({\tt LSFire+} in the present study) for the $k$th set of input parameters 
$\boldsymbol{\theta}^{(k)}$. 

Two types of surrogate models are compared in the following: the gPC-expansion that retrieves the global forward model behavior on the one hand, the GP regression that is a local interpolator of the forward model behavior at the training points on the other hand. Different types of surrogate are tested to determine what is the best choice in the present application. For gPC-expansion, the user needs to determine the appropriate total polynomial order of the expansion as well as the appropriate type and number of basis functions $\Psi_{\boldsymbol{\alpha}}$. There are also different projection strategies to compute the coefficients $\gamma_{\boldsymbol{\alpha}}$. For GP regression, the user needs to choose the type of correlation structure and to estimate its associated hyperparameters.

\subsection{Generalized Polynomial Chaos (gPC) Expansion}\label{sec:uq_pc}

$\boldsymbol{\theta}$ is defined in the input physical space and its counterpart in the standard probabilistic space is noted $\boldsymbol{\zeta} = (\zeta_1, \cdots, \zeta_d)$, with $\zeta_i$ the random variable associated with the $i$th uncertain parameter $\theta_i$ in $\boldsymbol{\theta}$ characterized by its marginal PDF $\rho_{\theta_i}$. $\boldsymbol{\theta}$ is thus rescaled in the standard probabilistic space to which the gPC framework applies. 

\subsubsection{Polynomial Basis}\label{sec:uq_pc_basis}

$\boldsymbol{\theta}$ is projected onto a stochastic space spanned by the orthonormal polynomial functions $\lbrace\Psi_{\boldsymbol{\alpha}}(\boldsymbol{\zeta})\rbrace_{\boldsymbol{\alpha} \in \mathcal{A}}$. The basis functions are orthonormal with respect to the joint PDF $\boldsymbol{\rho}_{\boldsymbol{\zeta}}(\boldsymbol{\zeta})$, i.e.
\begin{equation}
\langle \Psi_{\boldsymbol{\alpha}}(\boldsymbol{\zeta}),\Psi_{\boldsymbol{\beta}}(\boldsymbol{\zeta})\rangle 
= \int_{Z}\,\Psi_{\boldsymbol{\alpha}}(\boldsymbol{\zeta})\,\Psi_{\boldsymbol{\beta}}(\boldsymbol{\zeta})\,\boldsymbol{\rho}_{\boldsymbol{\zeta}}\,d\boldsymbol{\zeta} = \delta_{\boldsymbol{\alpha}\boldsymbol{\beta}},
\label{eq:pc_innerproduct}
\end{equation}
with $\delta_{\boldsymbol{\alpha}\boldsymbol{\beta}}$ the Kronecker delta-function and $Z \subseteq \mathbb{R}^d$ the space in which $\boldsymbol{\zeta}$ evolves. In practice, the orthonormal basis is built using the tensor product of one-dimensional polynomial functions, $\Psi_{\boldsymbol{\alpha}} = \phi_{\alpha_1}\ldots\phi_{\alpha_d}$ with $\phi_{\alpha_i}$ the one-dimensional polynomial function. The choice for the basis functions depends on the probability measure of the random variables. According to Askey's scheme, the Jacobi polynomials form the optimal basis for random variables following Beta-distribution, and the Legendre polynomials are the counterpart for uniform distribution~\cite{xiu2002}.

Assuming that the solution of the fire spread model is of finite variance, each quantity of interest $\mathrm{y}$ (see Sec.~\ref{sec:model_var_out}) can be considered as a random variable for which there exists a gPC expansion of the form
\begin{equation}
\widehat{\mathrm{y}}\left(\boldsymbol{\theta}\right) = \mathcal{M}_{\text{pc}}(\boldsymbol{\theta}(\boldsymbol{\zeta})) = \displaystyle\sum_{\boldsymbol{\alpha} \in \mathcal{A}}\,\gamma_{\boldsymbol{\alpha}}\,\Psi_{\boldsymbol{\alpha}}\left(\boldsymbol{\zeta}\right).
\label{eq:surrogate_form_pc}
\end{equation}
$\Psi_{\boldsymbol{\alpha}}$ is the $\boldsymbol{\alpha}$th multivariate basis function chosen in adequacy with the PDF $\boldsymbol{\rho}_{\boldsymbol{\theta}}$ associated with the parameters $\boldsymbol{\theta}$ (all random variables in $\boldsymbol{\theta}$ are assumed independent so that $\boldsymbol{\rho}_{\boldsymbol{\theta}}$ is the product of the marginal PDFs $\lbrace\rho_{\theta_i}\rbrace_{i=1,\cdots,d}$). $\boldsymbol{\alpha} = (\alpha_1, \cdots, \alpha_d)$ is a multi-index in $\mathcal{A}$, which identifies the components of the multivariate polynomial $\Psi_{\boldsymbol{\alpha}}$.

Note that Eq.~(\ref{eq:surrogate_form_pc}) represents how the normalized burnt area $A_t$ or the normalized MSR $S_t$ varies according to changes in the input vector $\boldsymbol{\theta}$. Once the PDF $\boldsymbol{\rho}_{\boldsymbol{\theta}}$ is chosen, $\lbrace\gamma_{\boldsymbol{\alpha}}\rbrace_{\boldsymbol{\alpha} \in \mathcal{A}}$ are the unknowns to determine to build the surrogate $\mathcal{M}_{\text{pc}}$.

\subsubsection{Truncation Strategy}\label{sec:uq_pc_truncation}

For computational purposes, the sum in Eq.~(\ref{eq:surrogate_form_pc}) is truncated to a finite number of terms $r$ that is associated with the total polynomial order $P$ of the gPC-expansion. There are several ways of choosing the number of terms $r$ referred to as the ``truncation strategy". Note that we will investigate the sensitivity of the surrogate performance to the choice of the truncation strategy for a given size $N$ of the training set $\mathcal{D}_N$ in Sec.~\ref{sec:results}. Note that we will use the concept of ``enumeration functions" in the following: a linear (or hyperbolic) enumeration function is a mapping $\mathfrak{I}$ from $\mathbb{N}$ to $\mathbb{N}^d$, which establishes a bijective mapping between a given integer $i$ and a multi-index $\boldsymbol{\alpha}$.

\paragraph{Linear Truncation Strategy.} The standard truncation strategy (referred to as ``linear") consists in retaining in the gPC-expansion all polynomials involving the $d$ random variables of total degree less or equal to $P$. Hence, $\boldsymbol{\alpha} = (\alpha_1, \cdots, \alpha_d) \in \lbrace 0, 1, \cdots, P \rbrace^d$. The number of terms $r$ is therefore constrained in this linear case by the number of random variables $d$ and by the total polynomial order $P$ so that 
\begin{equation}
r_{\text{lin}} = (d + P)!/(d!~P!).
\label{eq:pc_terms_linear}
\end{equation}
The set of selected multi-indices for the multi-variate polynomials $\mathcal{A}$ is defined as 
\begin{equation}
\mathcal{A}_{\text{lin}} \equiv \mathcal{A}_{\text{lin}}(d, P) = \lbrace \boldsymbol{\alpha} \in \mathbb{N}^d: |\boldsymbol{\alpha}| \leq P \rbrace \subset \mathbb{N}^d,
\label{eq:pc_multiindex_linear}
\end{equation}
where $|\boldsymbol{\alpha}| = \left|\left| \boldsymbol \alpha \right| \right|_1 = \alpha_1 + \cdots + \alpha_d$ is the ``total order" of the multi-index. In this case, we refer to the basis as the ``full basis" for a given total polynomial order $P$.

\paragraph{Hyperbolic Truncation Strategy.} According to the sparsity-of-effects principle, high-order interaction terms (i.e.~polynomial terms involving several uncertain parameters of $\boldsymbol{\theta}$) are often less important in physical problems and can be neglected with respect to main effects (i.e.~polynomial terms involving a single uncertain parameter of $\boldsymbol{\theta}$) and low-order interaction terms. As an alternative to the linear truncation strategy, the ``hyperbolic" truncation strategy consists in eliminating a priori high-order interaction terms. A more general way than Eq.~(\ref{eq:pc_multiindex_linear}) to define the number of terms $r$ in the gPC expansion consists in introducing $q$-quasi-norms:
\begin{equation}
\mathcal{A}_{\text{hyp}} \equiv \mathcal{A}_{\text{hyp}}(d, P, q) = \left\lbrace \boldsymbol{\alpha} \in \mathbb N^{d} : || \boldsymbol{\alpha} ||_q  \leq P \right\rbrace,
\end{equation}
where the $q$-semi-norm is given by
\begin{equation}
\left| \left| \boldsymbol{\alpha} \right|\right|_q \equiv \left( \sum_{i = 1}^{d}\,(\alpha_i)^q\right)^{1/q}.
\end{equation}
The number of terms in the gPC-expansion is expressed by the cardinality of $\mathcal{A}$, which varies according to $P$ and $q$ for a fixed dimension $d$. The adoption of such semi-norms penalizes high-rank indices and high-order interactions. The lower the value of $q$, the higher the penalty in the determination of $\mathcal{A}$. When $q = 1$ we retrieve the linear truncation strategy and therefore a full basis of cardinality $\mathcal{A}_{\text{lin}}(d,P)$. In the following, we will study how the performance of the surrogate depends on the choice of the hyperbolic parameter $q \in [0, 1]$.

\paragraph{Sparse Truncation Strategies.} There are alternatives to reduce the number of terms in the gPC-expansion. We will now schematically represent three of them, ordered by complexity: 1-``sequential strategy", 2- ``cleaning strategy", 3- ``least angle regression". \\
1- The sequential strategy~\cite{baudin2017} consists in constructing the gPC-expansion in an incremental way, starting from the first term $\Psi_0$ ($K_0 = \{0\}$) and adding one term at a time in the basis ($K_{i+1} = K_{i} \cup \{\Psi_{i+1}\}$). The terms that are sequentially added to the basis are ordered according to the adopted enumeration strategy (linear or hyperbolic). The response surface is therefore of increasing complexity, since the enumeration functions in both cases increase the polynomial complexity when increasing the index. In the present study, the construction process is stopped when a given accuracy is achieved, or when the number of terms in the gPC-expansion reaches the maximum size of the basis $r_{\text{max}}$ specified by the user. \\
2- An alternative to the sequential strategy is the cleaning strategy~\cite{baudin2017}, which builds a gPC-expansion containing at most $r_{\text{max}}$ significant coefficients, i.e.~at most $r_{\text{max}}$ significant basis functions, starting from the full basis (still retaining the constraint of hyperbolic truncation if selected). The key idea of the cleaning strategy is to discard from the active basis the polynomials $\Psi_{\boldsymbol{\alpha}}$ that are associated with coefficients of low magnitude, i.e.~satisfying
\begin{equation} 
|\gamma_{\boldsymbol{\alpha}}| \leq \epsilon\cdot\max_{\boldsymbol{\alpha}' \in \mathcal{A}'} |\gamma_{\boldsymbol{\alpha}'}|
\end{equation}
where $\epsilon$ is the significance factor set to $10^{-4}$, and where $\mathcal{A}'$ represents the current active basis. This selection procedure means that the terms in the gPC-expansion are not ordered according to the degree of the polynomial functions but instead according to the magnitude of the coefficients. \\
3- In complement to the sequential and cleaning strategies, there is a more advanced approach called least-angle regression (LAR) to select the active polynomial terms. The key idea of the LAR approach is to select at each iteration a polynomial among the $r$ terms of the full basis (or eventually the hyperbolic-truncated basis) based on the correlation of the polynomial term with the current residual. The selected term is added to the active set of polynomials. The coefficients of the active basis are computed so that every active polynomial is equicorrelated with the current residual until convergence is reached. Thus, LAR builds a collection of surrogates that are less and less sparse along the iterations. Iterations stop either when the full basis has been looked through or when the maximum size of the training set has been reached. When the iterations stopped, the polynomial coefficients are computed via the least-square algorithm presented below. More details can be found in~\cite{blatman2011,blatman2009phd,efron2004}.

\subsubsection{Projection strategy}
\label{sssec:proj_strat}

In this work, we focus on non-intrusive approaches based on $\ell_2-$minimization methods to numerically compute the coefficients $\lbrace\gamma_{\boldsymbol{\alpha}}\rbrace_{\boldsymbol{\alpha} \in \mathcal{A}}$ using the $N$ snapshots from the training set $\mathcal{D}_N$. 

\paragraph{Galerkin Pseudo-Spectral Projection.} This Galerkin-type projection relies on the orthonormality property of the polynomial basis. Using this approach, the $\boldsymbol{\alpha}$th coefficient $\gamma_{\boldsymbol{\alpha}}$ is computed using the definition of the inner product that is numerically approximated using tensor-based Gauss quadrature (referred to as ``quadrature" in the following) as follows
\begin{equation}
\gamma_{\boldsymbol{\alpha}} 
= \langle \mathrm{y},\Psi_{\boldsymbol{\alpha}} \rangle
\,\cong\,\displaystyle\sum_{k = 1}^{N}\,\mathrm{y}^{(k)}\,\Psi_{\boldsymbol{\alpha}}(\boldsymbol{\zeta}^{(k)})\,w^{(k)},
\label{eq:pc_quadrature}
\end{equation}
where $\mathrm{y}^{(k)} = \mathcal{M}(\boldsymbol{\theta}^{(k)})$ is the $k$th snapshot of the $\mathcal{D}_N$-database corresponding to the {\tt LSfire+} simulation for the $k$th quadrature root $\boldsymbol{\theta}^{(k)}$ of $\Psi_{\boldsymbol{\alpha}}$, and where $w^{k}$ is the weight associated with $\boldsymbol{\zeta}^{(k)}$ (corresponding to $\boldsymbol{\theta}^{(k)}$ in the standard probabilistic space). When considering a full basis, $(P+1)$ is the number of quadrature roots required in each uncertain direction to ensure an accurate calculation of the integral $\langle\mathrm{y},\Psi_{\boldsymbol{\alpha}}\rangle$. Hence, in our problem, we have $N = (P+1)^3$ simulations in the training set to build the PC surrogates through Galerkin pseudo-spectral projection. 

\paragraph{Least-Square Minimization.} With this approach, the estimation of the coefficients $\lbrace\gamma_{\boldsymbol{\alpha}}\rbrace_{\boldsymbol{\alpha} \in \mathcal{A}}$ is done by solving a least-square minimization problem, i.e.~by minimizing the approximation error between the (exact) {\tt LSfire+} model evaluations and the PC-surrogate estimations at the points of the training set $\mathcal{D}_N$. The least-square projection solves a minimization problem over the given basis as follows:
\begin{equation}
\widehat{\boldsymbol{\gamma}} = 
\underset{\boldsymbol{\gamma} \in \mathbb{R}^r}{\operatorname{argmin}}\,\displaystyle\sum_{k=1}^N\,\left(\mathrm{y}^{(k)} - \displaystyle\sum_{\boldsymbol{\alpha} \in \mathcal{A}^P}\,\gamma_{\boldsymbol{\alpha}}\,\Psi_{\boldsymbol{\alpha}}\left(\mathbf{x}^{(k)}\right)\right)^2
\end{equation}
which is achieved through classical linear algebra algorithms. Note that the sample size $N$ required by this strategy for the problem to be well posed is at least equal to $(r+1)$, where $r$ is the number of gPC-coefficients (i.e.~the cardinality of the set $\mathcal{A}$). Note also that least-square minimization is used here to compute the coefficients selected by the sparse truncation methods (sequential, cleaning or LAR). When using non-sparse truncation strategies, this projection method is referred to as the standard least-square (SLS) approach.  

\subsubsection{Workflow scheme for constructing the gPC-expansion}
A complete algorithm relative to the implementation of the gPC-surrogate can be summarized as follows:
\begin{enumerate}
\item choose the polynomial basis $\lbrace\Psi_{\boldsymbol{\alpha}}\rbrace_{\boldsymbol{\alpha} \in \mathcal{A}}$ according to the assumed marginal PDFs of the inputs $\boldsymbol{\theta} = \left(\left\lVert\mathrm{U}\right\rVert, I ,D \right)^T$ or $\boldsymbol{\theta} = \left(\mu,\sigma , D \right)^T$;
\item choose the total polynomial degree $P$ according to the complexity of the physical processes;
\item truncate the expansion to $r_{\text{lin}}$ or $r_{\text{hyp}}$ terms corresponding to the multi-index set $\mathcal{A}_{\text{lin}}$ or $\mathcal{A}_{\text{hyp}}$ using linear or hyperbolic truncation ($r_{\text{lin}}$ depends on $d$, $P$; $r_{\text{hyp}}$ depends on $d$, $P$ and $q$ with $q$ the hyperbolic factor satisfying $0 < q \leq 1$);
\item in the case of a sparse strategy (sequential, cleaning or LAR), find a suitable set of multi-indices $\mathcal{A} \subset \mathcal A_{\text{lin,hyp}}$ with a cardinality $r \leq r_{\text{lin, hyp}}$, otherwise skip this step;
\item apply a projection strategy (quadrature or least-square) to compute the coefficients $\lbrace\gamma_{\boldsymbol{\alpha}}\rbrace_{\boldsymbol{\alpha} \in \mathcal{A} \subset \mathbb{N}^d}$ using $N = (P+1)^{d}$ snapshots from the simulation database $\mathcal{D}_{N_{\text{ref}}}$; 
\item formulate the surrogate model $\mathcal{M}_{\text{pc}}$, which can be evaluated for any new pair of parameters $\boldsymbol{\theta}^* = \left(\left\lVert\mathrm{U}\right\rVert^*, I^* , D^* \right)^T$ or $\left(\mu^*,\sigma^* , D^* \right)^T$.
\end{enumerate}

\subsection{Gaussian Process (GP) surrogate model}\label{sec:GP}

As stated by~\cite{rasmussen2006}, a GP is a random process (here the observable from the fireline evolution $\mathrm{y}$) indexed over a domain (here $\mathbb{R}^d$), for which any finite collection of process values (here $\left\{\mathrm{y}(\boldsymbol{\theta}^{(k)})\right\}_{1\leq k \leq N}, \boldsymbol{\theta}^{(k)} \in \Theta$) has a joint Gaussian distribution. Concretely, let $\widetilde{\mathrm{y}}$ be a Gaussian random process fully described by its zero mean and its correlation $\pi$:
\begin{equation}
\widetilde{\mathrm{y}}(\boldsymbol{\theta})\sim \text{GP}\left(0, \sigma_{\text{gp}}^2\,\pi(\boldsymbol{\theta},\boldsymbol{\theta}')\right),
\end{equation}
with $\pi(\boldsymbol{\theta},\boldsymbol{\theta}') = \mathbb{E}\left[\widetilde{\mathrm{y}}(\boldsymbol{\theta})\widetilde{\mathrm{y}}(\boldsymbol{\theta}')\right]$. In the present case, the correlation function $\pi$ (or kernel) is chosen as a squared exponential (also known as ``RBF kernel", RBF standing for radial basis function):
\begin{equation}
\pi(\boldsymbol{\theta}, \boldsymbol{\theta}') = \exp\left(-\frac{\|\boldsymbol{\theta} - \boldsymbol{\theta}'\|^2}{2\,\ell_{\text{gp}}^2}\right),
\label{eq:RBF}
\end{equation}
where $\ell_{\text{gp}}$ is a length-scale representing the model output dependency between two inputs $\boldsymbol{\theta}$ and $\boldsymbol{\theta}'$, and where $\sigma_{\text{gp}}^2$ is the variance of the observable. The surrogate model is thus the mean of the GP, resulting of conditioning $\widetilde{\mathrm{y}}$ on the training set $\mathcal{Y} = \left\{\mathrm{y}\left(\boldsymbol{\theta}^{(k)}\right)\right\}_{1\leq k \leq N}$. The quantity of interest provided by the GP-surrogate for any given $\boldsymbol{\theta}^*\in\mathbb{R}^d$ satisfies
\begin{equation}
\mathrm{y}_{\text{gp}}(\boldsymbol{\theta}^*)=\sum_{k = 1}^N\,\beta_{k}\,\pi\left(\boldsymbol{\theta}^*,\boldsymbol{\theta}^{(k)}\right),
\end{equation}
where 
\begin{equation}
\beta_{k} = \left(\mathbf{\Pi} + \tau_{\text{gp}}^2\,\mathbf{I}_N\right)^{-1}\left(\mathrm{y}(\boldsymbol{\theta}^{(1)}) \ldots \mathrm{y}(\boldsymbol{\theta}^{(N)})\right)^T,
\end{equation}
\begin{equation}
\mathbf{\Pi}= \left(\pi(\boldsymbol{\theta}^{(j)},\boldsymbol{\theta}^{(k)})\right)_{1\leq j,k \leq N},
\end{equation}
and where $\tau_{\text{gp}}$ (referred to as the ``nugget effect") is used to avoid ill-conditioning issues for the matrix $\mathbf{\Pi}$. The hyperparameters $\left\{\ell_{\text{gp}}, \sigma_{\text{gp}}, \tau_{\text{gp}}\right\}$ are optimized through maximum likelihood applied to the dataset $\mathcal{D}_N$ using a basin hopping technique~\cite{wales1997}.

\subsection{Design of Experiments}\label{sec:doe}

We build several datasets to analyze the performance of the gPC- and GP-surrogates in an extensive way in Section~\ref{sec:results}; these datasets are summarized in Table~\ref{tbl:databases}. Note that estimating the generalization error of the surrogate model requires the use of an independent dataset, that is why we use a Monte Carlo random sampling including $N = 216$ members for validation. Note also that the Halton's low-discrepancy sequence is involved in this work in order to explore the hypercube defined by the distribution of the uncertain parameters. This design of experiment will be compared to a tensor-based Gauss quadrature in terms of performance of the surrogate model. The reader shall refer to Section~\ref{sec:model_var_in} for more details on the range of variation and the marginal PDFs of each uncertain parameter. 
\begin{table}[h]
\centering
\caption{Datasets $\mathcal{D}_N$ of {\tt LSfire+} simulations used in this work for building surrogates (``training") or for validating them (``validation").}
\begin{tabular}{ccc}
\hline
Sampling Strategy & Purpose & Sample size \\
\hline 
    & $\boldsymbol{\theta} = \left(\left\lVert\mathrm{U}\right\rVert, I , D \right)^T$  &  \\
Halton's sequence & Training & 216 \\
Monte Carlo random sampling & Validation & 216 \\
\hline
    & $\boldsymbol{\theta} = \left(\mu,\sigma , D \right)^T$  & \\
Halton's sequence & Training  & 216  \\
Quadrature rule & Training & 216 \\
Monte Carlo random sampling & Validation & 216 \\
\hline
\end{tabular} 
\label{tbl:databases}
\end{table}

\subsection{Error Metrics}\label{sec:error}

In the present study, two error metrics are used to assess the quality of the surrogate predictions: the empirical error between the surrogate prediction and the {\tt LSfire+} model prediction (also known as ``training error") on the one hand, and the $Q_2$ predictive coefficient~\cite{marrel2009} on the other hand. 

\subsubsection{Empirical Error $\epsilon_{\text{emp}}$} 
The truncation of the gPC-expansion can eventually introduce an approximation error at the training points, which can be computed posterior to the surrogate construction. This empirical error denoted by $\epsilon_{\text{emp}}$ reads
\be
\epsilon_{\text{emp}} = \frac{1}{N}\sum_{k=1}^{N} \left( \mathrm{y}^{(k)} - \widehat{\mathrm{y}}^{(k)} \right)	,
\ee
with $\mathrm{y}^{(k)}$ the $k$th element of the training set $\mathcal{D}_N$ (either the Halton's low discrepancy sequence or the quadrature database, see Table~\ref{tbl:databases}) and $\widehat{\mathrm{y}}^{(k)}$ the corresponding value predicted by the surrogate for the same element of the training set. 

However, this error estimator has several drawbacks. First, the GP-model (built without noise in the kernel) is an interpolator so that the approximation error is expected to be $\epsilon_{\text{emp}} = 0$. Second, this estimator may severely underestimate the magnitude of the mean square error. When the size of the training set $N$ comes closer to the cardinality of the gPC-expansion $\mathcal{A}$, $\epsilon_{\text{emp}}$ may tend to zero, while the actual mean square error does not; this issue is known as ``overfitting". 

\subsubsection{Predictive coefficient $Q_2$}

We require a more robust error estimator suitable for both gPC-expansion and GP-model. In this work, we use the $Q_2$ predictive coefficient based on cross-validation. The computation of $Q_2$ relies on two distinct datasets: the current training set $\mathcal{D}_{N}$ (either the Halton's sequence or the quadrature database) and a Monte Carlo sample $\mathcal{D}_{N_{\text{ref}}}$ that is independent of the surrogate construction and that is therefore referred to as the ``validation dataset". $Q_2$ is computed as  
 \begin{align} 
Q_{2} = 1 - \frac{\displaystyle\sum_{k = 1}^{N_{\text{ref}}}\,\left(\mathrm{y}^{(k)} - \widehat{\mathrm{y}}^{(k)}\right)^2}{\displaystyle\sum_{k = 1}^{N_{\text{ref}}}\,\left(\mathrm{y}^{(k)} - \overline{\mathrm{y}}_{\text{ref}}\right)^2},
\end{align}
with $\mathrm{y}^{(k)}$ the $k$th element of the Monte Carlo sample $\mathcal{D}_{N_{\text{ref}}}$, $\widehat{\mathrm{y}}^{(k)}$ the surrogate prediction for the same element of $\mathcal{D}_{N_{\text{ref}}}$ and $\overline{\mathrm{y}}_{\text{ref}}$ the empirical mean over the Monte Carlo sample $\mathcal{D}_{N_{\text{ref}}}$. Note that computing $Q_2$, the training set $\mathcal{D}_{N}$ is only used to construct the surrogate model and to obtain the estimation $\widehat{\mathrm{y}}$ of the quantity of interest $\mathrm{y}$. The target value for $Q_2$ is 1.

\subsection{Statistical Analysis}\label{sec:SA}

Once the surrogates are available for the different observables ($A_1$, $A_2$, $S_1$, $S_2$ -- see Section~\ref{sec:model_var_out}), the statistics of the quantities of interest can be obtained. For the gPC-expansion, they can be derived analytically from the coefficients. For the GP-surrogate, we evaluate the surrogate predictions over a new dataset $\mathcal{D}_{N_{\text{sample}}}$ of size $N_{\text{sample}} = 10,000$ that is a subset of $\mathbb{R}^3$ and that is obtained using a standard Monte Carlo random sampling; this dataset is only used as input to the surrogate model and not to {\tt LSfire+}. 

\subsubsection{Estimation of Statistical Moments}

The mean value and STD of the observable $\mathrm{y}$ can be estimated as
\begin{eqnarray}
\mu_{\widehat{\mathrm{y}}} &=& \frac{1}{N_{\text{sample}}}\,\displaystyle\sum_{k = 1}^{N_{\text{sample}}}\,\widehat{\mathrm{y}}^{(k)}, \label{ref:mean} \\
\sigma_{\widehat{\mathrm{y}}} &=& \sqrt{\frac{1}{N_{\text{sample}}-1}\,\displaystyle\sum_{k = 1}^{N_{\text{sample}}}\,\left(\widehat{\mathrm{y}}^{(k)} - \mu_{\widehat{\mathrm{y}}}\right)^2}, \label{ref:std}
\end{eqnarray}
with $\mathbf{\widehat{\mathrm{y}}}^{(k)}$  the $k$th element of the dataset $\mathcal{D}_{N_{\text{sample}}}$ containing the surrogate evaluations over the aforementioned Monte Carlo sampled points. 

Using the gPC-surrogate, the statistical moments can be derived analytically from the coefficients $\lbrace \gamma_{\boldsymbol{\alpha}} \rbrace_{\boldsymbol{\alpha} \in \mathcal{A} \subset \mathbb{N}^d}$ such that the mean and the STD read:
\begin{eqnarray}
\mu_{\widehat{\mathrm{y}}_{\text{pc}}} &=& \gamma_{0}, \label{eq:pcmean} \\
\sigma_{\widehat{\mathrm{y}}_{\text{pc}}} &=& \sqrt{\displaystyle\sum_{\boldsymbol{\alpha} \in \mathcal{A} \subset \mathbb{N}^d\atop \boldsymbol{\alpha} \neq 0}\,\gamma_{\boldsymbol{\alpha}}^2}. \label{eq:pcstd}
\end{eqnarray}

\subsubsection{Sensitivity Analysis Diagnostics}
\label{sec:methodo_sa}

Sobol' indices~\cite{sobol1993,saltelli2007} are commonly used for sensitivity analysis based on variance analysis. They provide the quantification of how much of the variance in the quantity of interest is due to the variance in the input parameters assuming (1)~these input random variables are independent and (2)~the random output is squared integrable. 

For the GP-surrogate approach, Sobol' indices are stochastically estimated using Martinez' formulation since this estimator is stable and provides asymptotic confidence intervals for first-order and total-order indices~\cite{baudin2016}. 

For the gPC-expansion approach, Sobol' indices can be directly derived from the gPC-coefficients. For the $i$th component of the input random variable 
$\boldsymbol{\theta}$, the Sobol' index $\mathbb{S}_{\text{pc},i}$ reads:
\begin{equation}
\mathbb{S}_{\text{pc},i} = \frac{1}{{(\sigma_{\widehat{\mathrm{y}}_{\text{pc}}})^2}}\,\displaystyle\sum_{\boldsymbol{\alpha} \in \mathcal{A}_i \subset \mathbb{N}^d \atop \boldsymbol{\alpha} \neq 0} \left(\gamma_{\boldsymbol{\alpha}}\right)^2,
\end{equation}
where $\sigma_{\widehat{\mathrm{y}}_{\text{pc}}}$ is the STD computed in Eq.~\eqref{eq:pcstd}, and where $\mathcal{A}_{i}$ is the set of multi-indices selected in $\mathcal{A}$ such that the computation of $\mathbb{S}_{\text{pc},i}$ only includes terms that depend on the input variable $\theta_i$, namely
\begin{equation}
\mathcal{A}_{i} = \lbrace \boldsymbol{\alpha} \in \mathbb{N}^d, |\boldsymbol{\alpha}| \leq P \; | \; \alpha_i > 0, \alpha_{k \neq i} = 0 \rbrace.
\end{equation}

\subsection{Numerical Implementation}\label{sec:tools}

The GP implementation relies on the Python package \emph{scikit-learn}~\cite{pedregosa2011} (see http://scikit-learn.org/). The gPC-implementation relies on the Python package \emph{OpenTURNS}~\cite{baudin2017} (see www.openturns.org). The \emph{batman}~\cite{roy2018b} Python package is used to build datasets and perform statistical analysis.

\section{Results}\label{sec:results}

The objective of this study is two-fold. First, we provide an extensive comparison of the performance of different surrogate strategies for a given budget (i.e.~a given size $N$ of the training set $\mathcal{D}_N$, $N = 216$); the different types of surrogate are given in Table~\ref{tbl:surrogates}. We evaluate their impact on the predicted quantities of interest $A_t$ and $S_t$ in terms of mean value and STD, but also their impact on the predicted Sobol' sensitivity indices. This extensive analysis is carried out for the case $\boldsymbol{\theta} = \left(\mu, \sigma, D \right)^T$, meaning that we only consider uncertainty in the fluctuating part of the forward model {\tt LSFire+}. Second, we use this framework to rank the uncertain parameters, either $\boldsymbol{\theta} = \left(\left\lVert\mathrm{U}\right\rVert, I, \tau\right)^T$ or $\boldsymbol{\theta} = \left(\mu, \sigma, D \right)^T$, by order of importance and identify the most influential input parameters for the problem of turbulence and fire-spotting.

\subsection{Comparison of surrogate performance}\label{sec:SensAlgResults}

\subsubsection{Error assessment}

Table~\ref{table:MSD_Q2_err} presents the error metrics (i.e.~the $\epsilon_{\text{emp}}$ empirical error and the $Q_2$ predictive coefficient) obtained for different types of surrogate (gPC on the one hand, and GP on the other hand) with respect to $\boldsymbol{\theta} = \left(\mu, \sigma, D \right)^T$ but for a given size of the training set $N = 216$. The performance of the gPC-surrogate is analyzed in details for varying truncation and projection schemes summarized in Table~\ref{tbl:surrogates}; the GP-surrogate is obtained using a standard RBF kernel and is considered here as a basis for comparison in order to evaluate the quality of the gPC-surrogates. For each approach, one surrogate model is built for each of the four observables $\lbrace A_1, A_2, S_1, S_2 \rbrace$ corresponding to the two quantities of interest $A_t$ and $S_t$ at times $t_1 = 26~\text{min}$ and $t_2 = 34~\text{min}$. 

In Table~\ref{table:MSD_Q2_err} we first focus on the results obtained with linear truncation ($q = 1$), meaning that the basis of polynomial functions is full for a given total polynomial order $P$. Table~\ref{tbl:sparsity1} (right column) presents corresponding scatter plots (referred to as ``adequacy plots") of the surrogate predictions with respect to the physical model predictions. These plots quantify the adequacy of the surrogate to the physical model at the training points in terms of predicted burnt area ratio $A_2$. It is found that the $Q_2$ predictive coefficient is over 0.9 only for the LAR and cleaning sparse methods for all observables. The empirical error is of the same order of magnitude, varying between $10^{-3}$ for the MSR ratio $S_t$ and $10^{-4}$ for the burnt area ratio $A_t$. Note that for a given observable at a given time, there is no significant difference among the surrogate strategies in terms of empirical error. We therefore focus the following analysis on the standalone $Q_2$ predictive coefficient. Note also that the performance of each surrogate is time independent since for a given observable, the $Q_2$ predictive coefficient is similar at times $t_1$ and $t_2$. We therefore focus on results at time $t_2$ in the following.

When moving to hyperbolic truncation schemes ($q = 0.75$ or $q = 0.5$), we reduce a priori the number of coefficients to compute in the gPC-expansion, while the size of the training set remains the same ($N = 216$). The lower the value of $q$, the smaller the number of gPC-coefficients $r$. Table~\ref{tbl:sparsity05} (right column) presents adequacy plots for hyperbolic truncation with $q=0.5$; this is to compare to the adequacy plots obtained for linear truncation in Table~\ref{tbl:sparsity1} (right column). Results show that the performance of the quadrature approach does not improve when $q$ decreases. In the opposite, the performance of the SLS approach improves and features a $Q_2$ predictive coefficient over 0.9 for $A_2$ and over 0.8 for $S_2$ when using hyperbolic truncation. This improvement is also noticeable in Table~\ref{tbl:sparsity05} (right column), where hyperbolic truncation allows to better represent the model response for low values of the burnt area ratio ($A_2 < 0.03$). The sequential sparse method also provides better results for a hyperbolic coefficient $q = 0.5$. The performance of LAR and cleaning sparse methods remains similar as in the linear case $q = 1$. 

LAR appears as the most accurate gPC strategy and has a $Q_2$ predictive coefficient that is similar to that obtained with the GP-model based on RBF kernel. Hyperbolic truncation does not add much value to the results compared to linear truncation, except for the SLS strategy. This may be explained by the fact that the terms that are important to retain in the gPC-expansion are not located in an isotropic way in the three dimensions ($d = 3$). It is therefore of interest to identify which polynomial terms are important to keep in the basis in order to obtain a good performance of the surrogate in each of the three dimensions. 

\begin{table}[h]
\caption{Types of surrogate used in this work. Recall that $q$ is the hyperbolic parameter for truncation ($q =1$ corresponds to linear truncation) and $N$ is the size of the training set.}
\hspace{-1.3truecm}
\begin{tabular}{lccc}
\hline
Name & Truncation & Sparse & Training set \\
\hline 
Quad. (Quadrature) & $q = 1, 0.75, 0.5$ & No & Gauss quadrature, $N = 216$ \\
SLS (Standard Least-Squares) & $q = 1, 0.75, 0.5$ & No & Halton, $N = 216$ \\
LAR (Least-Angle Regression) & $q = 1, 0.75, 0.5$ & Yes & Halton, $N = 216$ \\
Cleaning & $q = 1, 0.75, 0.5$ & Yes & Halton, $N = 216$ \\
Sequential & $q = 1, 0.75, 0.5$ & Yes & Halton, $N = 216$ \\
\hline
RBF kernel & -- & -- & Halton, $N = 216$ \\
\hline
\end{tabular} 
\label{tbl:surrogates}
\end{table}

\begin{table}[h]
\caption{Error metrics $\epsilon_{\text{emp}}$ and $Q_2$ for gPC-expansions and GP-model detailed in Table~\ref{tbl:surrogates}. The size of the training set is $N = 216$. One type of surrogate is built for each of the four observables, $A_1$, $A_2$, $S_1$ and $S_2$.}
\begin{tabular}{c|c|c|c|c|c|c|c|c|}
\cline{2-9}\cline{2-9} & \multicolumn{8}{|c|}{\textbf{gPC expansion -- Linear truncation} ($q = 1$)} \\
                                 &  \multicolumn{2}{|c|}{$A_1$}  &\multicolumn{2}{ |c| }{$A_2$} & \multicolumn{2}{ |c| }{$S_1$} & \multicolumn{2}{ |c| }{$S_2$}   \\ \cline{2-9}
\cline{2-9}\cline{2-9} & $\epsilon_{\text{emp}} $ & $Q_2$ &$\epsilon_{\text{emp}} $ & $Q_2$ &$\epsilon_{\text{emp}} $ & $Q_2$ &$\epsilon_{\text{emp}} $ & $Q_2$ \\                              
\cline{1-9}\cline{1-9}  \multicolumn{1}{ |c| }{Quad.}  & $1.4 \cdot 10^{-4}$ & $0.84$ & $2.7 \cdot 10^{-4}$ & $0.86$  &  $5.5 \cdot 10^{-4}$ & $0.77$&  $4.6 \cdot 10^{-4}$& $0.83$ \\ 
\multicolumn{1}{ |c| }{SLS}   & $3.0 \cdot 10^{-4}$ & $0.83$ & $6.3 \cdot 10^{-4}$ & $0.88$ &  $1.0 \cdot 10^{-3}$ & $ 0.74$ &$2.3 \cdot 10^{-3}$ &$0.75$ \\  
\multicolumn{1}{ |c| }{LAR}  & $1.0 \cdot 10^{-4}$ & $0.99$ &  $4.2 \cdot 10^{-4}$ & $0.970$ & $5.0 \cdot 10^{-4}$ &$0.96$&$2.3 \cdot 10^{-3}$& $0.95$\\  
\multicolumn{1}{ |c| }{Cleaning}  & $1.0 \cdot 10^{-4}$ & $0.96$ & $4.1 \cdot 10^{-4}$ &$0.95$  & $ 5.5 \cdot 10^{-4}$ &$0.96$& $1.2 \cdot 10^{-3}$ & $0.95$\\  
\multicolumn{1}{ |c| }{Sequential}  & $3.3 \cdot 10^{-4}$ & $0.85$ & $6.7 \cdot 10^{-4}$ &$0.89$  & $ 1.1 \cdot 10^{-3}$& $0.77$ &$2.5 \cdot 10^{-3}$& $0.85$ \\ \hline \cline{2-9}
\cline{2-9}\cline{2-9}& \multicolumn{8}{ |c| }{\textbf{gPC expansion -- Hyperbolic truncation} ($q = 0.75$)} \\
                               &  \multicolumn{2}{ |c| }{$A_1$}  &\multicolumn{2}{ |c| }{$A_2$} & \multicolumn{2}{ |c| }{$S_1$} & \multicolumn{2}{ |c| }{$S_2$}   \\ \cline{2-9}
\cline{2-9}\cline{2-9} & $\epsilon_{\text{emp}} $ & $Q_2$ &$\epsilon_{\text{emp}} $ & $Q_2$ &$\epsilon_{\text{emp}} $ & $Q_2$ &$\epsilon_{\text{emp}} $ & $Q_2$ \\ 
\cline{1-9}\cline{1-9} \multicolumn{1}{ |c| }{Quad.}  & $3.7 \cdot 10^{-4}$ & $ 0.76$ & $8.6 \cdot 10^{-4}$ & $0.77$  &  $1.6 \cdot 10^{-3}$ & $0.67$&  $3.7 \cdot 10^{-4}$& $0.66$ \\  
\multicolumn{1}{ |c| }{SLS}   & $1.5 \cdot 10^{-4}$ & $0.93$ & $1.8 \cdot 10^{-4}$ &$ 0.93$ &  $1.0 \cdot 10^{-3}$ & $ 0.84$ &$2.5 \cdot 10^{-3}$ &$0.84$ \\  
\multicolumn{1}{ |c| }{LAR}  & $2.0 \cdot 10^{-4}$ & $0.94$ &  $5.6 \cdot 10^{-4}$ & $0.95$ & $1.0\cdot 10^{-3}$ &$0.84$&$2.6 \cdot 10^{-3}$& $0.86$\\  
\multicolumn{1}{ |c| }{Cleaning}  & $9.9 \cdot 10^{-5}$ & $0.94$ & $3.3 \cdot 10^{-4}$ &$0.90$  & $ 5.0 \cdot 10^{-4}$ &$0.96$& $1.1 \cdot 10^{-3}$ & $0.96$\\  
\multicolumn{1}{ |c| }{Sequential}  & $1.9 \cdot 10^{-4}$ & $0.94$ & $4.7 \cdot 10^{-4}$ &$0.94$  & $ 8.7 \cdot 10^{-4}$& $0.86$ &$1.9 \cdot 10^{-3}$& $0.92$ \\ \hline \cline{2-9}
\cline{2-9}\cline{2-9}& \multicolumn{8}{ |c| }{\textbf{gPC expansion -- Hyperbolic truncation} ($q = 0.5$)} \\
                                &  \multicolumn{2}{ |c| }{$A_1$}  &\multicolumn{2}{ |c| }{$A_2$} & \multicolumn{2}{ |c| }{$S_1$} & \multicolumn{2}{ |c| }{$S_2$}   \\ \cline{2-9}
\cline{2-9}\cline{2-9} & $\epsilon_{\text{emp}} $ & $Q_2$ &$\epsilon_{\text{emp}} $ & $Q_2$ &$\epsilon_{\text{emp}} $ & $Q_2$ &$\epsilon_{\text{emp}} $ & $Q_2$ \\ 
\cline{1-9}\cline{1-9} \multicolumn{1}{ |c| }{Quad.}  & $1.8 \cdot 10^{-4}$ & $0.83$ & $2.0 \cdot 10^{-4}$ & $0.87$  &  $6.2 \cdot 10^{-4}$ & $0.74$&  $3.6 \cdot 10^{-4}$& $0.83$ \\  
\multicolumn{1}{ |c| }{SLS}   & $1.4 \cdot 10^{-4}$ & $0.96$ & $9.6 \cdot 10^{-5}$ &$ 0.95$ &  $7.4 \cdot 10^{-4}$ & $ 0.86$ &$1.9\cdot 10^{-3}$ &$0.86$ \\  
\multicolumn{1}{ |c| }{LAR}  & $1.5 \cdot 10^{-4}$ & $0.97$ &  $4.3 \cdot 10^{-4}$ & $0.97$ & $6.5\cdot 10^{-4}$ &$0.93$&$1.6 \cdot 10^{-3}$& $0.94$\\  
\multicolumn{1}{ |c| }{Cleaning}  & $8.8 \cdot 10^{-5}$ & $0.95$ & $3.3 \cdot 10^{-4}$ &$0.94$  & $ 4.5 \cdot 10^{-4}$ &$0.92$& $9.2 \cdot 10^{-4}$ & $0.98$\\  
\multicolumn{1}{ |c| }{Sequential}  & $1.3 \cdot 10^{-4}$ & $0.97$ & $4.2 \cdot 10^{-4}$ &$0.96$  & $ 6.4 \cdot 10^{-4}$& $0.93$ &$1.5 \cdot 10^{-3}$& $0.95$ \\ \hline \cline{2-9}
\cline{2-9}\cline{2-9}& \multicolumn{8}{ |c| }{\textbf{GP model}} \\
\cline{1-9}\cline{1-9}
\multicolumn{1}{ |c| }{RBF}  & $ -- $ & $0.99$ & $ -- $ &$0.98$  & $ -- $& $0.88$ &$ -- $& $0.99$ \\  \hline
\end{tabular}
\label{table:MSD_Q2_err}
\end{table}

\begin{table}
\caption{Comparison between quadrature, SLS and sparse (LAR, cleaning, sequential) methods to build the gPC-expansion for the burnt area ratio $A_2$ using linear truncation. Left: sparsity plots representing the magnitude of the coefficients with respect to the three-dimensional input space ($d = 3$). Right: adequacy scatter plots comparing surrogate ($x$-axis) and model ($y$-axis) predictions at the training points. For SLS and LAR, results are obtained with the best fit obtained for varying $P$.}
\centering
\begin{tabular}{l|cc}
Quad. &  \includegraphics[width=0.4\linewidth]{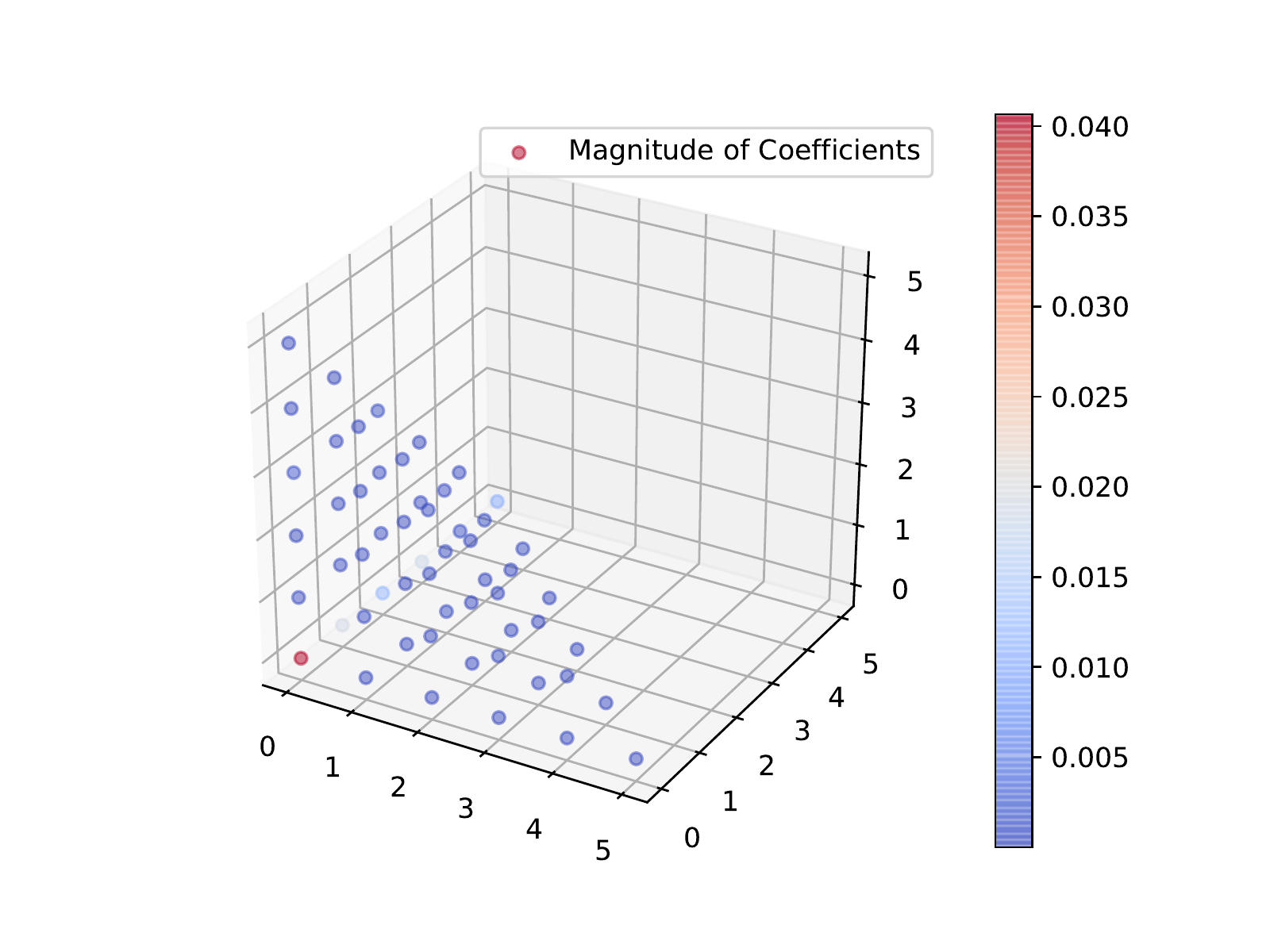} & 
                \includegraphics[width=0.4\linewidth]{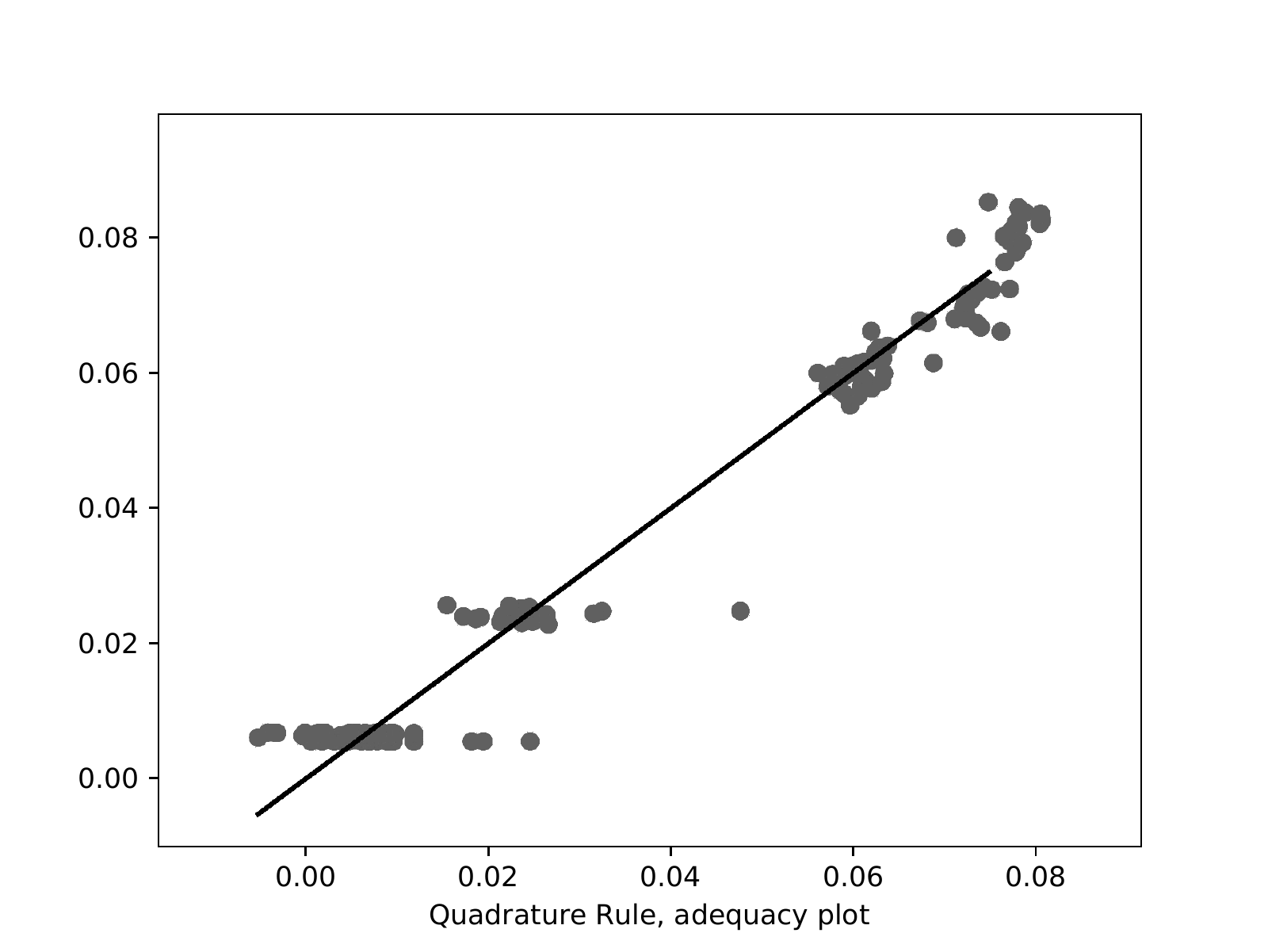} \\
SLS &  \includegraphics[width=0.4\linewidth]{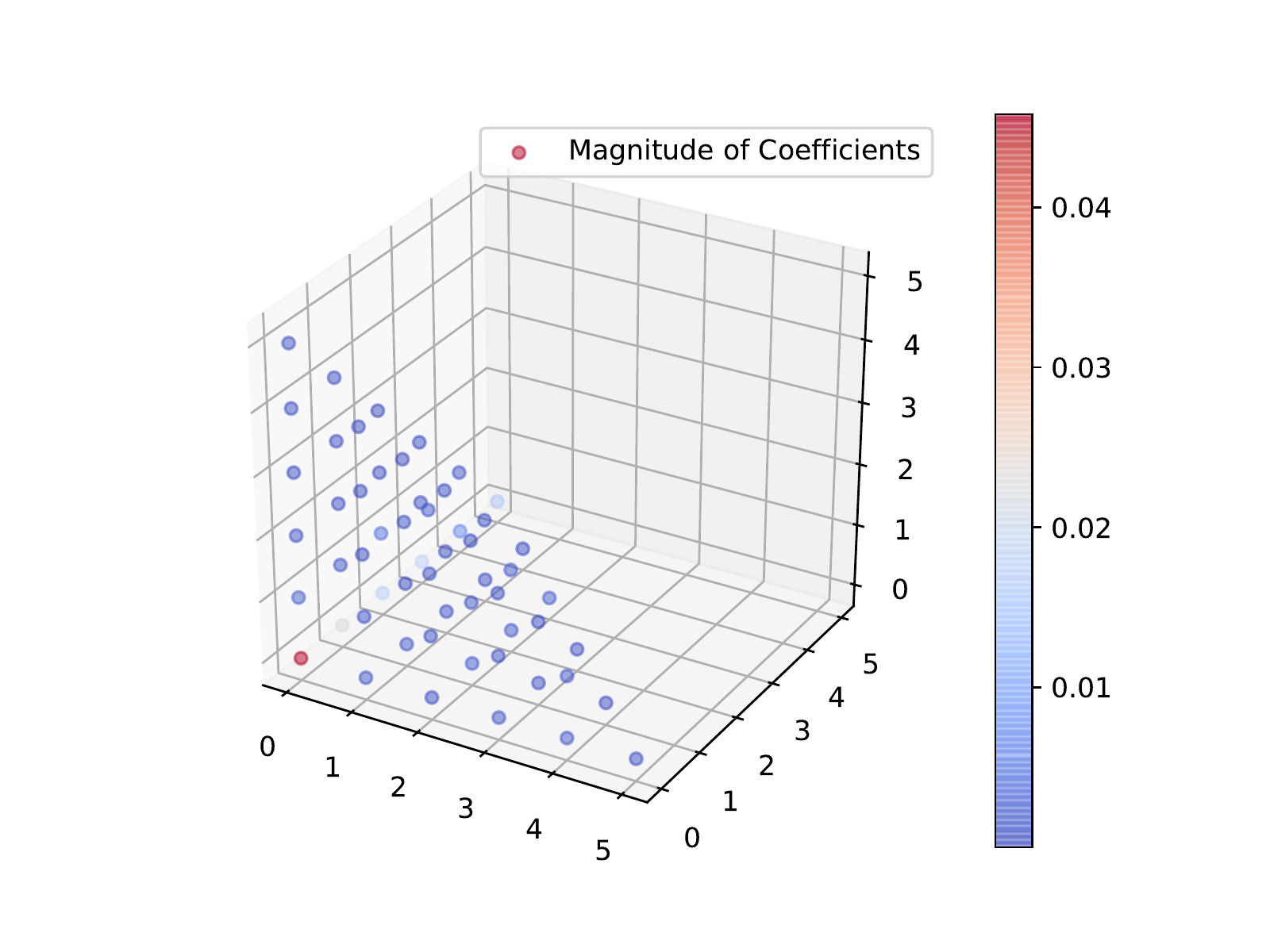} & 
            \includegraphics[width=0.4\linewidth]{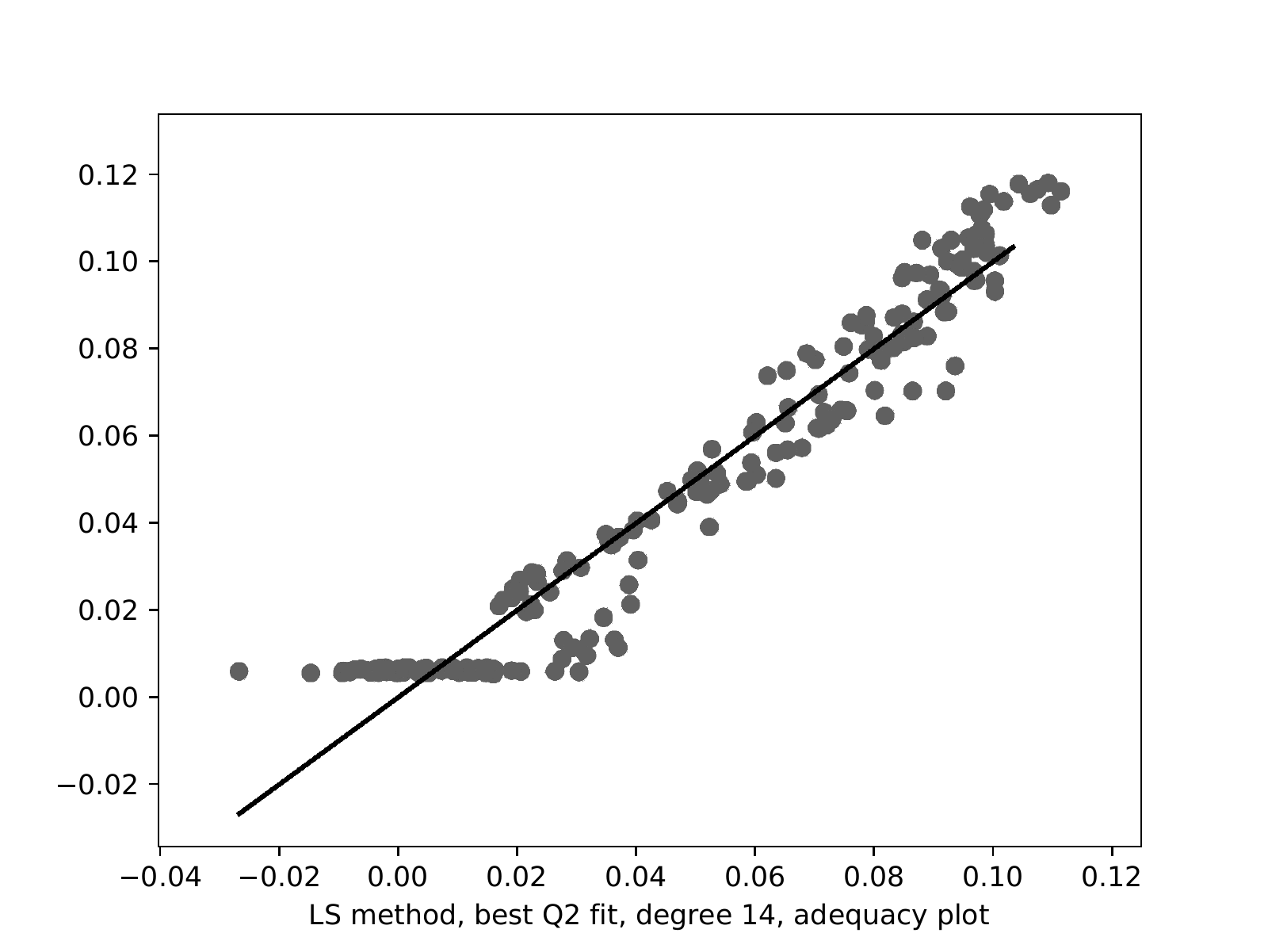} \\
LAR &  \includegraphics[width=0.4\linewidth]{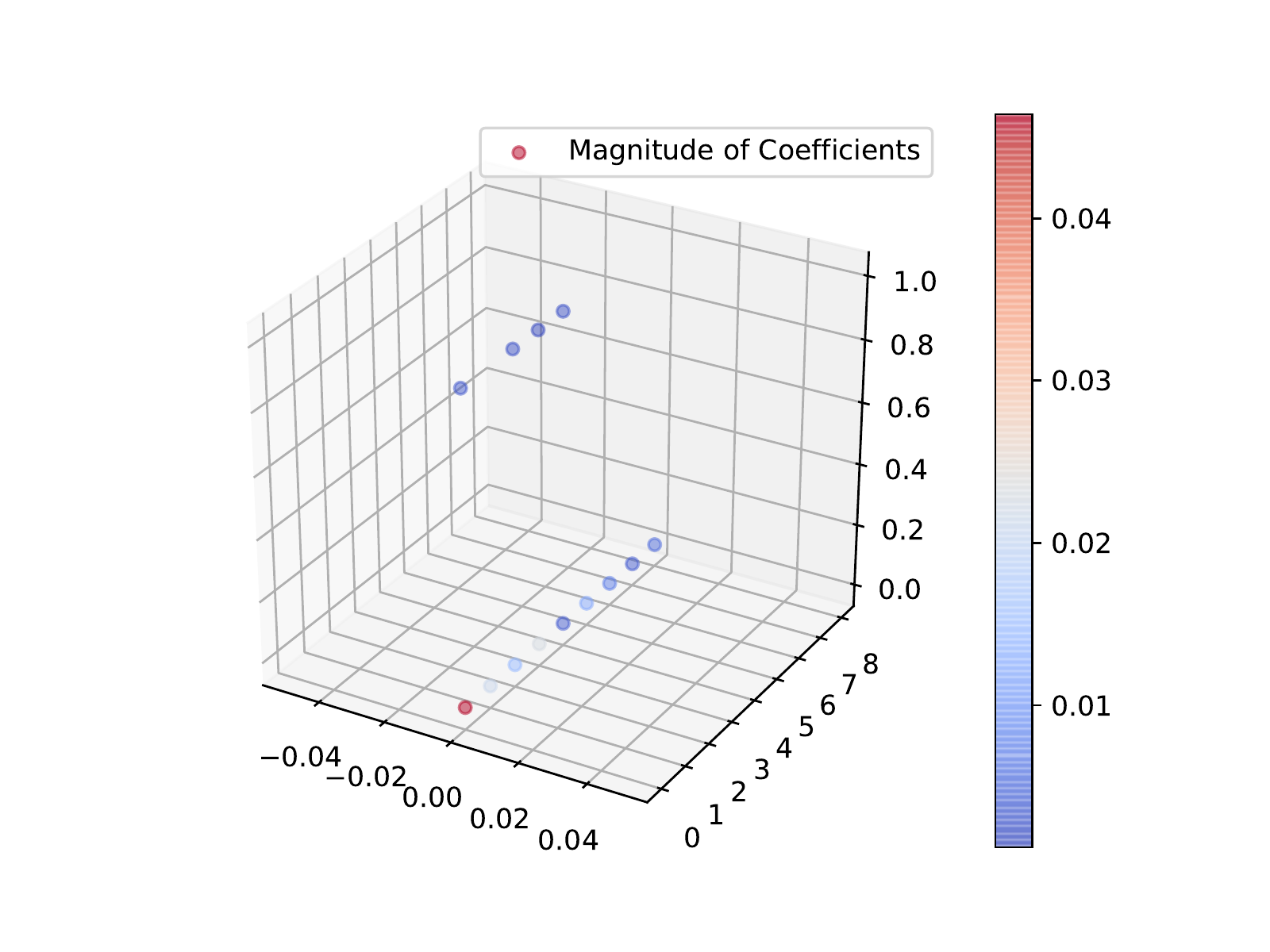} & 
            \includegraphics[width=0.4\linewidth]{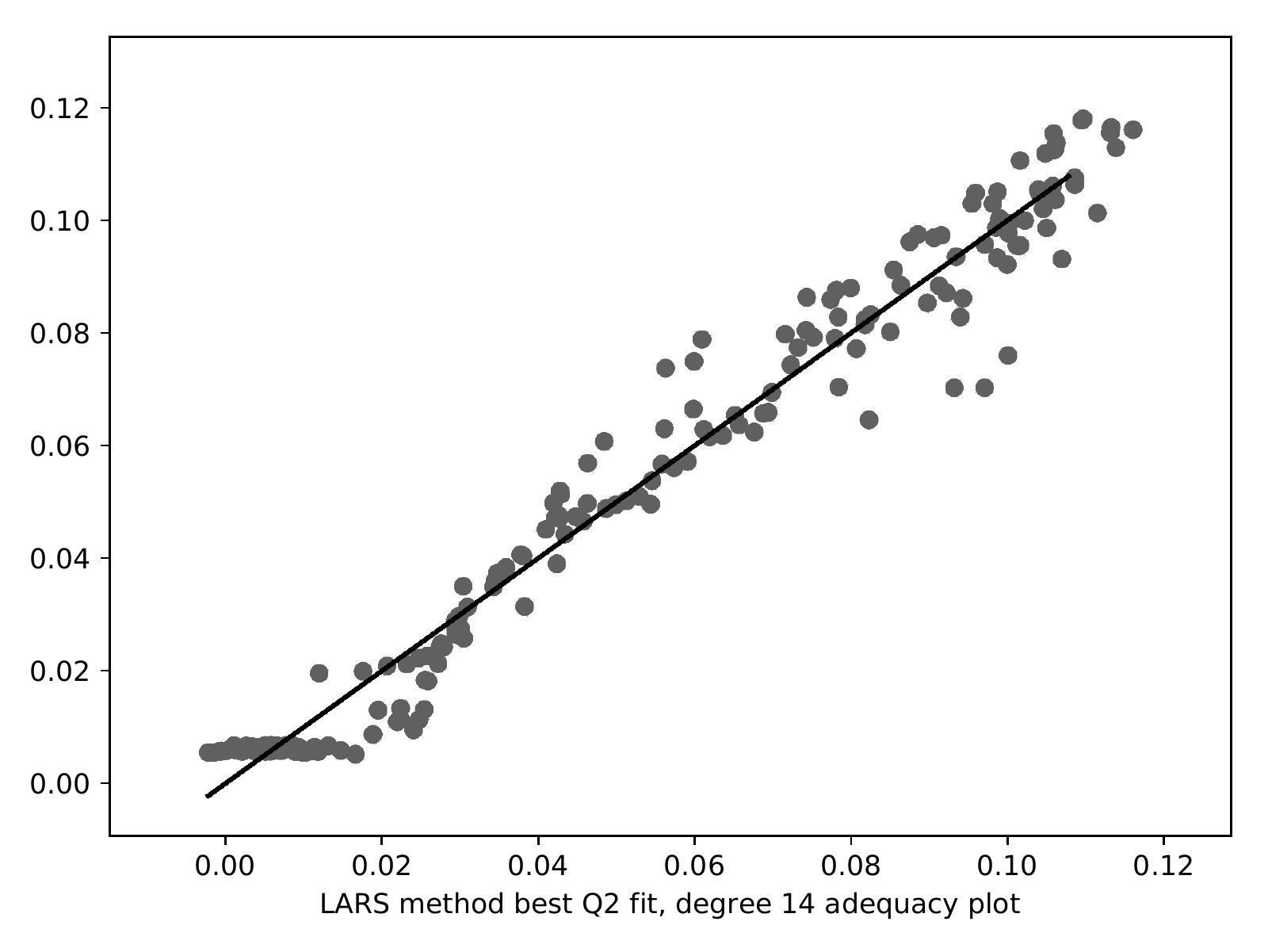} \\
Cleaning & \includegraphics[width=0.4\linewidth]{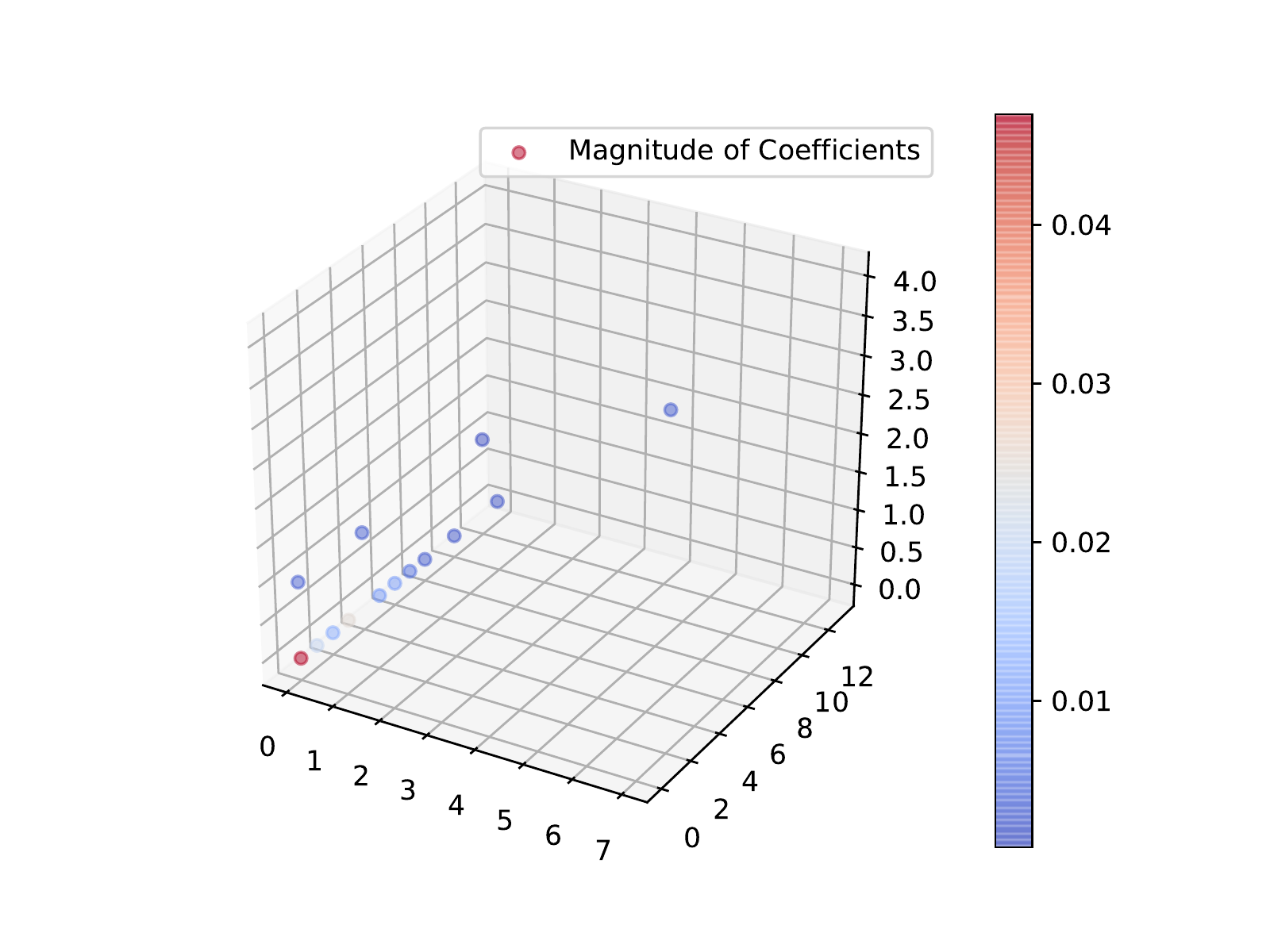} & 
            \includegraphics[width=0.4\linewidth]{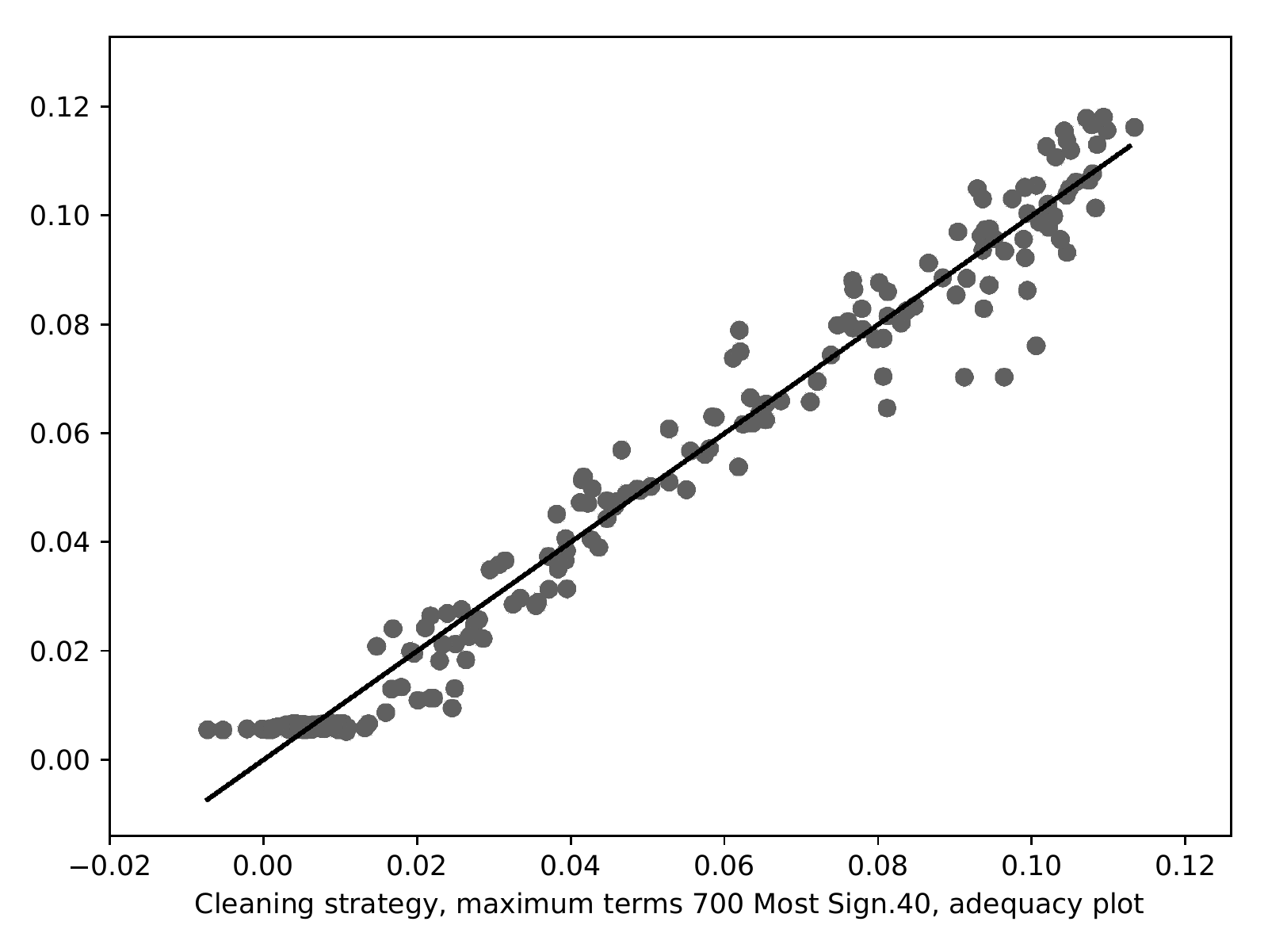} \\
Sequential & \includegraphics[width=0.4\linewidth]{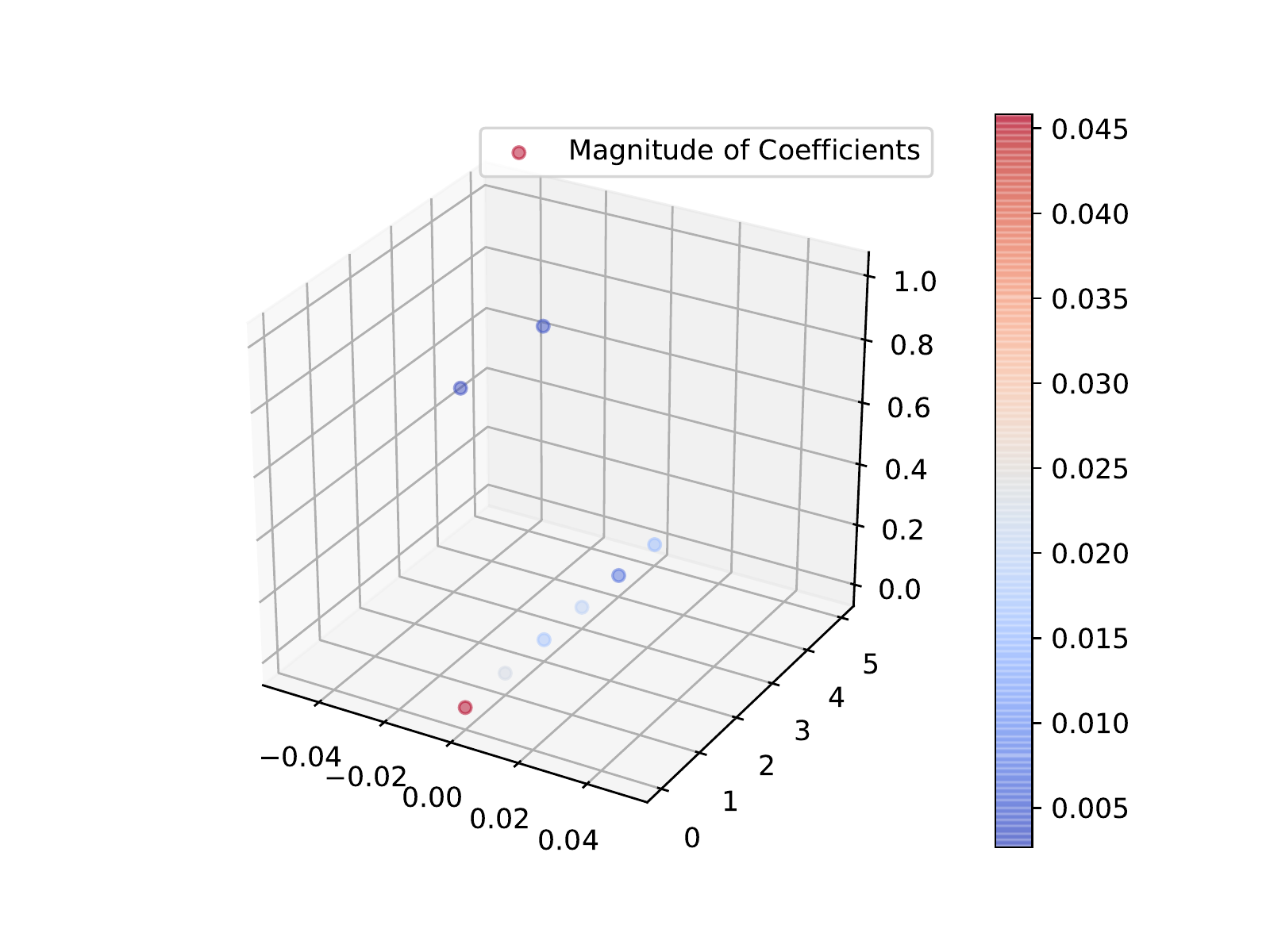} & 
            \includegraphics[width=0.4\linewidth]{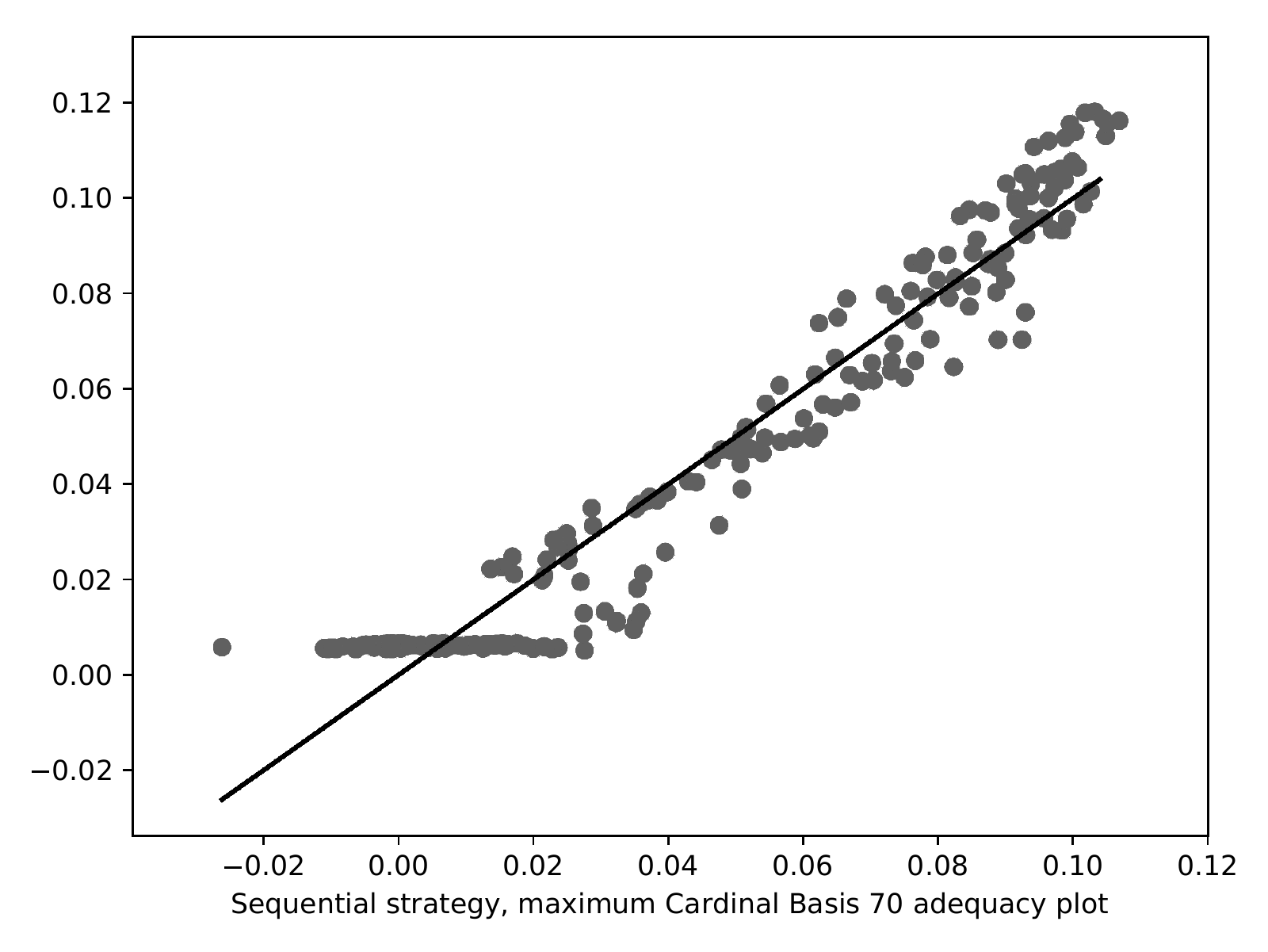} \\
\end{tabular}
\label{tbl:sparsity1}
\end{table}

\begin{table}
\centering
\caption{Same caption as Fig.~\ref{tbl:sparsity1} but for hyperbolic truncation with $q = 0.5$.}
\begin{tabular}{l|cc}
Quad. & &  \\  
      &    \includegraphics[width=0.4\linewidth]{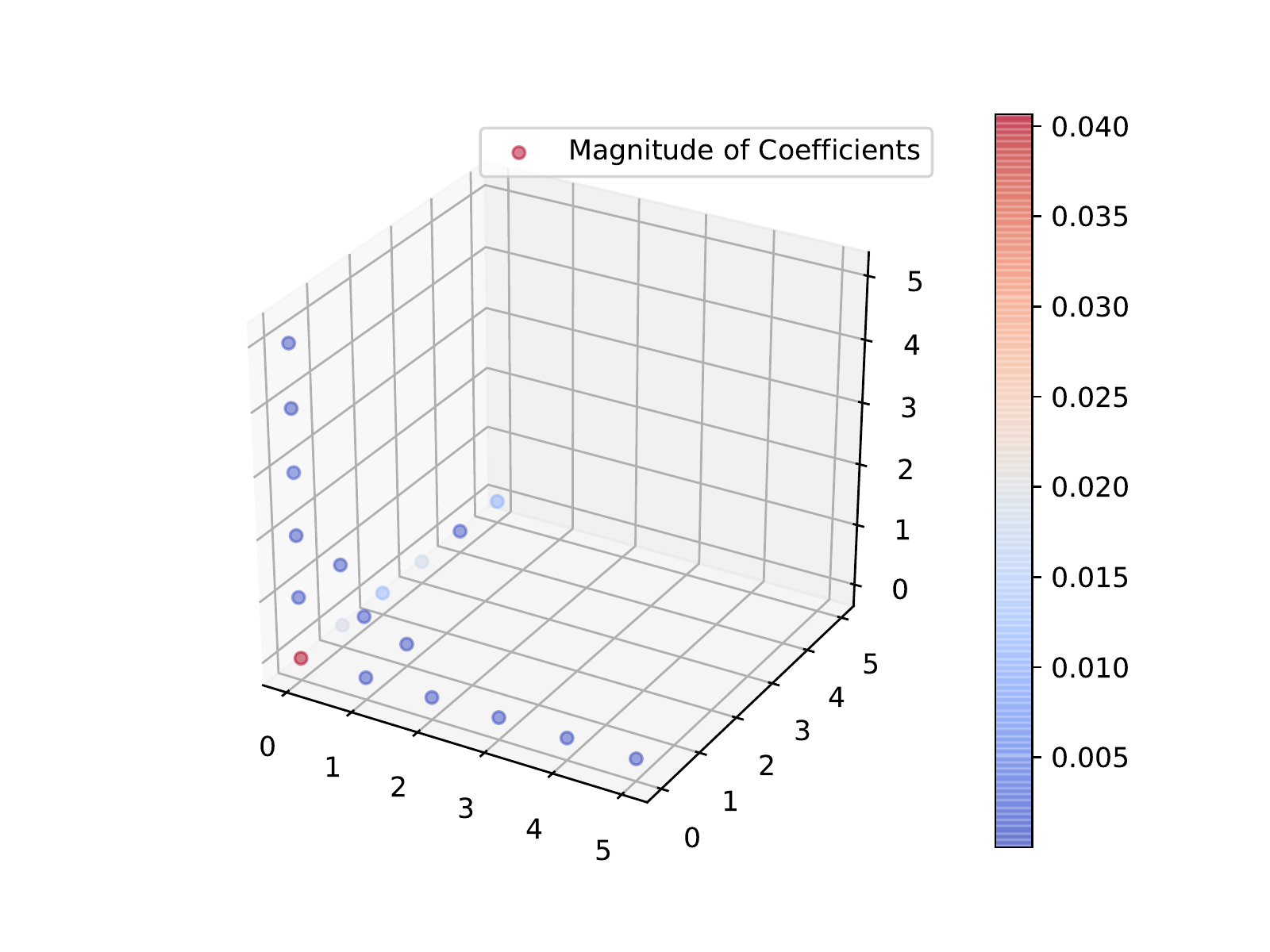} & 
            \includegraphics[width=0.4\linewidth]{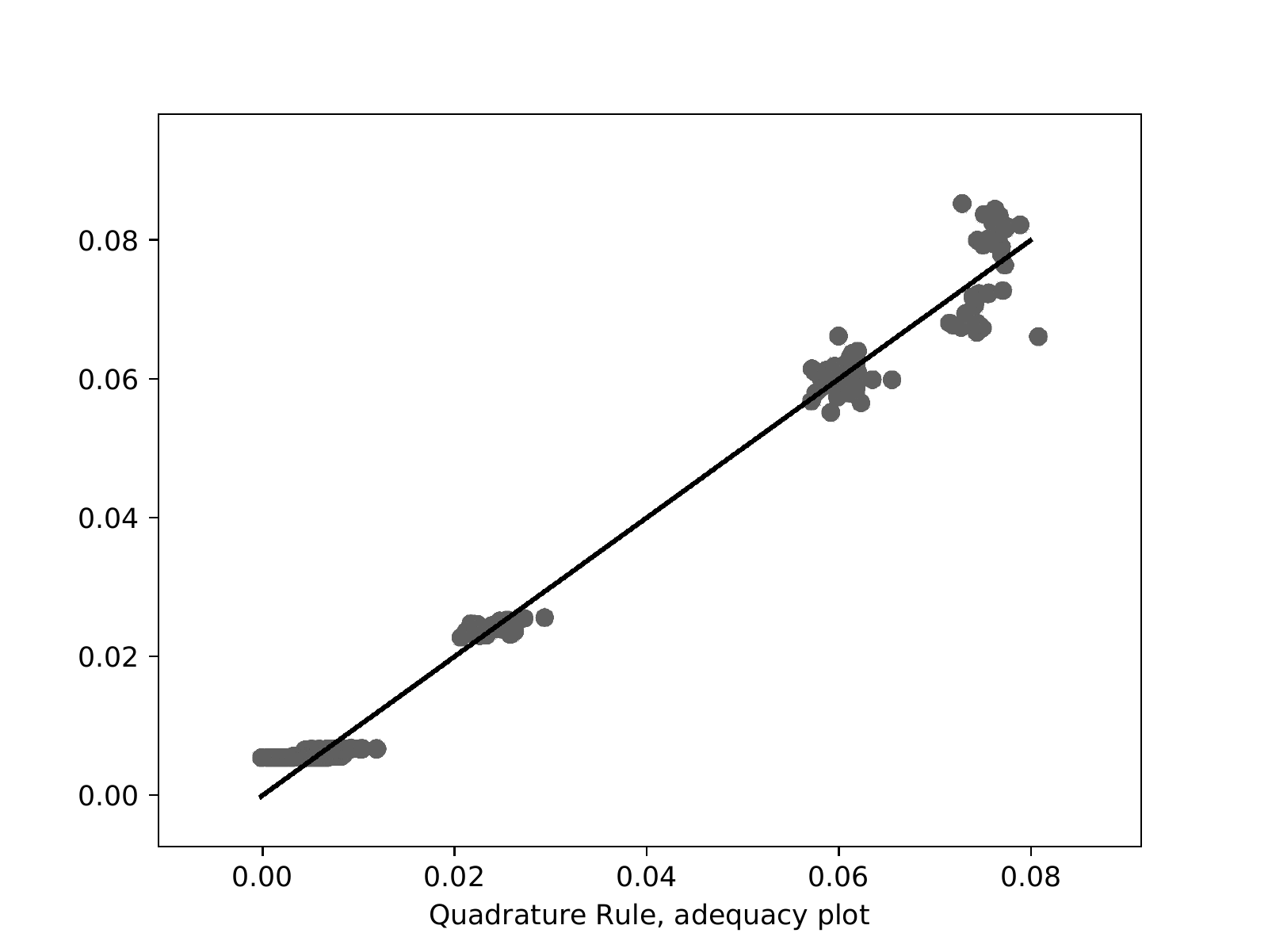} \\
SLS && \\ 
        &     \includegraphics[width=0.4\linewidth]{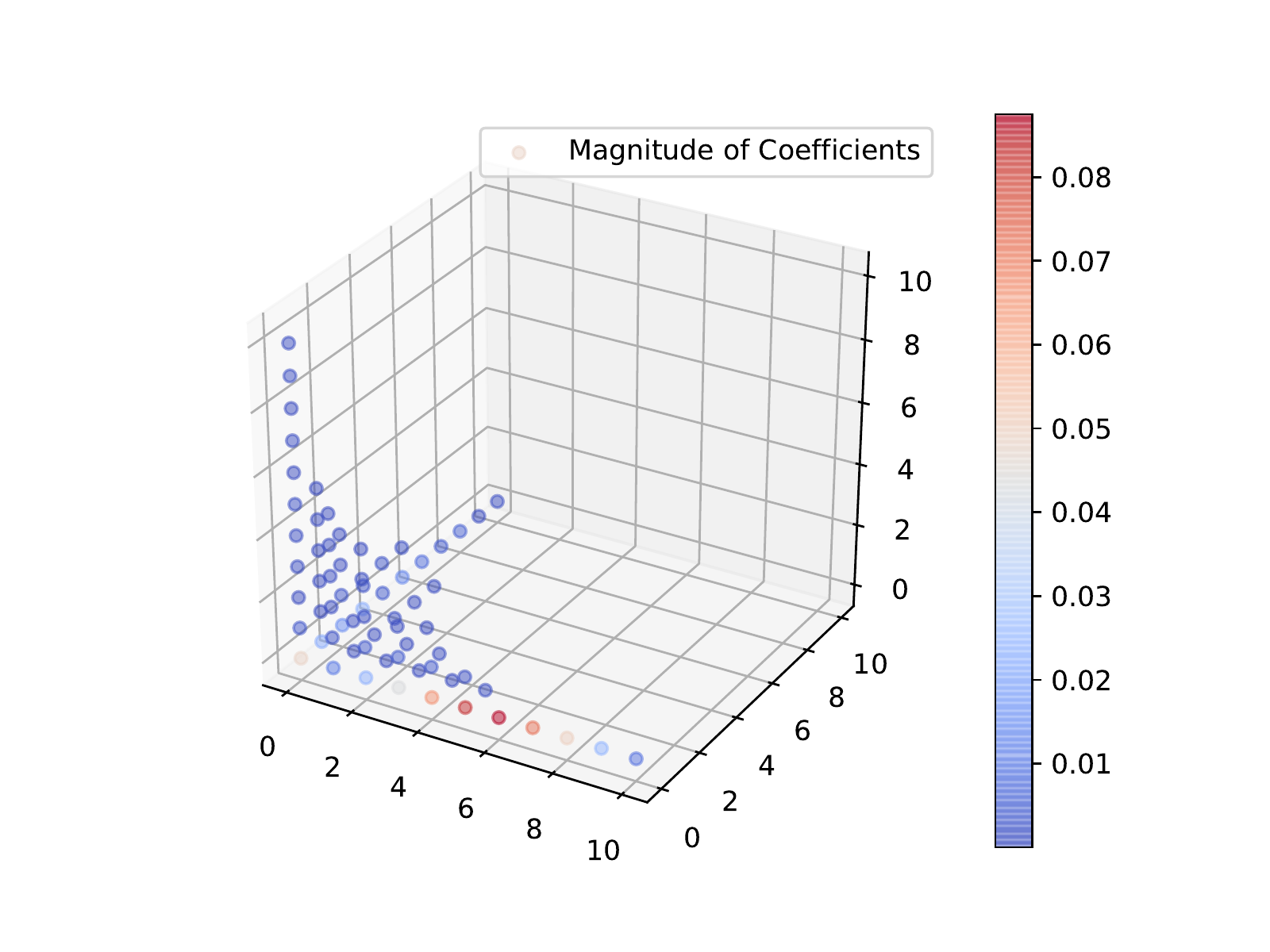} & 
            \includegraphics[width=0.4\linewidth]{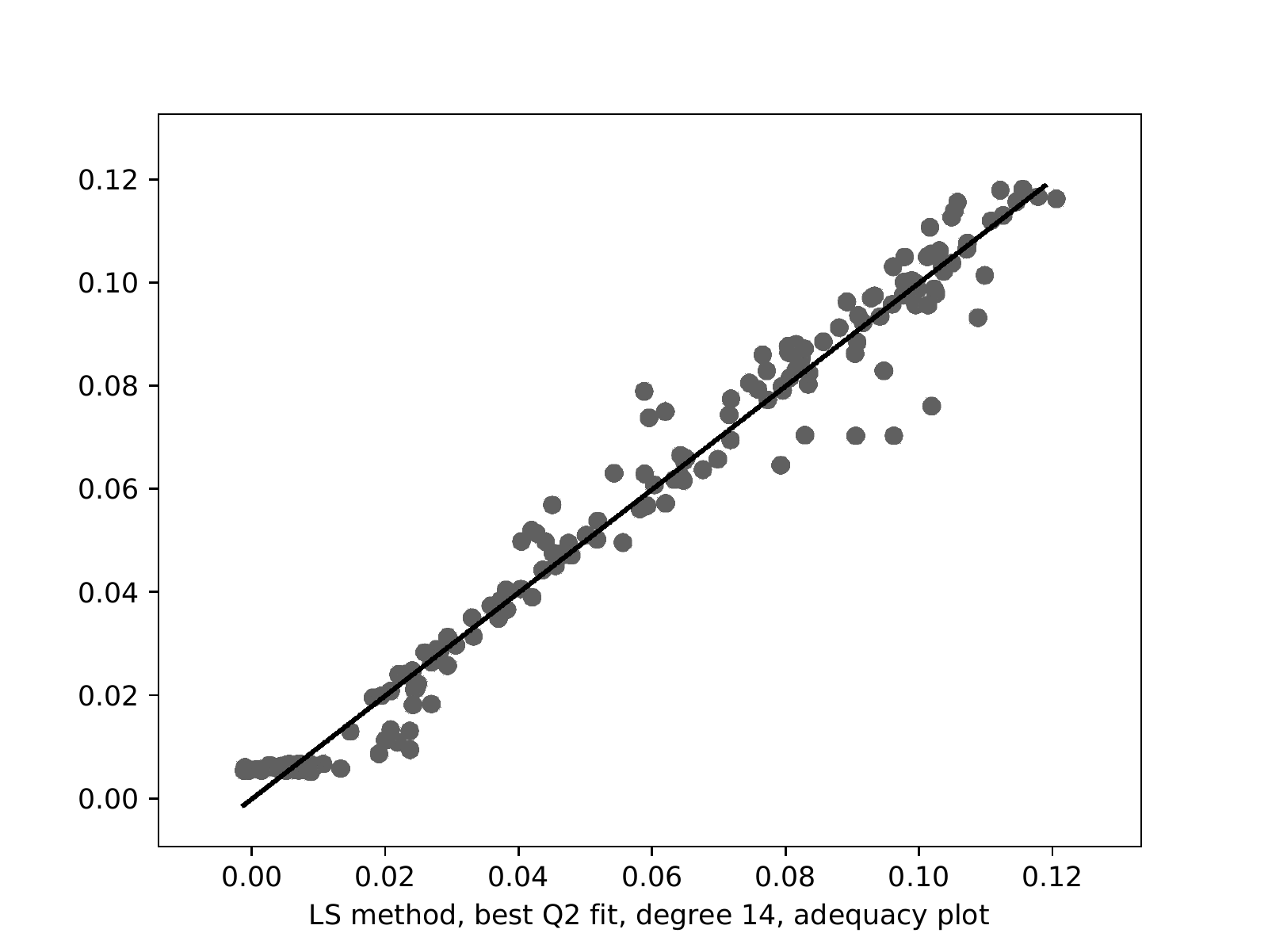} \\
LAR && \\   
      &      \includegraphics[width=0.4\linewidth]{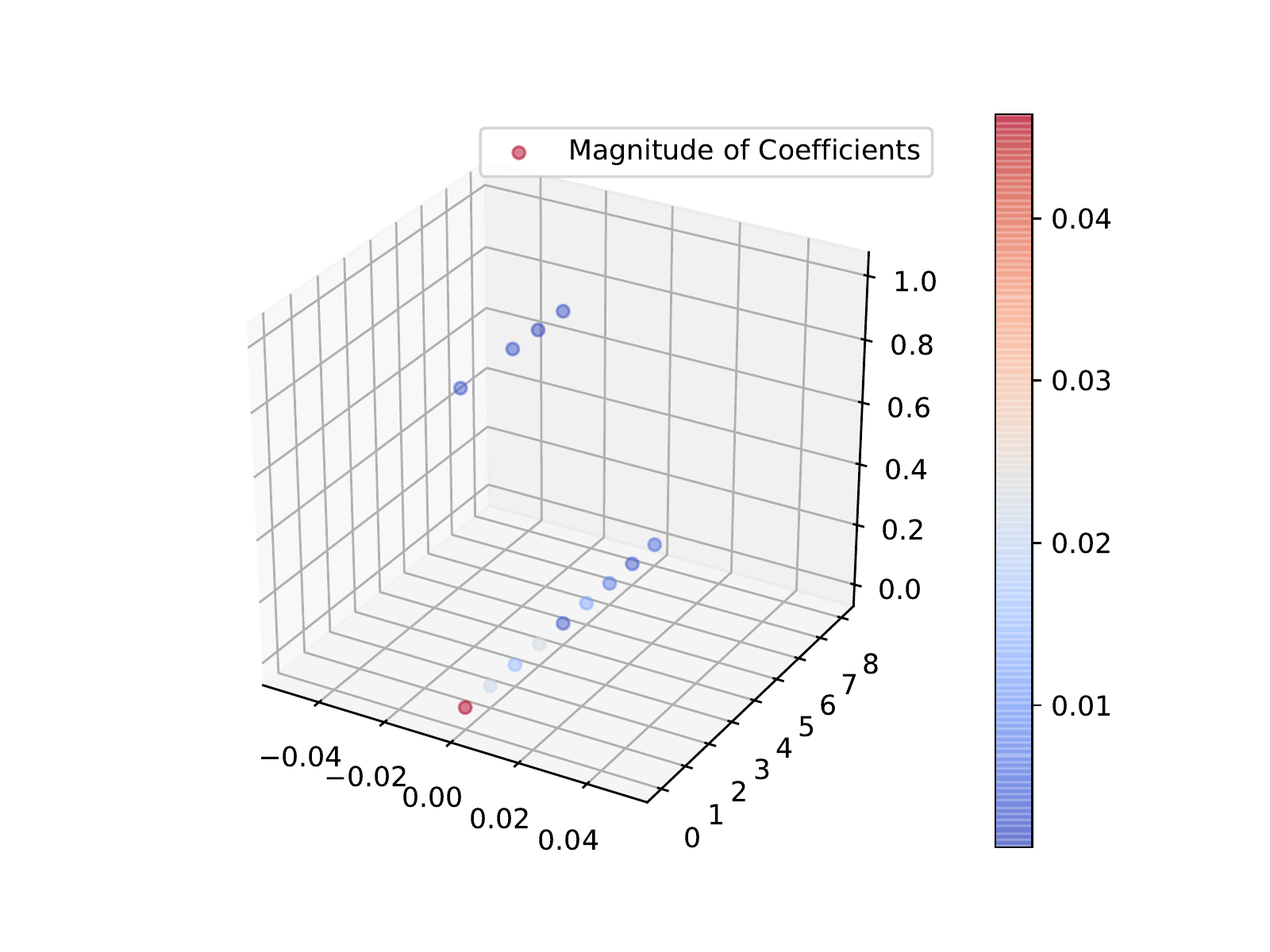} & 
            \includegraphics[width=0.4\linewidth]{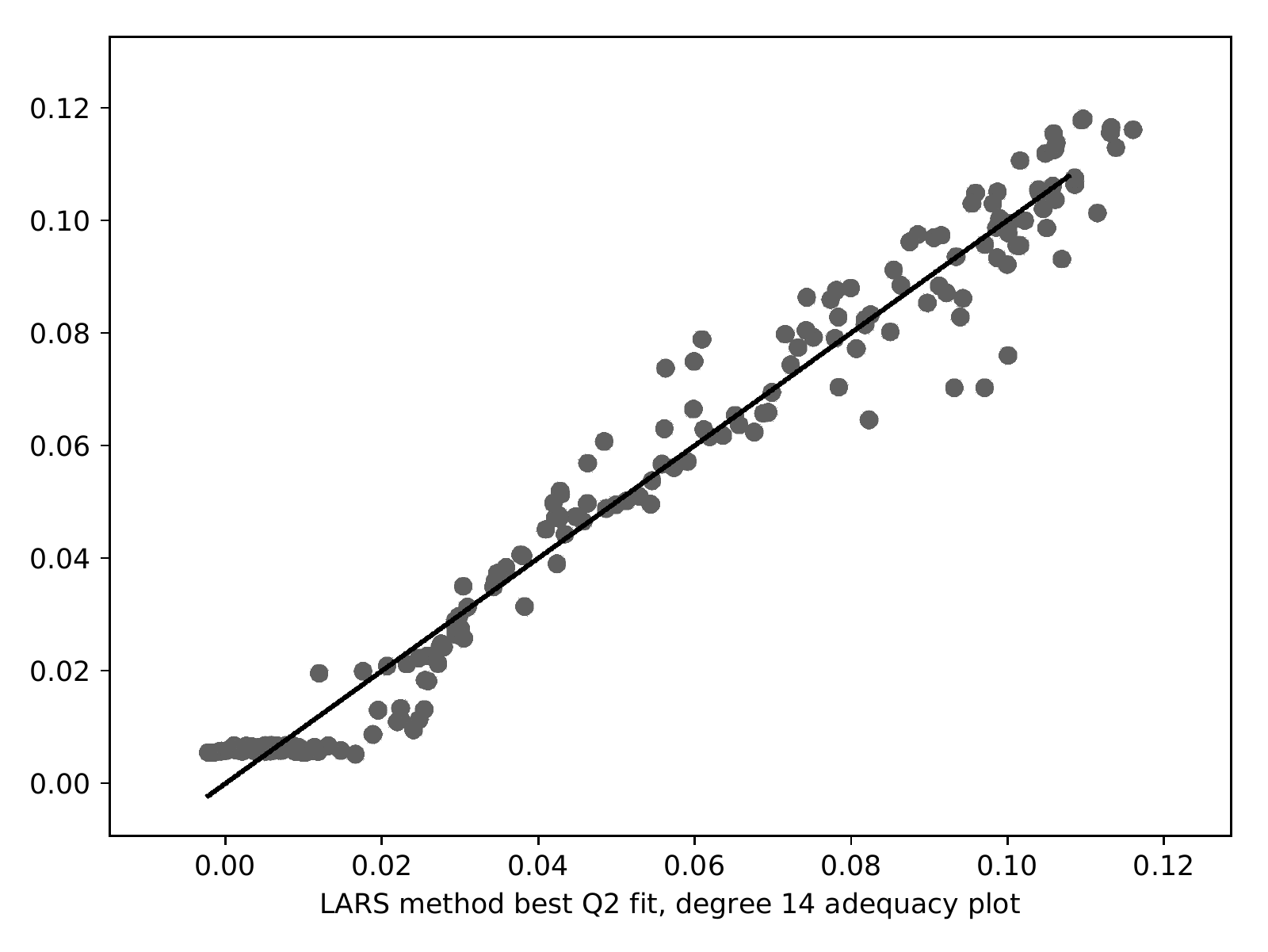} \\
Cleaning && \\      
        &      \includegraphics[width=0.4\linewidth]{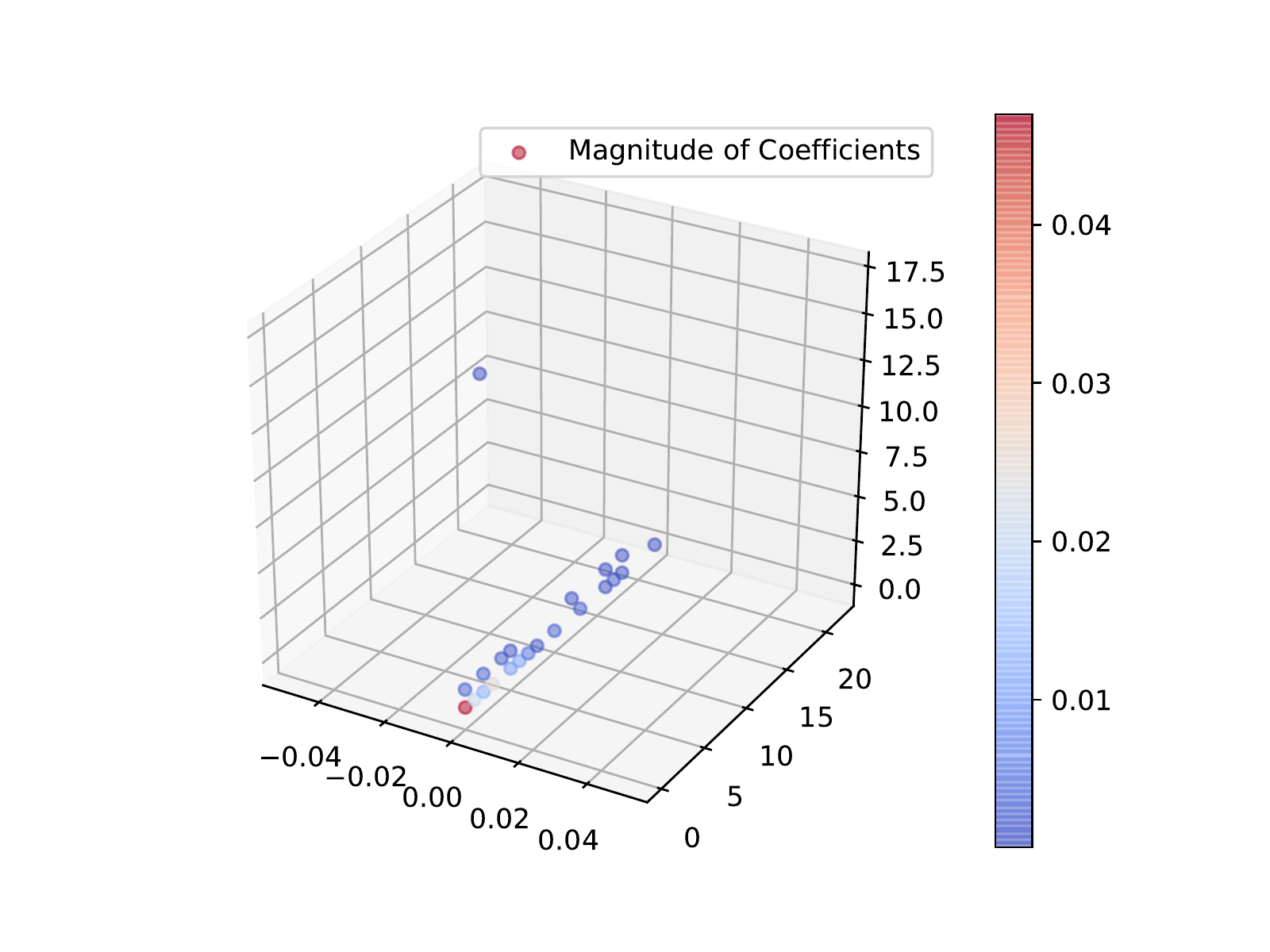} & 
            \includegraphics[width=0.4\linewidth]{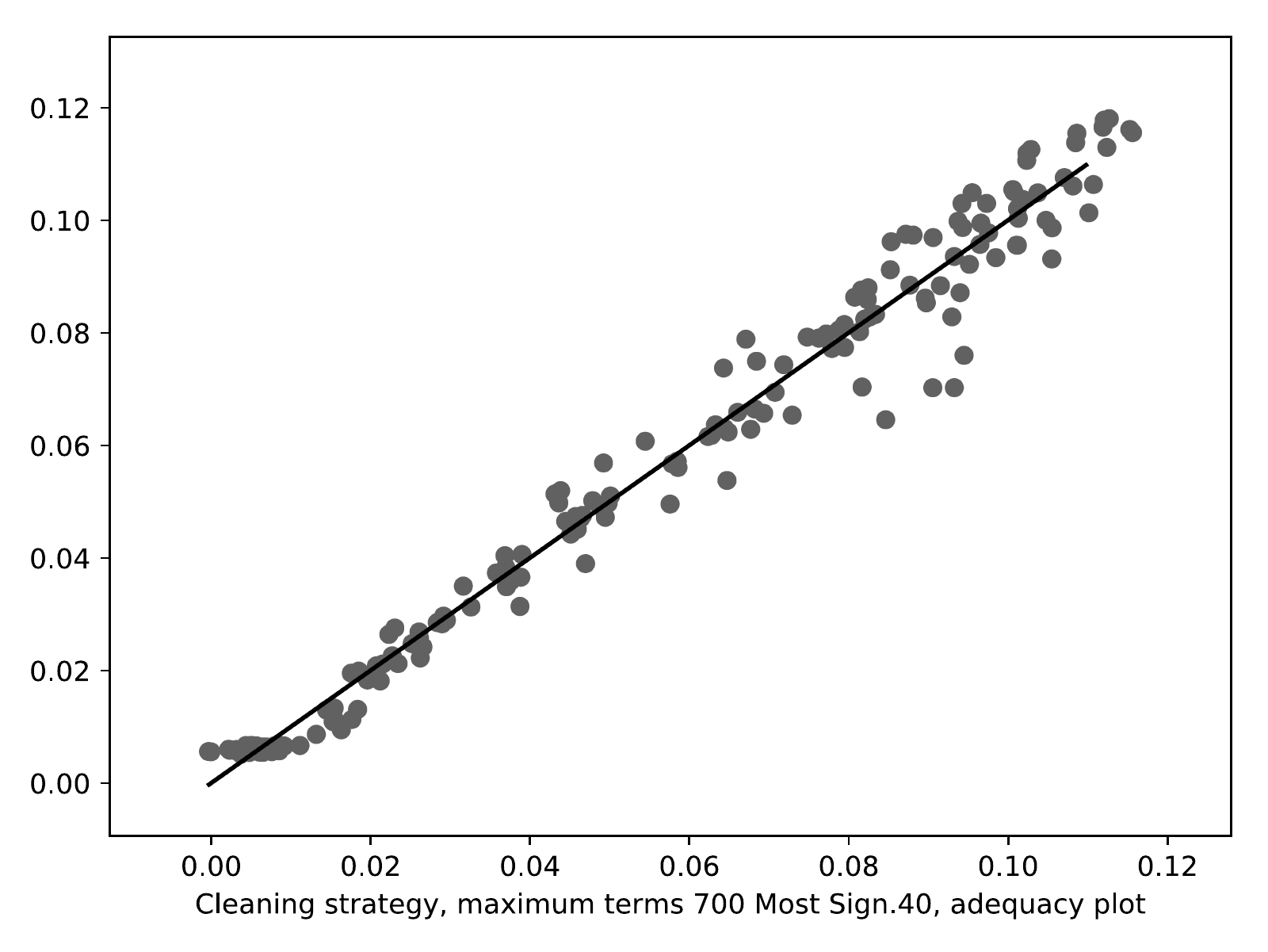} \\
Sequential && \\      
       &     \includegraphics[width=0.4\linewidth]{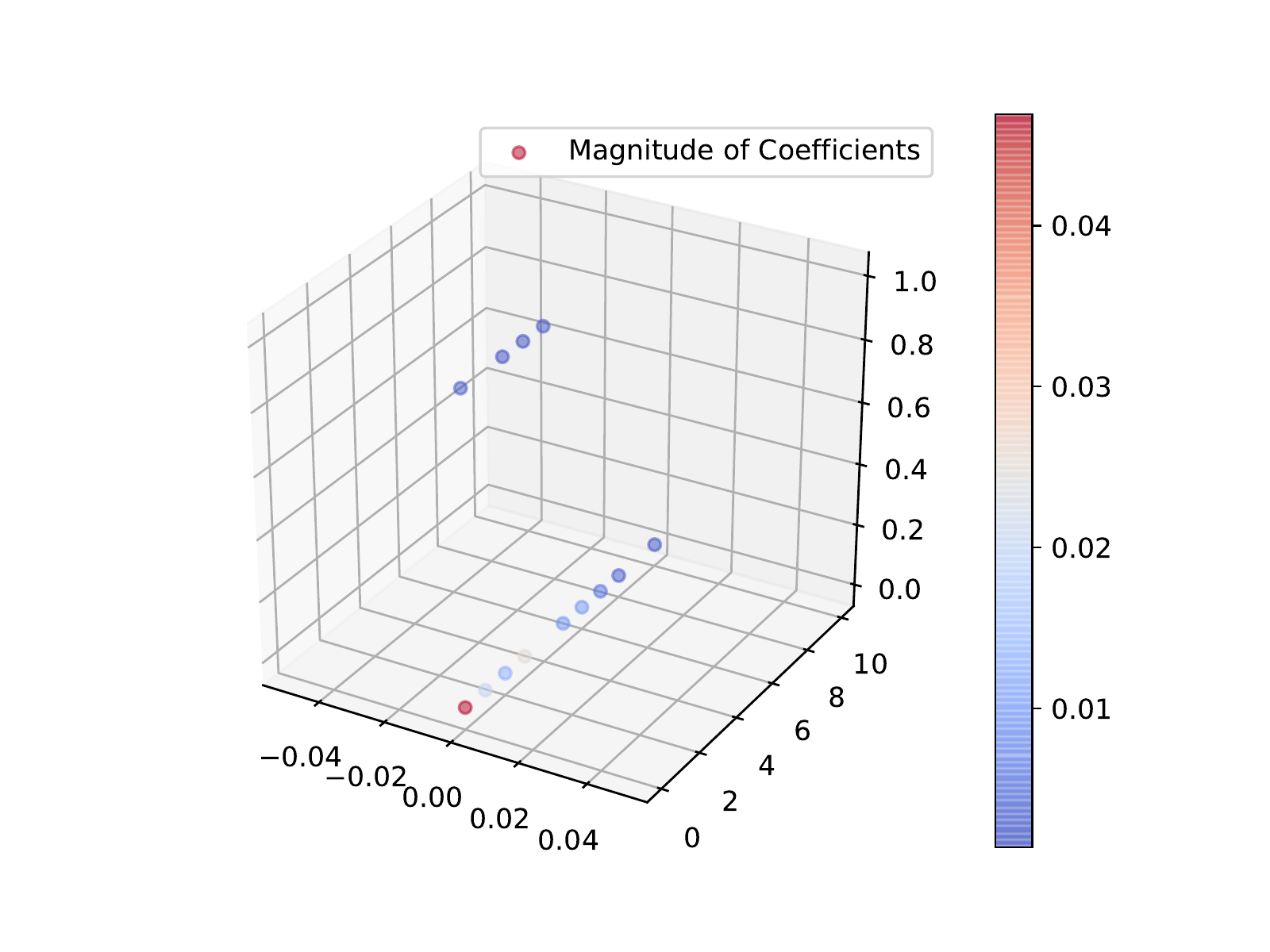} & 
            \includegraphics[width=0.4\linewidth]{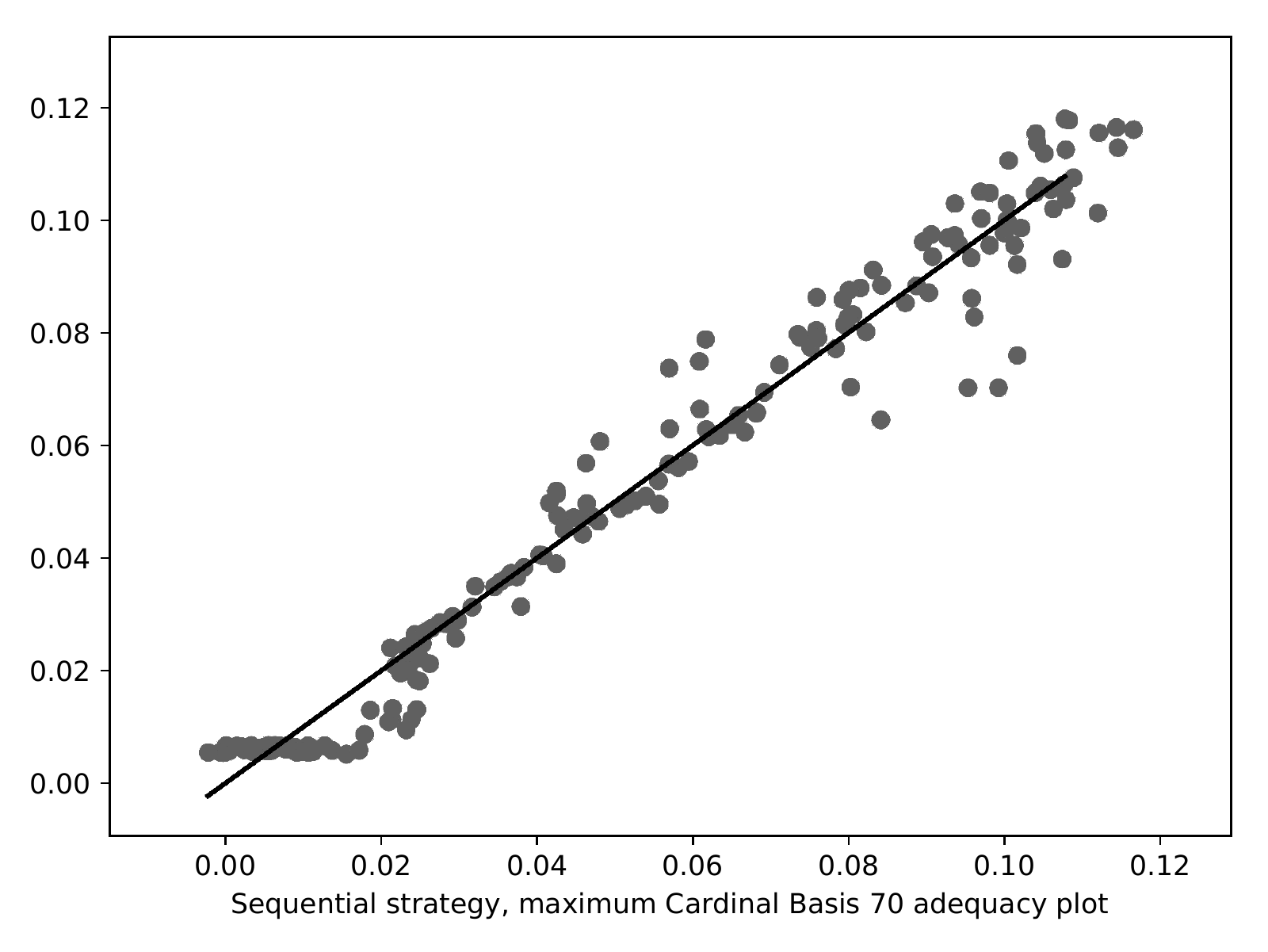} \\
\end{tabular}
\label{tbl:sparsity05}
    \end{table}

\subsubsection{Sensitivity of gPC-surrogates to total polynomial order $P$}

In Table~\ref{table:MSD_Q2_err}, the results for SLS and LAR methods are obtained by choosing the optimal value of the total polynomial order $P$ in the sense that the surrogate was obtained by finding the value of $P$ that maximizes the $Q_2$ predictive coefficient; $P$ varying between 1 and 14. Recall that the total polynomial order $P$ determines the size of the full basis used to construct the surrogate when using linear truncation. The SLS method considers the full basis, while the LAR method selects the most influential terms among the full basis. Since the size of the training set is fixed to $N = 216$ and since $(P +1)^3 = 216$ for $P = 5$, we know that the problem becomes ill-posed for a full basis when the total polynomial order is over 5. This is not an issue for LAR since it selects inline the influential coefficients in the basis. It is therefore of interest to investigate if the LAR method features an improved performance when $P > 5$.

\begin{figure}[h!]
\centering
\begin{subfigure}[b]{0.4\linewidth}
\includegraphics[width=\linewidth]{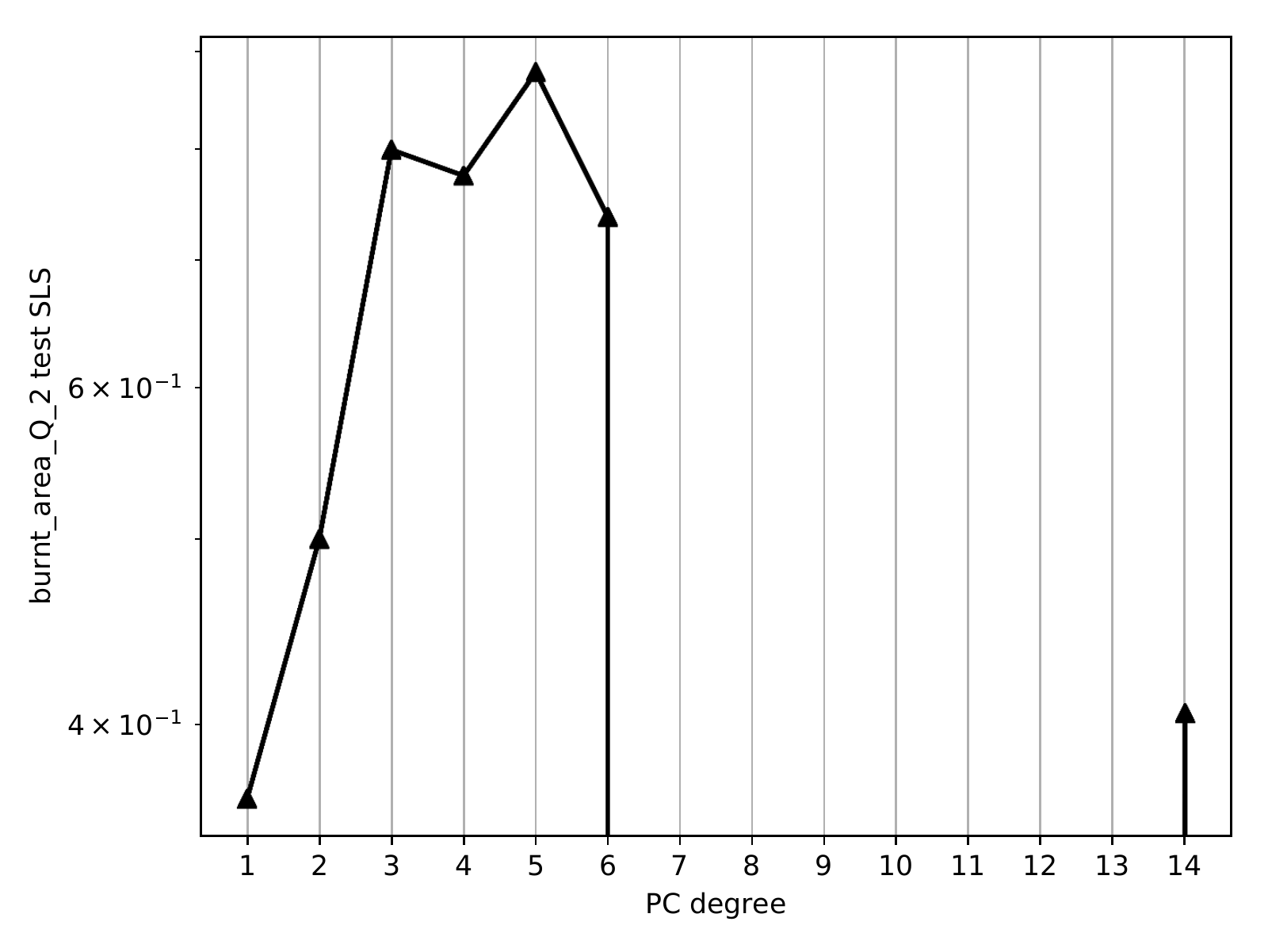}
\caption{SLS, $q = 1$.}
\end{subfigure}
\begin{subfigure}[b]{0.4\linewidth}
\includegraphics[width=\linewidth]{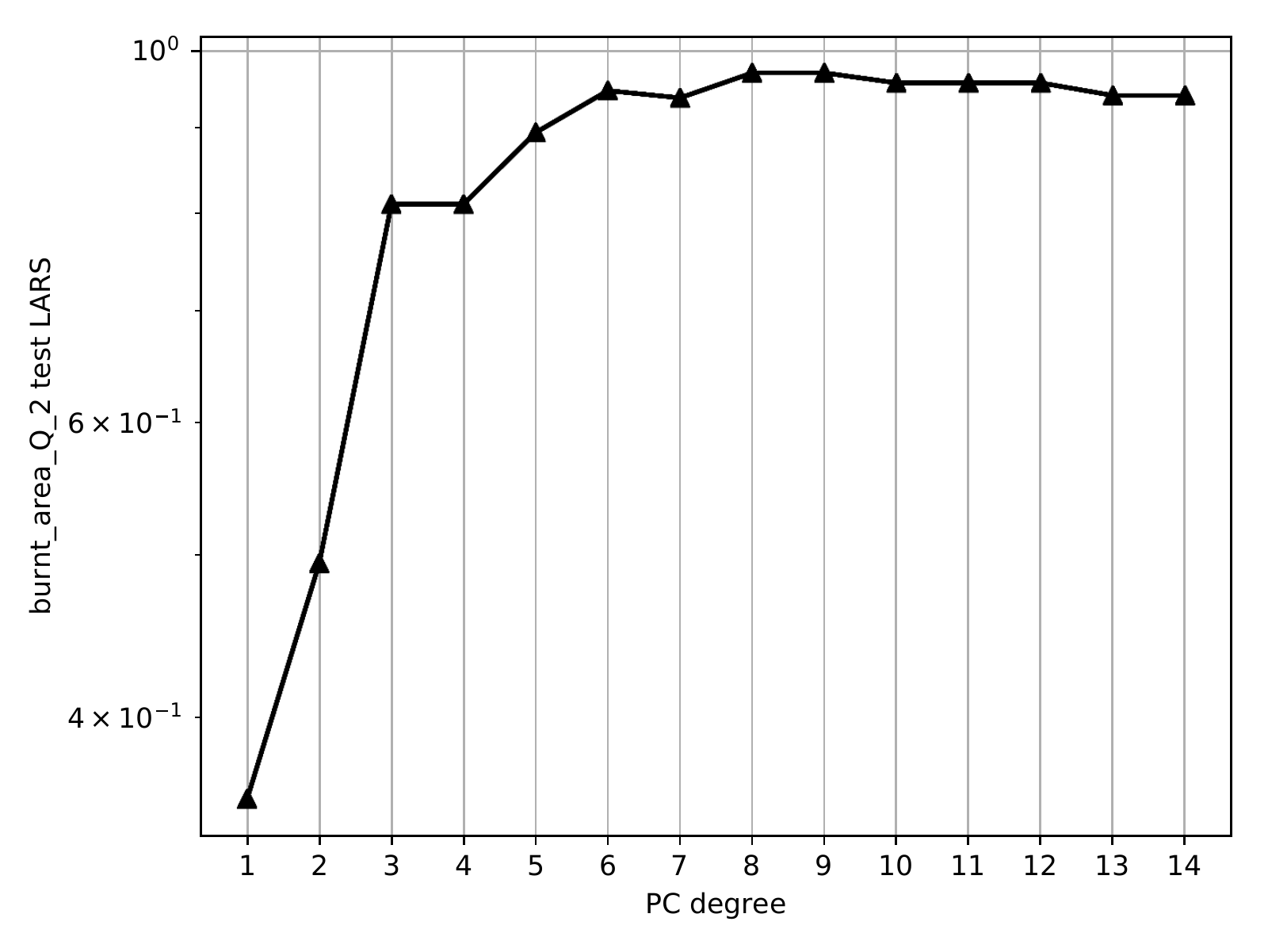}
\caption{LAR, $q = 1$.}
\end{subfigure}
\begin{subfigure}[b]{0.4\linewidth}
\includegraphics[width=\linewidth]{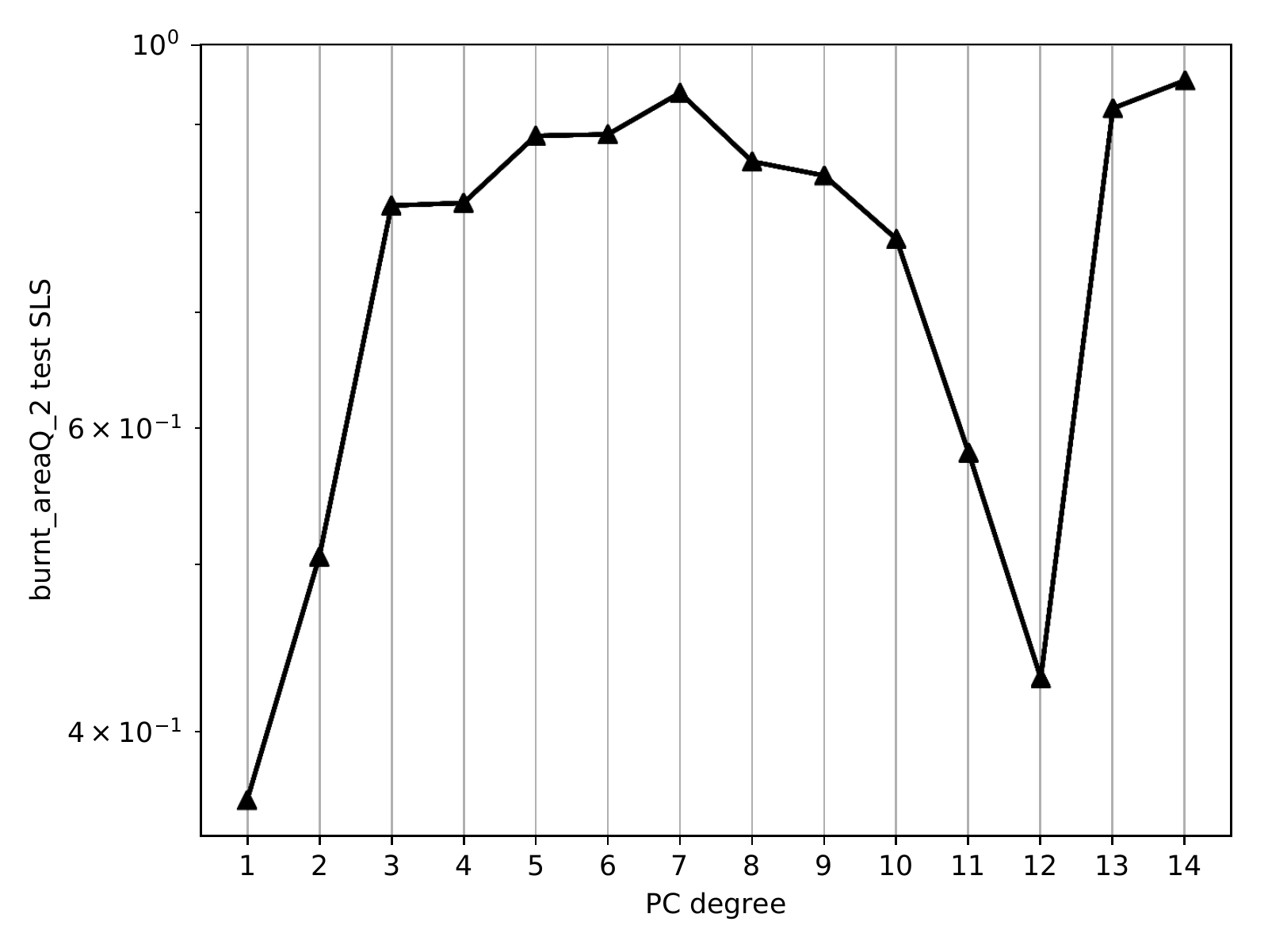}
\caption{SLS, $q = 0.5$.}
\end{subfigure}
\begin{subfigure}[b]{0.4\linewidth}
\includegraphics[width=\linewidth]{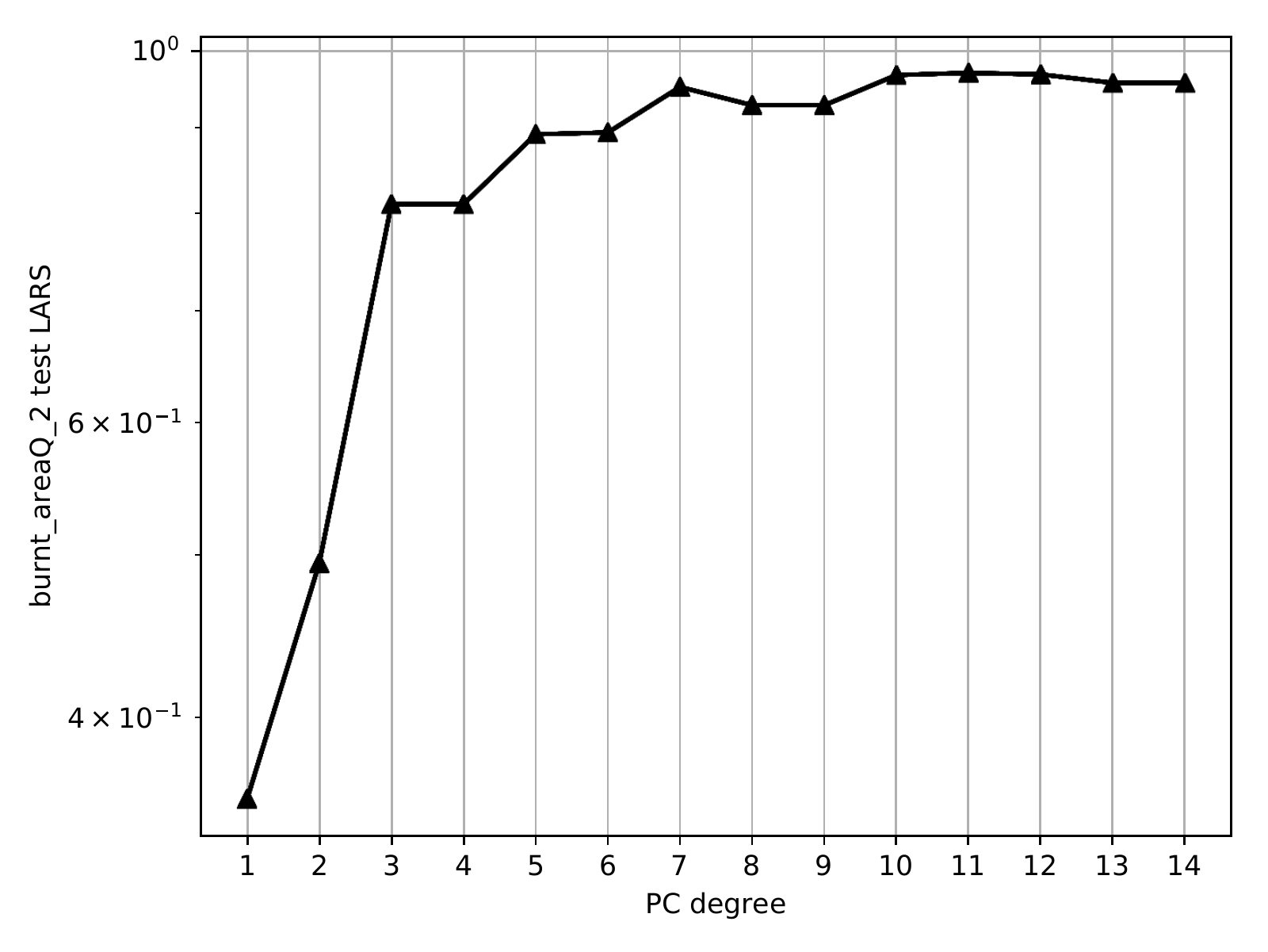}
\caption{LAR, $q = 0.5$.}
\end{subfigure}
\caption{Sensitivity of the $Q_2$ predictive coefficient with respect to the total polynomial order $P$. Comparison of the SLS (a)--(c) and LAR (b)--(d) surrogate methods for linear truncation (top panels) and hyperbolic truncation with $q = 0.5$ (bottom panels) for $1 \leq P \leq 14$.}
\label{fig:degreeSensitivity}
\end{figure}

Figure~\ref{fig:degreeSensitivity} presents the $Q_2$ predictive coefficient for $P$ varying between 1 and 14 for SLS and LAR surrogates obtained for the burnt area ratio $A_2$. As expected, Fig.~\ref{fig:degreeSensitivity}a shows that the best performance of the SLS method with linear truncation is obtained for $P = 5$ and that it degrades very fast when increasing $P$ (the $Q_2$ predictive coefficient is below 0.4 for $P > 6$). When moving to hyperbolic truncation with $q = 0.5$, Fig.~\ref{fig:degreeSensitivity}c shows that the $Q_2$ predictive coefficient remains over 0.4 for $P > 5$. The resulting surrogate is therefore improved in this configuration as already pointed out in Table~\ref{table:MSD_Q2_err}. Hyperbolic truncation allows the SLS approach to include high-order polynomials in the basis without generating an ill-posed problem (i.e.~without having more coefficients to compute than the size $N$ of the training set). Still, results show that the $Q_2$ predictive coefficient does not follow a monotonically increasing function toward the target value $1$ in this hyperbolic configuration; this configuration is therefore not robust. In the opposite, the LAR method shows a monotonic convergence towards the target value $1$ when increasing $P$ in Figs.~\ref{fig:degreeSensitivity}b--d. A good performance of LAR is obtained for $P = 10$ for both linear and hyperbolic truncation schemes.

This sensitivity study shows that a total polynomial order $P$ higher than~5 is required to build the response surface of the burnt area ratio. Similar results are obtained for the MSR ratio (not shown here). This demonstrates the benefits from sparse schemes when having a fixed and limited training set $\mathcal{D}_N$. Improving the performance of the SLS approach using linear truncation would require a higher total polynomial order $P$ and therefore a larger training set.

\subsubsection{Identification of the influential gPC-coefficients}

Table~\ref{tbl:sparsity1} (left column) presents a three-dimensional schematic (referred to as ``sparsity plot") of the coefficients retained in the gPC-expansion using linear truncation, each dimension corresponding to one stochastic/uncertain dimension. The three dimensions are here the turbulent diffusion coefficient $D$ and the lognormal parameters $\mu$ and $\sigma$. This is useful to visualize the polynomial degree associated with the active coefficients as well as the magnitude of the coefficients given by the colormap (recall that there is a direct link between the coefficients and the statistical moments of the predicted quantity of interest for gPC-expansion). 

Quadrature and SLS methods have the same full basis for a given polynomial order $P$ (here $P = 5$ since the size of the training set is $N = 216$); they are associated with a typical ``pyramidal" sparsity plot, where the first coefficient corresponding to the mean estimate of the burnt area ratio $A_2$ has the highest magnitude (approximately equal to 0.04). For sparse methods (LAR, cleaning, sequential), the number of coefficients is significantly reduced since the terms with the least impact are automatically filtered out of the sparse basis. The sparsity plot has no longer a ``pyramidal" shape. LAR and sequential strategies feature instead a two-dimensional structure (along the vertical plane) indicating that the burnt area ratio $A_2$ is not sensitive to the third dimension, here the lognormal parameter $\mu$, but only to the lognormal parameter $\sigma$ and to the turbulent diffusion coefficient $D$. Only the cleaning strategy retains a three-dimensional structure by accounting for interaction terms involving the lognormal parameter $\mu$. This highlights the presence of influential interaction terms involving several parameters. However, all sparse strategies indicate that one direction is dominant since the number of coefficients in this direction is high and the basis terms can go up to a total polynomial order $P = 12$ in the case of cleaning and $P = 8$ in the case of LAR (instead of the constrained $P = 5$ for quadrature and SLS). This dominant direction corresponds to the lognormal parameter $\sigma$. 

Note that Table~\ref{tbl:sparsity05} (left column) presents similar plots as Table~\ref{tbl:sparsity1} (left column) but for hyperbolic truncation with $q = 0.5$. The magnitude of the coefficients does not change for quadrature, explaining why hyperbolicity does not improve the performance of the surrogate based on quadrature. This is not the case of SLS, which now features high magnitude for the coefficients along the direction $D$ for polynomial terms having a degree between 4 and 8. This highlights the need to have polynomials of higher degree to capture underlying physical processes. Still, SLS with hyperbolicity is not sufficient to capture the same structure as sparse methods. Note that sparse methods converge to the same structure using linear or hyperbolic truncation schemes, indicating the robustness of these methods. 

The influence of the three parameters on the behavior of the burnt area ratio $A_2$ can be quantified using Sobol' sensitivity indices. Table~\ref{table:MSD_SOBOL_A2} presents the Sobol' indices using sparse methods and linear truncation for the burnt area ratio $A_2$ (same results are obtained using hyperbolic truncation with $q = 0.5$ -- not shown here). Table~\ref{table:MSD_SOBOL_S2} presents similar quantities for the MSR ratio $S_2$. Results confirm that the lognormal parameter $\sigma$ is the most influential one for both quantities of interest $A_2$ and $S_2$ with a first-order sensitivity index above 0.98 for $A_2$ and above 0.92 for $S_2$. This means that more than 90~\% of the variance in $A_2$ and $S_2$ is explained by uncertainties in the lognormal parameter $\sigma$. Results also show interaction effects are limited but still present between the lognormal parameter $\sigma$ and the turbulent diffusion parameter $D$ as foreseen in sparsity plots. Note that all sparse gPC-surrogates as well as the GP-model exhibit the same global trend. The main differences lie in the relevance of the lognormal parameter $\mu$. LAR and sequential strategies cut out any contribution of $\mu$ in the variability of the predicted quantities of interest. This is not the case of the cleaning strategy that has a non-zero total Sobol' index for $\mu$ as the GP-model.

We can evaluate the impact of the choice in the surrogate strategy on the predicted mean and STD estimates of the quantities of interest. Note that the coefficients of the gPC-expansion can be interpreted in a statistical way with the first coefficient being the mean estimate and the squared sum of the other coefficients being its corresponding variance estimate (see Section~\ref{sec:SA}). Table~\ref{table:MSD_Mean_Variance} presents the mean and STD estimate of the burnt area ratio $A_2$ and of the MSR ratio $S_2$ obtained for different gPC- and GP-surrogates. Results show the consistency of the statistical moments obtained using sparse gPC-expansions and GP-model for both $A_2$ and $S_2$. The SLS approach using linear truncation is able to retrieve accurate mean and STD estimates (about 1~\% deviation with respect to GP-model predictions). In the opposite, the quadrature approach provides mean and STD estimates with more than 10~\% deviation with respect to GP-model predictions. 

This highlights the importance of having high-order polynomial terms in some uncertain directions to build an accurate gPC-expansion and have accurate estimate of the statistical moments in the present study. These directions can be identified using Sobol' sensitivity indices. Sparse gPC-strategies are relevant to address such issues due to the flexibility of selecting the most influential polynomial terms during the construction of the surrogate (linear and hyperbolic schemes are defined a priori).

\begin{table}[h]
\caption{Comparison of Sobol' sensitivity indices associated with the burnt area ratio $A_2$ and obtained for Halton's low discrepancy sequence.}
\begin{tabular}{c|c|c|c|c|c|c|}
\cline{2-7}
& {$S_\mu$}  & $S_\sigma$ & $S_D$ &$S_{T,\mu}$ &$ S_{T,\sigma}$ &  $ S_{T,D}$ \\
\cline{2-7}
\cline{2-7}
& \multicolumn{6}{ |c| }{\textbf{gPC expansion -- Linear truncation} $q = 1$} \\
\cline{1-7}
\multicolumn{1}{ |c| }{LAR}            &  0.      &  0.986 & $5.67\cdot 10^{-3}$ & 0. & 0.994 & $1.35 \cdot 10^{-2}$ \\  \multicolumn{1}{ |c| }{Cleaning}     &	0.      &  0.984 & $5.89\cdot 10^{-3}$ & $4.70 \cdot 10^{-3}$ & 0.994 & $1.62 \cdot 10^{-2}$ \\  
\multicolumn{1}{ |c| }{Sequential}  &  0.      &  0.987 & $4.84 \cdot 10^{-3}$ & 0.       & 0.995 & $1.33 \cdot 10^{-2}$  \\  
\cline{1-7}\cline{1-7} & \multicolumn{6}{ |c| }{\textbf{GP model}} \\
\cline{1-7}\cline{1-7}
\multicolumn{1}{ |c| }{RBF kernel}  &  $4.59 \cdot 10^{-4}$ & 0.982 & $5.97 \cdot 10^{-3}$ & 0.001 & 0.992 & 0.012 \\  \hline
\end{tabular}
\label{table:MSD_SOBOL_A2}
\end{table}
\begin{table}[h]
\caption{Same caption as Table~\ref{table:MSD_SOBOL_A2} but for the MSR ratio $S_2$.}
\begin{tabular}{c|c|c|c|c|c|c|}
\cline{2-7}
& {$S_\mu$}  & $S_\sigma$ & $S_D$ &$S_{T,\mu}$ &$ S_{T,\sigma}$ &  $ S_{T,D}$ \\
 \cline{2-7}
   \cline{2-7}
& \multicolumn{6}{ |c| }{\textbf{gPC expansion -- Linear truncation} $q = 1$} \\
\cline{1-7}
\multicolumn{1}{ |c| }{LAR}            & 0.   & 0.948 & $1.49\cdot 10^{-2}$ & 0. & 0.985 & $5.22 \cdot 10^{-2}$ \\ 
\multicolumn{1}{ |c| }{Cleaning}     & 0.   & 0.925 & $1.66\cdot 10^{-2}$ & $2.66\cdot 10^{-3}$ & 0.983 & $7.18\cdot 10^{-2}$ \\
\multicolumn{1}{ |c| }{Sequential}  & 0.   & 0.954 & $1.45\cdot 10^{-2}$ & $7.15\cdot 10^{-3}$ & 0.978 & $4.63\cdot 10^{-2}$  \\ 
\cline{1-7}\cline{1-7} & \multicolumn{6}{ |c| }{\textbf{GP model}} \\
\cline{1-7}\cline{1-7}
\multicolumn{1}{ |c| }{RBF kernel} &  $5.43\cdot 10^{-4}$ & 0.941 & $9.89\cdot 10^{-3}$ & 0.002 & 0.975 & 0.047 \\  \hline
\cline{1-7}
\end{tabular}
\label{table:MSD_SOBOL_S2}
\end{table}

\begin{table}[h]
\caption{Mean and STD estimate of the burnt area ratio $A_2$ (left column) and of the MSR ratio $S_2$ (right column) using linear truncation scheme ($q=1$), Halton's low discrepancy sequence and gPC or GP surrogate approach.}
\begin{tabular}{c|c|c|}
\cline{2-3}\cline{2-3} &  \multicolumn{1}{ |c| }{$A_2$} & \multicolumn{1}{ |c| }{$S_2$} \\ \cline{2-3}
\cline{2-3}\cline{2-3}& \multicolumn{2}{ |c| }{\textbf{gPC expansion -- Linear truncation} ($q=1$)} \\
& mean $\pm$ STD & mean $\pm$ STD \\ \hline
\multicolumn{1}{ |c| }{Quad.}         & 0.0406 $\pm$ 0.175  & 0.102 $\pm$ 0.322 \\ 
\multicolumn{1}{ |c| }{SLS}            & 0.0458 $\pm$ 0.198  & 0.114 $\pm$ 0.333 \\  
\multicolumn{1}{ |c| }{LAR}            & 0.0464 $\pm$ 0.194  & 0.114 $\pm$ 0.324 \\  
\multicolumn{1}{ |c| }{Cleaning}     & 0.0469 $\pm$ 0.194  & 0.115 $\pm$ 0.327 \\  
\multicolumn{1}{ |c| }{Sequential}  & 0.0458 $\pm$ 0.196  & 0.113 $\pm$ 0.319 \\  \hline
\cline{2-3}\cline{2-3}& \multicolumn{2}{ |c| }{\textbf{GP model}} \\
& mean $\pm$ STD & mean $\pm$ STD \\ \hline
\multicolumn{1}{ |c| }{RBF kernel}  & 0.0463 $\pm$ 0.194 & 0.114 $\pm$ 0.327 \\  \hline
\end{tabular}
\label{table:MSD_Mean_Variance}
\end{table}

\subsubsection{Sensitivity to the size of the training set}

\begin{figure}[h!]
\centering
\includegraphics[width=0.6\linewidth]{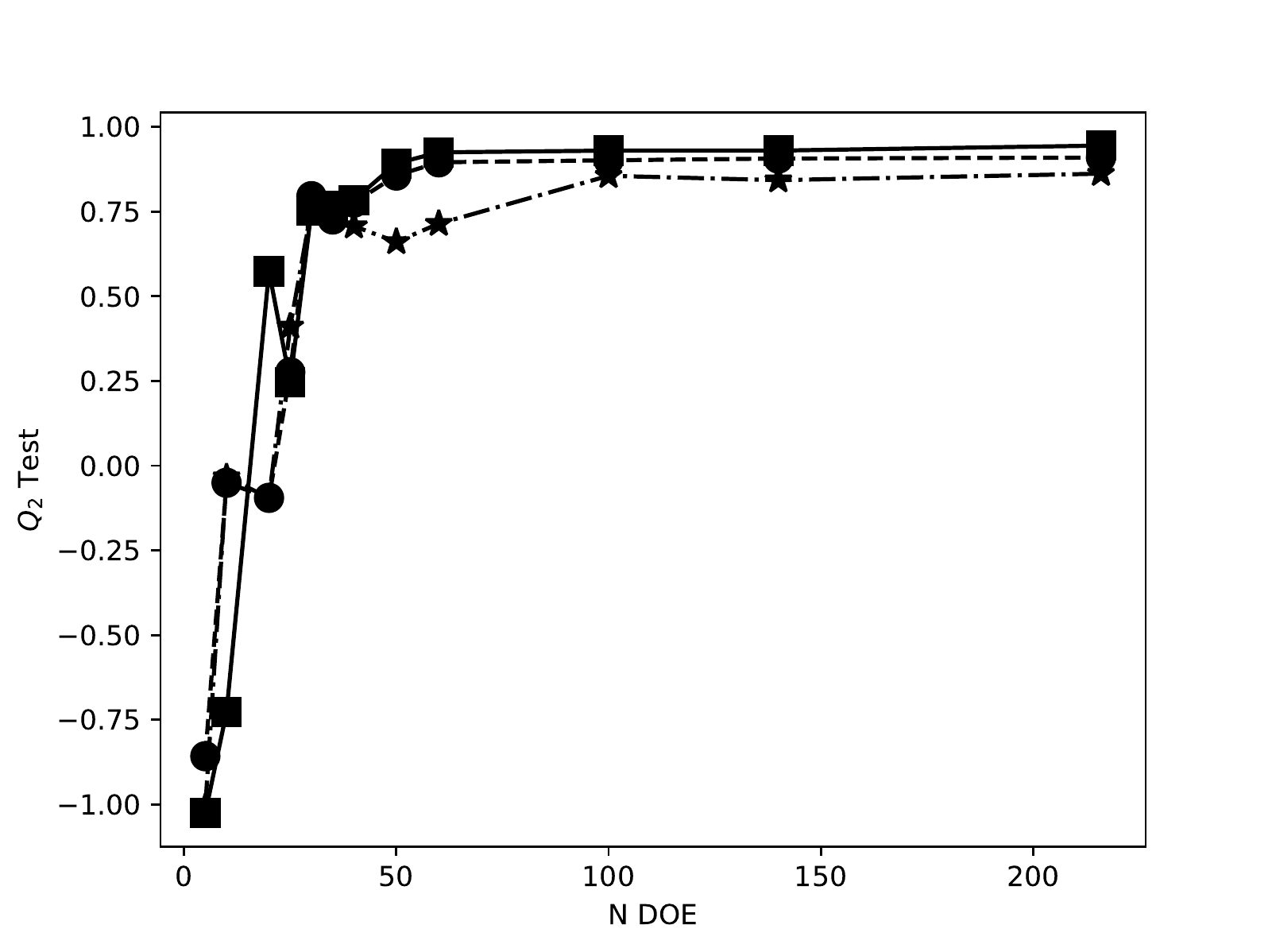} 
\caption{Convergence test with respect to $Q_2$ predictive coefficient for the LAR gPC-surrogate built using Halton's low discrepancy sequence (cross-validated using the Monte Carlo random sampling). Solid line with square symbols corresponds to linear truncation; dash-dotted line with star symbols corresponds to hyperbolic truncation with $q = 0.75$; and dashed line with circle symbols corresponds to hyperbolic truncation with $q = 0.5$.}
\label{fig:convergence_doe}
\end{figure}

So far the analysis was obtained for a fixed training set of size $N~=~216$ (generated using Halton's low discrepancy sequence or tensor-based Gauss quadrature in the case of quadrature). It is of interest to study if the same level of accuracy could be obtained for sparse gPC-surrogates built with a reduced training set ($N < 216$). To answer this question, we provide a convergence test for a training size $N$ varying between 10 and 216 with respect to the observable $S_2$.  For each size of the training set, a LAR gPC-surrogate is built and cross-validated using the available Monte Carlo database (Table~\ref{sec:doe}) through the computation of the $Q_2$ predictive coefficient. We carry out this convergence test for different truncation strategies, i.e.~ for different levels of hyperbolicity $q \in \lbrace 1, 0.75, 0.5\rbrace$. Figure~\ref{fig:convergence_doe} presents the evolution of $Q_2$ with respect to the size of the training set $N$. Results show the convergence of $Q_2$ to a constant value for $N > 100$. Linear truncation and hyperbolic truncation ($q = 0.5$) provide similar performance for $N > 100$. As before, we note that the hyperbolic solution obtained using $q = 0.75$ is not the best option.

\subsection{Analysis of the physical model predictions}\label{sec:PhysResults}

Results show that the LAR gPC-strategy features a good performance. In the following, we will use this strategy to further analyze the fire-spotting and turbulence submodel included in {\tt LSFire+}. We summarize in Table~\ref{tab:res_x1} and Table~\ref{tab:res_x2} the error metrics as well as the mean and STD estimate of the burnt area ratio $A_2$ and of the MSR ratio $S_2$ at time $t_2$ for the two sets of uncertain parameters $\boldsymbol{\theta} = \left(\left\lVert\mathrm{U}\right\rVert, I, \tau \right)^T$ and $\boldsymbol{\theta} = \left(\mu ,\sigma, D\right)^T$, respectively. Table~\ref{tbl:modelresultsUIT} and Table~\ref{tbl:table_of_figures0} present the corresponding Sobol' Indices. Note that the following analysis holds for any time $t$ since we show that results can be considered as time-independent. Note also that the empirical error $\epsilon_{\text{emp}}$ and the $Q_2$ predictive coefficient are in acceptable range for all tested configurations; we focus here on the physics of the problem.

Sobol' sensitivity indices order by relevance each parameter. In the case $\boldsymbol{\theta} = \left(\left\lVert\mathrm{U}\right\rVert, I, \tau \right)^T$, a clear dominance of the wind speed $\left\lVert\mathrm{U}\right\rVert$ is observed for the considered range of the fireline intensity $I$. This is a rather interesting result, since the normalization performed on the ROS model (i.e.~parameter $\alpha_{\text{w}}$ in Eq.~\ref{eq:byram}) makes the propagation of the deterministic fireline depending solely on the orientation of the wind vector and not on its magnitude. This means that the wind has a more general and fundamental role as reflected also in the enhancement of fire-spotting and secondary fire generation.

The ballistic term $\sigma$ in Eq.~\eqref{eq:lognormal} strongly depends on the value of $\left\lVert\mathrm{U}\right\rVert$. This is in line with the results of the second set of input parameters. In the case $\boldsymbol{\theta} = \left(\mu, \sigma, D \right)^T$, $\sigma$ is the most influential parameter when considering Sobol' indices, far above $D$ and $\mu$ (in order of relevance). The trend for the observables $A_t$ and $S_t$ is comparable, still $S_t$ gives slightly more relevance to $\mu$ and $D$ inputs than $A_t$. As expected, for both parameter sets, the mean of the $S_2$-observable is larger than that of $A_2$. Its STD is also larger. Uncertainties in $\left\lbrace \left\lVert\mathrm{U}\right\rVert, I, \tau \right\rbrace$ induce a more significant spread of the fireline position and shape compared to uncertainties in $\left\lbrace \mu, \sigma, D \right\rbrace$. This is due to the fact that in the first case we also vary the ember ignition time scale. 

In summary, these results highlight the importance of the mean wind factor, on the main fire propagation but also on the generation of secondary fires. This is consistent with the phenomenology of wildland fires and with the process of fire-spotting. In particular, fire-spotting refers to independent ignitions located far away from the main fireline. This occurs when the convective column lofts firebrands, the wind transports them up to their falling into the downwind vegetation and the firebrands ignite. The stronger the wind, the larger distance firebrands can be transported. This process is accounted in the model via the lognormal parameter $\sigma$. The importance of $\sigma$ is a proper mathematical feature of 
the adopted lognormal PDF for firebrand landing distance, since it controls the tail of the density function, the kurtosis of the lognormal density being  equal to $\displaystyle{\rm{e}^{4\sigma^2}+2\rm{e}^{3\sigma^2}+3\rm{e}^{2\sigma^2}-3}$. Hence this study shows that the new submodel correctly includes the double role of the mean wind, enhancing the propagation of the main fireline on the one hand, and carrying away firebrands for secondary ignitions on the other hand.

\begin{table}[h!]
\caption{Sobol' indices (first-order in black and total-order in gray) using LAR gPC-surrogate and linear truncation; $\boldsymbol{\theta} = \left( U, I, \tau \right)^T$; $N = 216$. Left: Sobol' indices associated with the burnt area ratio $A_2$. Right: Sobol' indices associated with the MSR ratio $S_2$.}
\centering
\begin{tabular}{cc}
\includegraphics[width=0.5\linewidth]{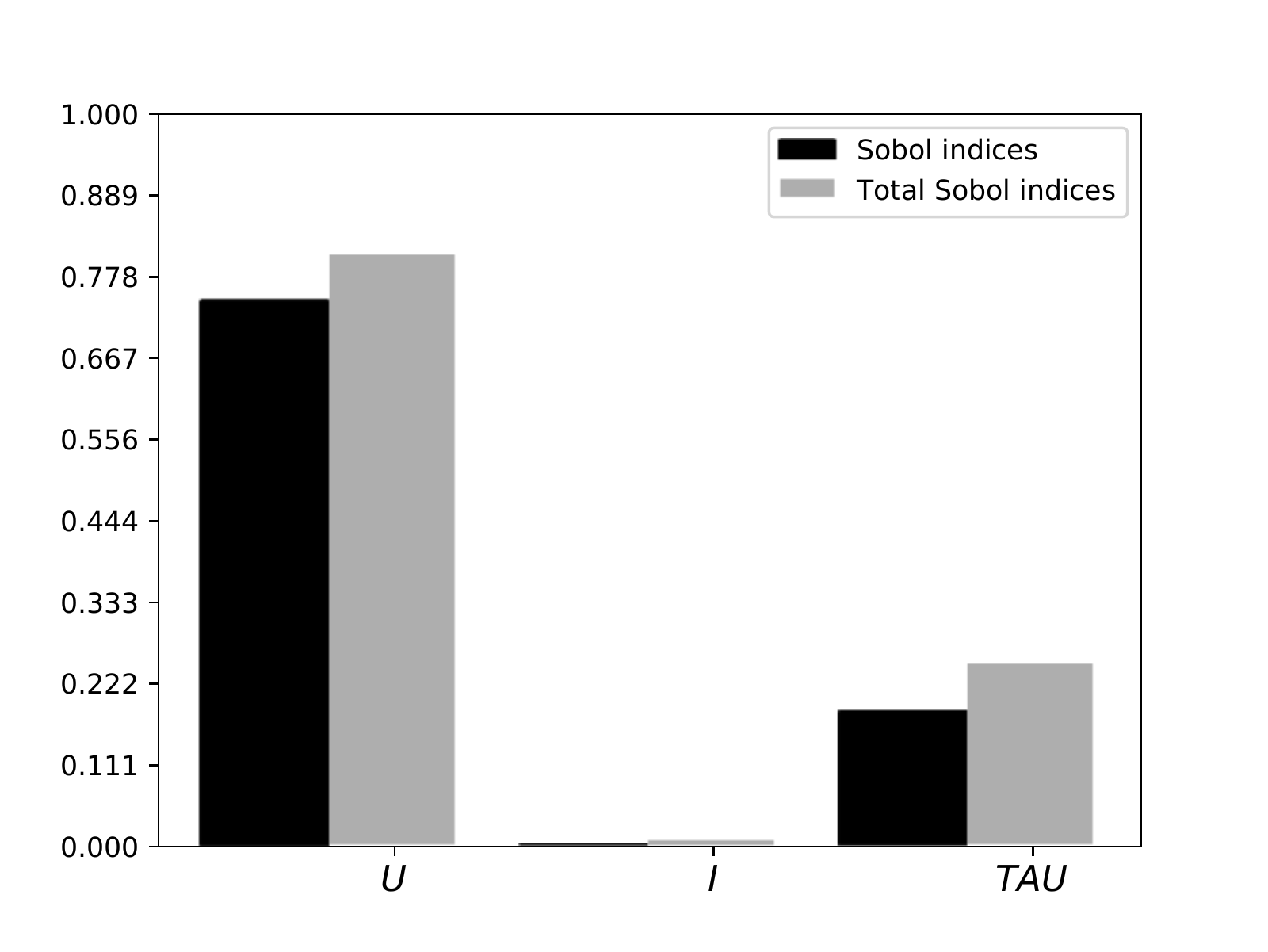} & 
\includegraphics[width=0.5\linewidth]{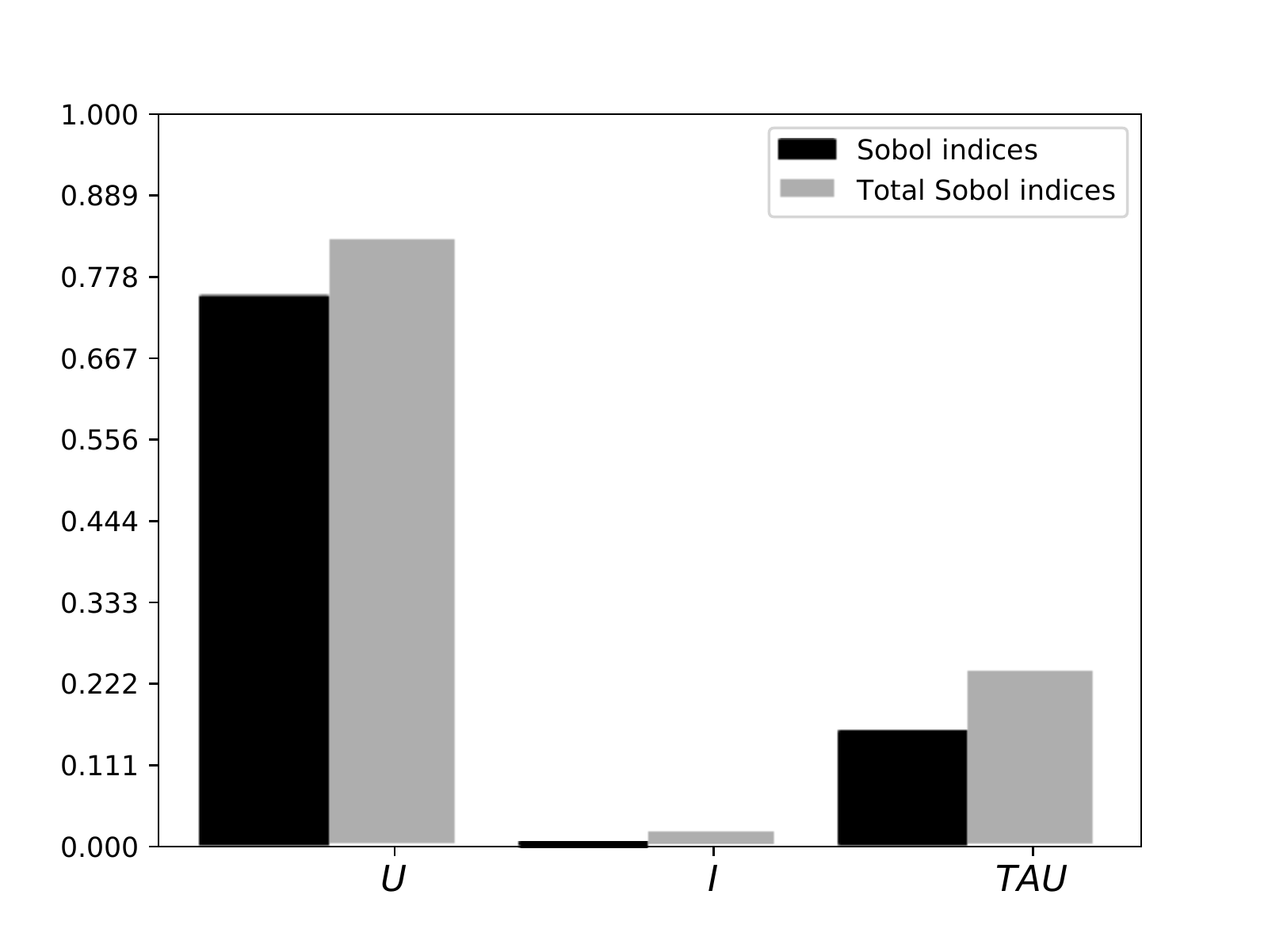}
\end{tabular}
\label{tbl:modelresultsUIT}
\end{table}

\begin{table}[h!]
\caption{Mean and STD of observables $A_2$ and $S_2$ as well as error metrics $\epsilon_{\text{emp}}$ and $Q_2$ using LAR gPC-surrogate and linear truncation; 
$\boldsymbol{\theta} = \left(U, I, \tau \right)^T$; $N = 216$.}
\begin{tabular}{c|cccc}
\hline
Quantity of interest & Mean & STD & $\epsilon_{\text{emp}}$ & $Q_2$ \\ \hline
$A_2$                     & 0.07 & 0.06 & $9 \cdot 10^{-4}$ & 0.95 \\ 
$S_2$                     & 0.19  & 0.13 & $2 \cdot 10^{-3}$ & 0.96 \\ 
\hline
\end{tabular}
\label{tab:res_x1}
\end{table}
\begin{table}[h!]
\centering
\caption{Same caption as in Table~\ref{tbl:modelresultsUIT} but for $\boldsymbol{\theta} = \left(\mu ,\sigma, D \right)^T$.}
\begin{tabular}{cc}
\includegraphics[width=0.5\linewidth]{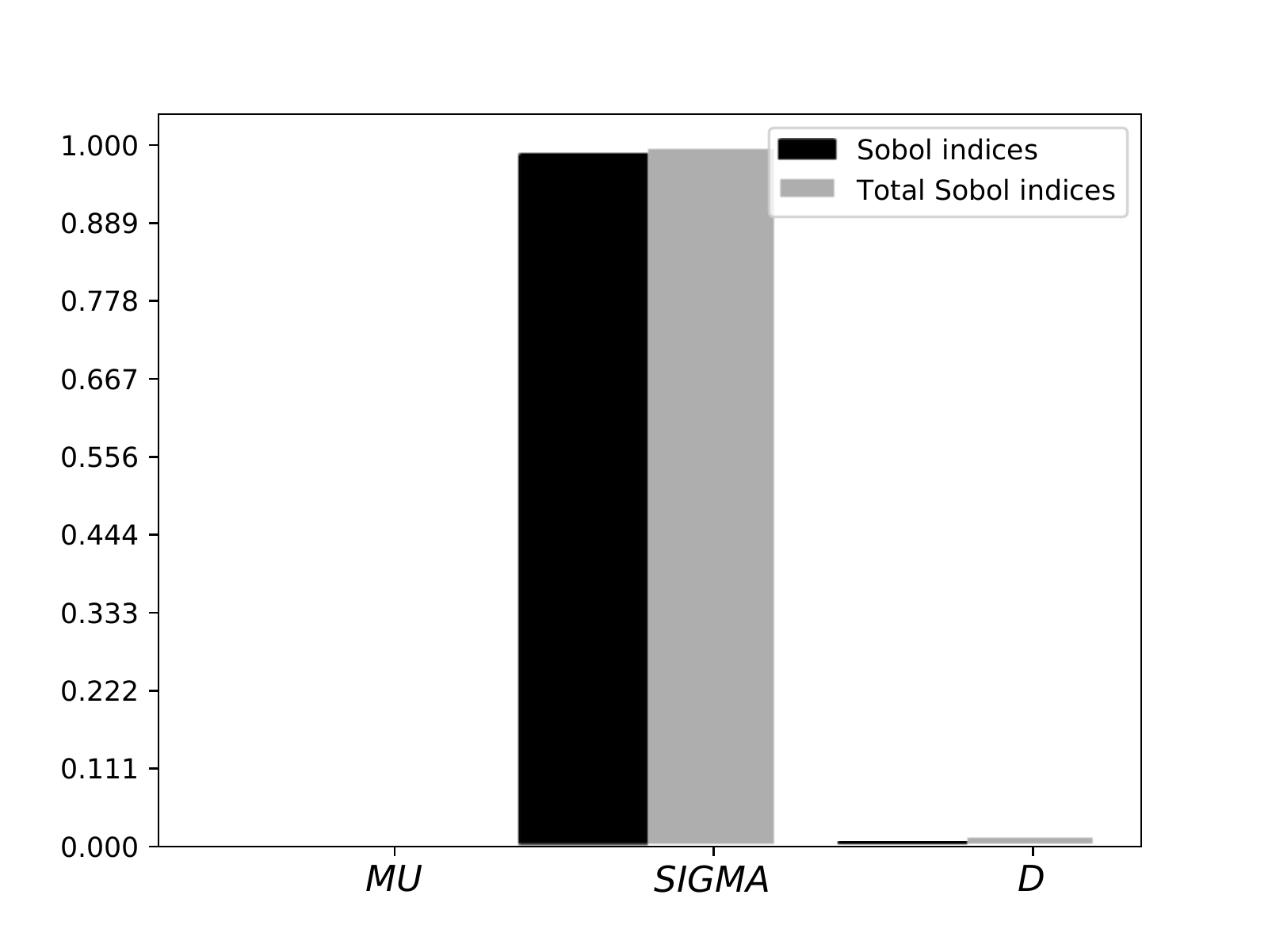} & 
\includegraphics[width=0.5\linewidth]{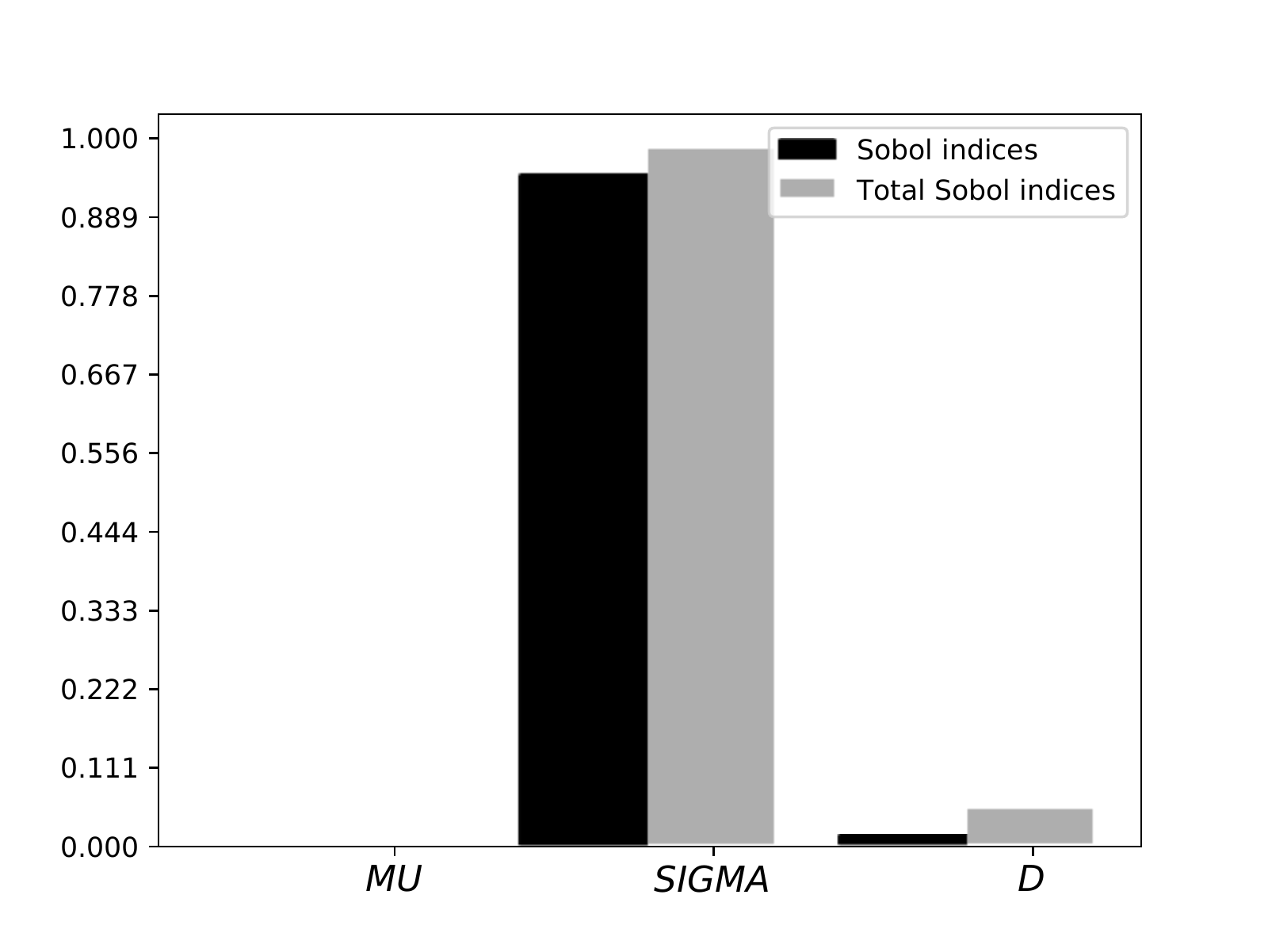} 
\end{tabular}
\label{tbl:table_of_figures0}
\end{table}

\begin{table}[h!]
\caption{Same caption as in Table~\ref{tab:res_x1} but for $\boldsymbol{\theta} = \left(\mu ,\sigma, D \right)^T$.}
\begin{tabular}{c|cccc}
\hline
Quantity of interest & Mean & STD & $\epsilon_{\text{emp}}$ & $Q_2$ \\ \hline
$A_2$ & 0.05 & 0.04 & $4 \cdot 10^{-4}$ & 0.97 \\  
$S_2$ & 0.11 & 0.11 & $2 \cdot 10^{-3}$ & 0.95 \\  
\hline
\end{tabular}
\label{tab:res_x2}
\end{table}

\section{Discussion and Conclusions}\label{sec:ccl}

This study presents an extensive comparative study of surrogate approaches to the nonlinear and multi-scale problem of turbulence and fire-spotting in wildland fire modeling, fire-spotting being a random process in which firebrand generation, emission and landing distance are intrinsically governed by the fire strength. A surrogate modeling approach is useful to analyze in a cost-effective way, how the fireline position and topology change according to variations in the input parameters for the new physical submodel introduced by Pagnini et al~\cite{pagnini2012,pagnini2014_smai,pagnini2014_nhess,kaur2016} based on a randomized representation of the fireline. Results are presented from both algorithmic and physical perspectives. From an algorithmic viewpoint, it is of interest to compare several approaches to carry out global sensitivity analysis and to select which ones are accurate and computationally efficient. From a wildland fire perspective, uncertainty quantification and sensitivity analysis is a good practice to analyze any new submodel, spot unimportant parameters and identify which parameters are dominant for obtaining a good representation of turbulence and fire-spotting.

In this work, fast surrogate models based on generalized Polynomial Chaos (gPC) and Gaussian Process (GP) were used to limit the required number of physical model evaluations to at least 100. We analyzed the performance of different formulations of the gPC-surrogate in terms of design of experiments (how to choose the training points? how many training points are required to achieve a certain accuracy?), polynomial basis structures (how to select the influential terms of the polynomial basis?) and projection schemes (how to compute the coefficients of the gPC-expansion?). The generalization error of these surrogates was classically estimated using the $Q_2$ predictive coefficient. Sparse gPC-methods have shown their accuracy in line with the GP model based on RBF kernel, but with a less cumbersome representation for Sobol' indices and statistical moments. Sparse methods provide more flexibility to select high-order polynomial terms in a given direction of the uncertain space, without requiring more physical model evaluations and therefore without increasing the computational cost of sensitivity analysis. The best performance for the gPC-surrogate was obtained using a sparse least-angle regression (LAR) with a training set built using a Halton's low discrepancy sequence. Using this approach, the new parametrization {\tt RandomFront 2.3b} for turbulence and fire-spotting was found to be a nonlinear model with a remarkable range of variations in the size and topology of the fire due to uncertainties in its input parameters. There is a clear dominance of the lognormal parameter $\sigma$ characterizing firebrand downwind transport and of the wind magnitude $\left\lVert\mathrm{U}\right\rVert$, which confirms that fire-spotting is a wind-driven, ballistic phenomenon. 

Several issues can be met when building a robust surrogate model. First, when the problem is multi-scale, i.e.~when uncertain parameters have correlation length-scales differing by several order of magnitudes. Sparse methods may filter out the less influential parameters. The LAR-based gPC surrogate was found to filter out the information coming from parameters with large length-scale. The cleaning-based surrogate proved to preserve these information, which may be important in a multi-scale problem such as fire-spotting. Second, when choosing how to sample the stochastic space and construct the training set. Standard projection schemes such as tensor-grid Gauss quadrature and standard least-square methods have shown their limitations: a large part of the training set was wasted in regions of the parameter space far from the nonlinear processes to be explored. In the opposite, sparse methods based on least-square projection were found to identify in which stochastic direction the physical processes are more complex and require higher order polynomials or high-order interaction terms. Using hyperbolic truncation was not flexible enough for this purpose. In any case, the present work shows the importance of comparing surrogate approaches for producing a reliable non-intrusive sensitivity analysis with a low budget (i.e.~a limited training set). Note that intrusive methods may also be powerful but require significant modifications of the model equations and thus of the legacy code, which is difficult to test for operationally-oriented simulators.  

The increasing strength and occurrence of megafires due to climate change calls for the development of new tools for the prediction of fire occurrence, growth and frequency at regional scales. Reliable wildland fire spread models are a promising approach to provide short-term variability of fire danger. Statistical methods such as uncertainty quantification and sensitivity analysis also have an important role to play~\cite{taylor_etal-ss-2013,sanmiguelayanz_etal-fem-2013,hernandez_etal-ag-2015}. Present work pushes toward the integration of fire-spotting into regional-scale operational wildland fire spread simulators. This is the main direction of the future developments of this research. Note that only a fire-atmosphere coupling system could provide the dynamics of secondary fires in agreement with the characteristics of the primary fires, which is important for emergency fire response. Future work will therefore include the integration of this turbulence and fire-spotting submodel into a coupled fire-atmosphere model. Future work will also include the extension of the surrogate approaches 
to vectorial inputs and outputs, in order to analyze the sensitivity of the fire behavior to a wind field and to describe the fire situation as a map instead of a scalar variable such as the burnt area or the minimum spanning rectangle.

\section*{Acknowledgements}
This research is supported by the Basque Government through the BERC 2014--2017 and BERC 2018--2021 programs, by the Spanish Ministry of Economy and Competitiveness MINECO through BCAM Severo Ochoa accreditation SEV-2013-0323 and
through project MTM2016-76016-R "MIP", and by the PhD grant "La Caixa 2014". The authors acknowledge EDF R\&D for their support on the OpenTURNS library. They also acknowledge Pamphile Roy and Matthias De Lozzo at CERFACS for helpful discussions on \emph{batman} and \emph{scikit-learn} tools.

\end{document}